\documentclass{article}
\usepackage[english]{babel}
\usepackage{amsmath}
\usepackage{amssymb}
\usepackage{varioref}
\usepackage{fullpage}
\usepackage{color}
\usepackage{graphicx}
\usepackage{float}

\newtheorem{lemma}{Lemma}

\usepackage{setspace}
\begin{document}

\title{\bf  Inflation  securities valuation with macroeconomic-based no-arbitrage dynamics }
\author{Gabriele  Sarais\\
  Department of Mathematics\\
    Imperial College London\\
    gs1608@imperial.ac.uk\\
  \\
 Damiano Brigo\\
  Department of Mathematics\\
    Imperial College London\\
    damiano.brigo@imperial.ac.uk\\
 Capco Institute \\
  \\
  \\
  Second version:
}
\maketitle

\begin{abstract}
We develop a model to price inflation and interest rates derivatives using continuous-time dynamics that have some links with macroeconomic monetary DSGE models equipped with a Taylor rule: in particular, the reaction function of the central bank, the bond market liquidity, inflation and growth expectations play an important role. The model  explains the effects of non-standard monetary policies (like quantitative easing or its tapering) and sheds light on how central bank policy can affect the value of inflation and interest rates derivatives. 

The model is built under standard no-arbitrage assumptions. Interestingly, the model yields short rate dynamics that are consistent with a time-varying Hull-White model, therefore making the calibration to the nominal interest curve and options straightforward. Further, we obtain closed forms for both zero-coupon and year-on-year inflation swap and options. The calibration strategy we propose is fully separable, which means that the calibration can be carried out in subsequent simple steps that do not require heavy computation.
A market calibration example is provided.

The advantages of such structural inflation modelling become apparent when one analysed the  risk of an inflation derivatives book: because the model explicitly takes into account economic variables, a trader can easily assess the impact of a change in central bank policy  on a complex book of fixed income instruments, which is normally not straightforward if one is using standard inflation pricing models.\end{abstract}

\\

\textbf{AMS classification codes}:
        60H10, 91B25, 91B51,  91B64, 91B70

\tableofcontents

\section{Monetary macroeconomic inflation models }

We consider a   macroeconomic  model  and propose a strategy to use it to price inflation derivatives: the main advantage of this approach is that the inflation dynamics are not taken exogenously but rather are the result of a well-established macroeconomic model with central bank policy. In particular, the co-movement of inflation and nominal interest rates is not taken as an input or modelled via a correlation structure (as it happens in many models currently used in the industry) but is the result of central bank policy, via  a well-known macroeconomic relationship (Taylor rule or some variation of it).

The task is not straightforward because
most macroeconomic literature is written in a somewhat less formalised way  compared to financial mathematics. Expectations are taken only with respect to the real-world econometric measure (which is known as  \(\mathbb{P}\) or as \textquotedblleft physical measure" in financial mathematics): therefore there is no apparent need to specify the measure used to take expectations, and  measure changes are not used. No mention is made
of filtrations, adapted processes, measurability. Distributional assumptions tend to be loose (randomness is usually introduced via some
so-called \textquotedblleft white noise", defined as a zero-mean process whose realisations are independent from each other over time). Stochastic processes tend to be assumed to reach a \textquotedblleft steady state", i.e. to converge to some equilibrium value in the long run: this level is always supposed to exist and to be finite. Sometimes variables are expressed as their percentage deviations with respect to their long term equilibrium level. Securities payoffs may be defined with  few details. 

Despite all these issues, this theory is the one that central bankers, economists, researchers, and market operators use and refer to: it can not be ignored.
The
challenges
that one faces to use this theory
in financial mathematics to carry out derivatives pricing are manifold: to complement the macroeconomic model with all the mathematical machinery that has been originally taken as a given in a way that the kernel of the model is not arbitrarily changed but is rather enhanced by an improved formalism. Further, when changes are made to the original model assumptions, these changes have to be not invasive and have to bear a clear advantage, especially in the calibration phase. At the same time, some approximations may be needed to derive some results that are essential for pricing (closed forms for nominal and inflation bonds, for example).

Further, on a recent article appeared on Bloomberg, finance professor Noah Smith suggests 
that financial institution should use more the DSGE macroeconomic model to fully understand the causality relationships in the economy (\textquotedblleft Wall Street Skips Economics Class", www.bloombergview.com/articles/2014-07-23/wall-street-skips-economics-class): this seems to confirm the intuition behind our work.

The article is structured as follows.
Firstly, we build
a general axiomatic framework
around the original macroeconomic discrete-time model: this entails specifying the feature of the time scale, probability space, and traded instruments. Then, for the benefit of the reader not expert in monetary macroeconomics, all the economic quantities and assumptions are  listed and defined. Secondly, we introduce  a standard monetary macroeconomics discrete-time model (DSGE model, or \textquotedblleft Dynamic Stochastic General Equilibrium" model) where some \(\mathbb{P}\)-dynamics for inflation are derived from optimality conditions and realistic market frictions.
Thirdly,
we derive the expression for the nominal rate and inflation rate volatility and higher order moments based on the DSGE model:
they turn out to be linear combinations for the volatilities and the higher order moments of the random processes used originally in the DSGE model; the advantage is clear, as one  chooses these parameters to match the moments of the distribution implied by market-traded options on interest rates and inflation. Fourthly, we obtain  approximated expressions for the nominal and inflation bonds, to calibrate the model to the observed nominal and inflation term structures.

To sum up, the first part of this article constitutes a  useful  attempt to bridge the gap between monetary macroeconomics and financial mathematics: as a result, we have built a toy pricing model based on the DSGE model. However, in the second part of this article we  suggest how the framework can be somewhat translated into continuous time to improve its analytical tractability and to take into account some very recent market features, like low interest rates and quantitative easing. Although there is no exact correspondence between the original discrete-time DSGE model and the newly-introduced continuous-time dynamics, the latter are clearly inspired by the former. Interestingly, we obtain Hull-White dynamics for the short rate in this model. We develop the theory in continuous time under no-arbitrage, and  provide some closed forms for common inflation and interest rates payoffs. Finally, we make an example of calibration of the continuous-time model to market data. To conclude, we discuss some pricing and risk management applications. \section{Introduction to DSGE models}
DSGE models are an essential tool for the working macro-economist: they are widely used both in academia and by central banks since they explain the short term real effects of  monetary policy. 
Strong empirical evidence supports the idea that money has real effects: DSGE models describe this effect by assuming a stochastic environment, optimizing behaviour and nominal rigidities in the economy. Consumers maximize their expected utility, which is based on consumption and real cash balances; firms maximize their expected profit stream but are not  able to change in each period the prices they charge.
The result is a discrete-time model where the macroeconomic variables are affected by their future expectations and some external shocks.
The short term nominal interest rate (\textquotedblleft short rate") is part of these dynamics. A further assumption is that the central bank uses a Taylor rule to set the short rate: this means that the short rate is changed in response to the other macroeconomic variables using a simple linear rule (see Taylor \cite{Label18}). This approach, albeit simple, has proven to be  powerful to explain the central bank behaviour.
 
Finally, we stress that so far we referred to DSGE models in plural as they can be regarded as a family of models that share the main features listed above:  consumer habits, capital, labour market rigidities, government, taxes, lagged variables, different central bank policies can be introduced in this framework, giving rise to more complex dynamics. Here we describe the baseline version of this model, which offers enough flexibility for our purposes.

We  present the  assumptions of a basic version of the DSGE  macroeconomic model, which  explains  the behaviour of the inflation rate \(p_{i}\) and the output gap \(x_{i}\) based on a general description of the economy. A  complete description of this model  can be found in this section or in  Walsh \cite{Label20}, which we follow to present the model.
However, before presenting the macroeconomic model, we specify the axiomatic foundations that are implicit  in the model and that  are normally not fully specified by economists.
\subsection{Axiomatic foundations}

\subsubsection{Time scale}The model is set in discrete time, where time is a non-negative variable: \(t_{i}\in
\mathbb{T}=\{t_{0},t_{1},t_{2},...,t_{n},...\}, n\in\mathbb{N}\).

Here \(t_{0}\) is the present time. To preserve generality, the discrete-time points are not required to be equally spaced. For a  variable \textit{y} at time \(t_{i}\) we often write \(y_{i}\)  to make the notation lighter: similarly, the discrete-time stochastic processes \(\left\{y_{t_{i}}\right\}_{i=0,1,   ...}\) can be denoted by \(\left\{y_{i}\right\}\). 
\subsubsection{Probability space}
We work with the  probability triplet \(\{\Omega, \mathcal{F}, \mathbb{P}\} \) and assume the existence of a market filtration \(\{\mathcal{F}_{t_{i}}\}_{i\geq0}\). In particular \(\mathbb{P}\) is the real-world (\textquotedblleft physical") probability measure. All filtration-related concepts (mainly the martingality property) are defined with respect to the market filtration. To simplify the notation, in discrete time we use the following notation for  conditional expectations: \(\mathbb{E}_{i}x_{i+j}=\mathbb{E}_{i}[x_{i+j}]=\mathbb{E}^\mathbb{P}[x_{t_{i+j}}\mid\mathcal{F}_{t_{i}}]\). 

Therefore, in this section if no probability measure is specified, the expectation is taken with respect to the real-world measure (\(\mathbb{P})\).  To perform a measure change from the physical measure \(\mathbb{P} \) to the risk-neutral measure \(\mathbb{Q}\), we  introduce the Radon-Nikodym derivative  \(({d\mathbb{Q}}/{d\mathbb{P}})\mid_{t_{i}}\), written \(\mu_{i}\) for brevity.
All regularity requests for the measure change process to exist and to be an \(L^{2}\) positive martingale are supposed to hold.
\subsubsection{Financial instruments}
We assume that the financial market is such that there are no transaction costs nor taxes: investors can take any position (either long or short) in any asset.  We assume the existence of the following:
\begin{enumerate}
\item 
The short term nominal interest rate \(n_{i}\) -- set by the central bank -- is the interest agreed at time \(t_{i-1}\) and paid at time \(t_{i}\) by the bank account on the balance at time \(t_{i-1}\)\footnote{Here we are assuming that the central bank lends money to the commercial banks at the same interest rate paid by these to the money market account holders. We are making the simplifying assumption that the central bank reviews its interest rate with the same time scale by which the interest are accrued in the money market account. This assumption let one  include in the pricing model a fairly realistic description of central policy.}. It is used to discount payments. The short term nominal interest rate process \(\left\{n_{i}\right \}_{i=0,1,...}\) is a previsible process, i.e. the short term nominal interest rate \(n_{i}\) is  \(\mathcal{F}_{i-1}\)-measurable.
\item 
The bank account \(B_{i}=\prod^{i}_{j=1}\)\(({1+\tau_{j}n_{j}}\)), with \(B_{0}=1\). Here \(\tau_{i}\) represents the year fraction between  times \(t_{i-1}\) and \(t_{i}\). Since the interest rate \(n_{i}\) is  \(\mathcal{F}_{i-1}\)-measurable, the bank account process \(\left\{B_{i}\right\}_{i=0,1,...}\) is a previsible process. At time  \(t_{i-1}\) the cash flow that  occurs at time \(t_{i}\) is already known: this is why the bank account is often referred to as the riskless asset.
\item The first two properties implicitly imply a lower bound on negative short nominal interest rates: in this model rates can be negative (as they have been in 2012, for example German Bunds up to 2 years maturity, and the European Central Bank has set the deposit rate to -0.1\% in 2014), however they can not be lower than -100\%, otherwise the nominal bank account would have a negative value, which is not possible. Rational agents would not put money into such account that turns assets into liabilities.
\item 
A system of discount bonds \(P(t_{i},t_{N})  \), that pay one unit of currency at time \textit{t\(_{N}\)} and have the following properties:
\begin{itemize}
\item 
\(P(t_{i},t_{N})=\mathbb{E}_{i}^{\mathbb{Q}}\left[    \prod^{N}_{j=i+1}{(1+\tau_{j}n_{j})}^{-1}\    \right]\)
\item 
\(P(t_{i},t_{i})=1,  \;\forall  i\)
\item 
\(P(t_{i},t_{N})>0, \;\forall i\leq N\  \)
\item 
\( \lim_{N\rightarrow+\infty} P(t_{i},t_{N})=0 \) 
\item \(P(t_{i},t_{i+1})=B_{i}/B_{i+1}\).
\end{itemize}\item 
The seasonality-adjusted price index process \(\left\{I_{i}\right\}_{i=0,1,...}\) that describes the evolution of the price level\footnote{Price indices time series clearly show  seasonality, mainly driven by sales in January and July and prices increases around Christmas. We do not model these patterns directly at this stage because  a seasonality correction can be easily introduced at the last stage. This can be done by assuming that the monthly inflation rate differs from the seasonality-adjusted inflation rate by a certain percentage. Intuitively, seasonality is more relevant for short-maturity inflation trades.}.

\item 
A system of zero-coupon inflation index swaps (ZCIIS), such that the floating leg pays   \(({I_{i+M-1}}/{I_{i}})-1 \) and the fixed leg pays \((1+X_{M})^{M\tau_{i}}-1\). Both payments happen at maturity. The  strikes \(X_{i}\) are quoted at time \(t_{i}\)  for all maturities \(t_{M}>t_{i}\).
Inflation payments are time-lagged in this model as it happens in reality: in fact the price index is subject to revisions and in practice ZCIIS pay the inflation lagged by one period.\item A system of index-linked zero-coupon bonds \(P^{I}(t_{i},t_{M})\), which pay at maturity \(t_{M}\) the cash equivalent of the price index \(I_{M-1}\). These bonds are priced consistently with the zero-coupon inflation swaps presented in the previous point. Inflation payments are time-lagged in this model as it happens in reality: in fact the price index is subject to revisions and in practice the inflation bonds pay the inflation lagged by one period. These bonds are quoted at time \(t_{i}\)  for all maturities \(t_{M}>t_{i}\).
We ignore the deflation floor that commonly traded inflation bonds have.\end{enumerate}

\textbf{Macroeconomic variables.} The inflation rate is defined by \(p_{i}=((I_{i}/I_{i-1})-1)/\tau_{i}\). This is the annualised percentage growth rate of the price index.

The output gap \(x_{i}\) is defined as the difference between the actual and the potential log-linearised growth rate of the economy\footnote{The reason why we are involving log-linearisation will become clear shortly. More information is also available later in this section. }: \(x_{i}= \hat y_{i}-\hat y_{i}^{f}\) . To provide a complete definition of the output gap, we introduce  the Gross Domestic Product (GDP) \(Y_{i}\) -- also known as output -- which is the value of all final goods and services produced in the economy between times \(t_{i-1}\) and \(t_{i}\). The GDP annualised growth rate is defined as: \(y_{i}=((Y_{i}/Y_{i-1})-1)/\tau_{i} \).
The growth rate \(y_{i}\) is assumed to have a long term equilibrium level \(\bar y\) such that \( \mathbb{E}(y_{i})\rightarrow \bar y\) as \(i\rightarrow+\infty\). The variable \(\hat y_{i}\) is defined as the percentage deviation between the GDP growth rate \(y_{i}\) and its long term equilibrium level \(\bar y\): \( \hat y_{i}=((y_{i}/ \bar y)-1).  \) Economists often refer to  it  as the log-linearised GDP growth rate, as \( \hat y_{i}=((y_{i}/ \bar y)-1)\cong log(y_{i}/ \bar y).  \)   

If we assume that the economy is subject to some \textquotedblleft inefficiencies", we can introduce the potential GDP \(Y^{f}_{i} \), which can be defined as the GDP  produced if there is no inefficiency: intuitively these inefficiencies prevent the actual GDP \(Y_{i}\) from reaching the \textquotedblleft full employment" GDP \(Y^{f}_{i}\). Therefore we  similarly derive the variables \(y_{i}^{f}\), \(\bar y_{i}^{f}\), and \(\hat y_{i}^{f}\), which complete the definition of the output gap \(x_{i}\).

We assume that the processes \(\left\{Y_{i}\right\}_{i=0,1,...}\), \(\{Y^{f}_{i}\}_{i=0,1,...}\), and \(\left\{I_{i}\right\}_{i=0,1,...}\) are adapted, therefore also the processes \(\left\{ x_{i} \right\}_{i=0,1,...}\) and \(\left\{ p_{i} \right\}_{i=0,1,...}\) are adapted too. To complete the formalisation, one needs to assume that all stochastic processes involved in the model converge in some sense to a finite equilibrium level when time tends to infinity. No further specification of such convergence is normally made in the macroeconomic model.

\textbf{Economic assumptions.} We list the microeconomic and macroeconomic assumptions:
\begin{enumerate}
\item 
The economy is closed, i.e. there is no exchange rate nor foreign market.
\item 
All markets are in equilibrium, i.e. demand matches supply for all goods and services markets.
\item 
The economy is a monetary one, i.e. there is no barter.
\item 
The representative consumer maximizes his utility function under an intertemporal budget constraint.
\item 
The representative consumer draws his utility from consuming and keeping cash balances for safety (money-in-utility approach).
\item 
There is no public sector, therefore there is no taxation.
\item 
Labour is the only production factor: no capital is required, therefore there are no investments.
\item 
Savings are invested in bonds that pay a nominal interest rate. \item 
The representative consumer consumes multiple goods, each  produced in a monopolistic market.
\item
The output  coincides with the private consumption, as there is no government expenditure, no import/export, no taxes nor investment.
\item 
Firms maximize  profits but are not free to modify in each period the prices they charge (sticky prices).
 \item 
The central bank sets the short rate as a linear function of  inflation and  output gap (Taylor rule).
The short rate moves around its equilibrium level.
\item The short rate can be negative in some circumstances.
\item 
There exists a system of expectations for the output gap and inflation.
\item 
Time is discrete.\item 
There is no credit risk.
\end{enumerate}
\subsection{Model derivation}
We follow Walsh \cite{Label20} to  introduce the main equations of the DSGE baseline model: as the material of this section is standard, we give a high level overview. Another interesting overview  can  be found in Clarida, Gali \& Gertler \cite{Label5}. The reader who is already familiar with this model can skip this section.\subsubsection{Economy description}
The baseline model we  work with represents a  simple closed economy, with no government and no tax system. The production function depends only on labour since capital is not considered: therefore there is no investment. From a macroeconomic  perspective we can therefore state that the output at time \(t_{i}\) equals the aggregate consumption at time \(t_{i}\):
\begin{displaymath}
Y_{i}=C_{i} \label{Output2}.
\end{displaymath}
The economy is a monetary one with money \(M_{i}\) and  price level  \(I_{i}\). \subsubsection{Consumers}
The representative household solves a two-steps optimisation problem. It first decides how to allocate its total consumption  between different goods -- produced by monopolistically competitive final goods producers (firms) -- and then chooses how much to consume in total, how much cash to hold, how much to  invest in bond holdings and how many hours to work.

In the first step we assume the existence of a continuum of goods \(c_{j}\)
produced by a continuum of monopolistic firms \textit{j} (\(j\in[0,1])\).
At time \(t_{i}\) the
household chooses  the combination of goods \(c_{ji}\) that minimizes the cost of the total consumption:
\begin{displaymath}
\textrm{min} \int^{1}_{0}p_{ji}c_{ji}dj
\end{displaymath}
by taking into account the constraint\begin{displaymath}
\left(\int^{1}_{0}(c_{ji})^\frac{\theta-1}{\theta}dj\right)^\frac{\theta}{\theta-1}\geq C_{i}.
\end{displaymath}
Here \(p_{ji}\) is the price of the good \textit{j} at time \(t_{i}\) and \(C_{i}\) is the total consumption time \(t_{i}\). The parameter \(\theta \) is used to model the price elasticity, i.e. how price-sensitive consumption is.
This standard optimisation problem is solved in Walsh \cite{Label20} (p. 233) and yields the optimal amount of consumption of good \textit{j} given the general price level \(I_{i}\), the total consumption \(C_{i}\) (to be determined in the next step) and the price of good \textit{j,} \(p_{ji}\):\begin{equation}
c_{ji}=\left(  \frac{p_{ji}}{I_{i}} \right)^{-\theta}C_{i}.
\label{Demand Curve}\end{equation}
The second step is modelled as an intertemporal maximisation of the expected utility under a budget constraint, and yields the usual Euler conditions.

 The representative household draws its utility from  consuming (\(C_{i}\)) and  holding real cash balances (\({M_{i}}/{I_{i}}\)) as insurance against uncertainty:
furthermore it has negative utility  from supplying labour \(N_{i}\) and can save money and purchase bonds \(B_{i}\) that pay a nominal interest rate \(n_{i}\) in each period. 
We assume a power utility function: the problem is therefore to find the sequences \(C_{i}\), \(M_{i}\), \(B_{i}\) and \(N_{i}\) that solve the problem
\begin{displaymath}
\textrm{max}\sum^{\infty}_{t_{i}=t_{0}}\beta^{t_{i}}\mathbb{E}_{0}\left[ \frac{C^{1-\sigma}_{i}}{1-\sigma}+\frac{\alpha}{1-d}\left( \frac{M_{i}}{I_{i}} \right)^{1-d}-\frac{N^{1+\eta}_{i}}{1+\eta} \right].
\end{displaymath}

The parameters \(\sigma\), \textit{d}, \(\alpha>0,\) \(\eta\)  
indicate how consumption, real cash balance and labour supply influence the utility function.
The expectation 
\(\mathbb{E}[\:\cdot]\:\)
is taken with respect to the physical measure 
\(\mathbb{P}\), as usual in any macroeconomic model: in this section when no measure is specified it is assumed that the physical measure is used. The parameter \(\beta\) \(\in(0,1] \) represents as usual a subjective discount factor over one period.
The parameter \(\sigma\), that is also known as  \textquotedblleft relative risk aversion", is used to model elasticity of utility to consumption in a constant relative risk aversion (CRRA) utility function. When \(\sigma\) is very high,  the agents are extremely risk-averse, as an increase in consumption creates an smaller increase in utility than the correspondent reduction in utility given the same absolute reduction in consumption; when \(\sigma\) is  zero, there is risk-neutrality, i.e. the utility grows linearly with consumption; when \(\sigma\) tends to 1, the utility function becomes a logarithmic utility function, which is moderately risk averse. 

The optimisation is carried out under the constraint that the total wealth at time \(t_{i}\) (which is allocated between consumption, real cash balance and bond holding) has been derived from the previous period or gained from supplying labour (\(W_{i}\) is the wage gained for 1 unit of labour at time \(t_{i}\)).  No wealth is introduced into the system \textit{ex nihilo}:
\begin{displaymath}
C_{i}+\frac{M_{i}}{I_{i}}+\frac{B_{i}}{I_{i}}=\frac{W_{i}N_{i}}{I_{i}}+\frac{M_{i-1}}{I_{i}}+\frac{B_{i-1}}{I_{i}}(1+n_{i-1}).
\end{displaymath}
The derivation of the Euler conditions is standard and can be found for example in the second chapter of Walsh \cite{Label20}. The first order conditions for this problem are the following:
\begin{equation}
C^{-\sigma}_{i}=(1+n_{i})\beta\mathbb{E}_{i}\left[ \frac{I_{i}}{I_{i+1}} \right]C^{-\sigma}_{i+1}
\end{equation}
\begin{equation}
\alpha\left(  \frac{M_{i}}{I_{i}}\right)C^{\sigma}_{i}=\frac{n_{i}}{1+n_{i}}
\end{equation}
\begin{equation}
\frac{N^{\eta}_{i}}{C^{-\sigma}_{i}}=\frac{W_{i}}{I_{i}}.
\end{equation}
Because the utility function is concave, the first order conditions are sufficient to find maxima. Since we assume that there is no government, no capital stock  (and therefore no investment) and the economy is closed, we  substitute the consumption with the output definition \(Y_{i}=C_{i}\), getting
\begin{displaymath}
Y^{-\sigma}_{i}=(1+n_{i})\beta\mathbb{E}_{i}\left[ \frac{I_{i}}{I_{i+1}} \right]Y^{-\sigma}_{i+1}.
\end{displaymath}
This condition may be rewritten
in log-linearized terms around a zero inflation equilibrium point, after making some approximations:\begin{equation}
\hat y_{i}=\mathbb{E}_{i}\hat y_{i+1}-\frac{1}{\sigma}(\hat n_{i}-\mathbb{E}_{i}p _{i+1}).
\label{GDP dev Dynamics}\end{equation}
 The inflation rate \(p_{i}\) is defined as the annualised relative change of the price level \(I_{i}\) from \(t_{i-1}\) to  \(t_{i}\).\footnote{It is worth explaining how the log-linearisation used above works.
Given a variable \(F_{i}\) at time  \(t_{i}\) we assume that its long term equilibrium level is  \(\bar F\) (i.e. that the limit of \(F_{i}\) when time goes to infinity is  \(\bar F\)).
With the lower case hat \(\hat f_{i}\) we  indicate the deviation at time  \(t_{i}\) of the variable  \(F_{i}\) from its long term equilibrium level \(\bar F\) in percentage terms: this  can be approximated with the natural logarithm of their ratio for small deviations. In formulas:
\begin{displaymath}
\hat f_{i} =\frac{F_{i}}{\bar F}-1\cong   \log(\frac{F_{i}}{\bar F}).\label{LogLin}
\end{displaymath}  
Uhlig \cite{Label19} gives extensive explanations and examples of this technique: given the analogy between this transformation and the natural logarithm, products can be approximated with sums, powers become multiplicative coefficients and constants disappear.
}\subsubsection{Firms}
The firm profit maximisation problem has to take into account three constraints: the demand curve, the production technology and  price stickiness.
It involves finding the optimal amount of labour \(N_{i}\) to minimize the production cost and the optimal good price \(p_{ji}\) to maximize the expected profit stream.
The demand curve is  \eqref{Demand Curve}. Secondly, technology is such that the output of the \textit{j}-th firm depends only on labour \(N_{ji}\)
\begin{displaymath}
c_{ji}=Z_{i}N_{ji}.\label{ProdFunction}
\end{displaymath}
\(Z_{i}\) is a positive random variable with mean 1 that represents a stochastic productivity shock.
Thirdly, firms are able to adjust their prices in each period only with probability 1 - \(\omega\).
This price stickiness assumption is the most interesting one and is essential to define the inflation dynamics of this model.

The first consequence is that 
the output \(Y_{i}\)  deviates from the output in flexible prices \(Y^{f}_{i}\): by making use of \eqref{LogLin}, we can then define their difference in log-linearized terms as the output gap
\begin{displaymath}
x_{i}=\hat y_{i}-\hat y_{i}^{f}.\label{OutputGap}
\end{displaymath}
We do not explain    the subsequent  details: instead we develop some intuition of the inflation mechanics. Since prices are sticky and firms are maximizing their expected profit stream, firms   increase their prices not only if  production costs rise (which would also happen in a flexible prices framework), but also to compensate for the expected losses they can face as they may not increase prices in the future (with probability \(\omega\)).

This
has two important consequences: firstly, as prices influence output via the demand curve
\eqref{Demand Curve} and the macroeconomic identity \(Y_{i}=C_{i}\), inflation is related to the output gap. The output gap increases with inflation. Secondly, if there are inflation expectations,  firms  raise  prices in the current period because they may not be able to do so in the future. Inflation is therefore a self-fulfilling prophecy.

The result, after some algebraic manipulations,  is the so-called neo-Keynesian Phillips curve, which states that the current level of inflation depends both on inflation expectations and the output gap:
\begin{equation}
p_{i}=\mathbb{\beta E}_{i}p_{i+1}+kx_{i}.
\label{Inflation Dynamics}\end{equation}
The parameter \(k\geqslant0\) can be regarded as a measure of the market price flexibility and is defined as
\begin{displaymath}
k=\frac{(1-\omega )(1-\beta \omega )(\sigma+\eta)}{\omega}.\label{kappa}
\end{displaymath}
It is worth stressing that if prices never change, \(\omega=1 \): therefore   \(k\) equals zero and inflation is only  driven by expectations. As before, the derivation can be found in Walsh \cite{Label20} (5.4, 5.7).
\subsubsection{Putting things together}
Equation \eqref{GDP dev Dynamics} can be rewritten in terms of output gap \(x_{i}=\hat y_{i}-\hat y_{i}^{f}\): defining
\begin{displaymath}
u_{i}=\mathbb{E}_{i}\hat y_{i+1}^{f}-\hat y_{i}^{f},
\end{displaymath}
we get to a final form for \eqref{GDP dev Dynamics} that can be put in a system with \eqref{Inflation Dynamics}
\begin{equation}
x_{i}=\mathbb{E}_{i} x_{i+1}-\frac{1}{\sigma}(\hat n_{i+1}-\mathbb{E}_{i}p _{i+1})+u_{i}.
\label{Gap Dynamics}\end{equation}
As we define the rate \(n_{i+1}\) as the rate set by the central bank at time \(t_{i}\) and paid at time  \(t_{i+1}\) \(\)we have written \(\hat n_{i+1}\) rather than  \(\hat n_{i}\): we reconcile the standard DSGE model with the request that the short rate is previsible: this has no significant impact on the following model construction. This curve can be interpreted as a neo-Keynesian demand curve, where the output gap shows a negative dependency on a function of the real interest rate. The process \(\{u_{i}\}_{i=0,1,...}\) can be thought as a  discrete-time stochastic process that relates the level of the log linearised flexible price output deviation from its expectations: this difference should depend somehow on the productivity shock \(Z_{i}\), but for our purposes we can think of it as a general stochastic process. Again, we stress that the original macroeconomic model does not make any further assumption on the shock processes: we  make the necessary assumptions this problem in the following sections, where the DSGE model is used for pricing purposes.
\subsubsection{The Taylor rule and the central bank}
Equations \eqref{Gap Dynamics} and \eqref{Inflation Dynamics} define a discrete-time, bi-dimensional, forward looking stochastic system which is influenced by two exogenous variables: the log-linearized short rate \(\hat n_{i+1}\) and the process \(\{u_{i}\}_{i=0,1,...}\), related to the productivity shock. We introduce the central bank, which uses the short rate as policy instrument.  In each period the central bank changes the short rate  in response to the  inflation and output gap with the following rule:
\begin{equation}
\hat n_{i+1}=\delta_{\pi}p_{i}+\delta_{x}x_{i}+v_{i}.\label{TaylorRule}
\end{equation}
This rule, proposed by Taylor \cite{Label18}, states that the central bank responds  to inflation and output gap by setting the short rate: a discrete-time stochastic process  \(\{v_{i}\}_{i=0,1,...}\), independent from the process \(\{u_{i}\}_{i=0,1,...}\), is added to increase the flexibility of the model. We remind the reader that the rate \(n_{i+1}\) is set by the central bank at time \(t_{i}\) and paid at time  \(t_{i+1}\): for this reason we allow a lag in the above form of the Taylor rule.  At this stage we also notice that the short rate can be negative in this formulation, which is consistent with the assumptions we have made earlier: however values of the nominal rate below -100\%, albeit theoretically possible, are to be ruled out under a reasonable model parametrisation.
Finally one notes that the Taylor rule has been defined for \(\hat n_{i+1}\), which, as explained for the other variables, is the percentage deviation of the nominal rate from its equilibrium level.

 Bullard \& Mitra \cite{Label4} analyse similar rules with more realistic timing assumptions (the central bank may be reacting to future expectations of gap and inflation, or may be looking at their lagged values instead).
In addition, the short rate can be smoothed as suggested by Woodford \cite{Label21}, essentially by combining \eqref{TaylorRule} with an autoregressive process. 
This framework is somehow simple, as the central bank is not optimizing any objective function: however, it has  explained the behaviour of the FED in the last decades, as shown by Clarida, Gali \& Gertler \cite{Label51}. Finally, this linear rule can be regarded as good linear approximation of the optimal policy solution.
\subsection{System stability}
If the Taylor rule \eqref{TaylorRule} is plugged into \eqref{Gap Dynamics} and \eqref{Inflation Dynamics}, we  obtain the following system:
\begin{equation}
\begin{bmatrix}x_{i} \\
p_{i}\ \\
\end{bmatrix}
=\frac{1}{\sigma+\delta_{x}+k\delta_{\pi}}
\left(
\begin{bmatrix}\sigma\ & 1-\beta\delta_{\pi}\ \\
k\sigma\ & k+\beta(\sigma+\delta_{x})\ \\
\end{bmatrix}
\mathbb{E}_{i}
\begin{bmatrix}x_{i+1} \\
p_{i+1}\ \\
\end{bmatrix}
+\begin{bmatrix}1 \\
k \\
\end{bmatrix}
\left(\sigma u_{i}-v_{i}\right)
\right).
\end{equation}\label{DSGE_full_dyn}

 The notation is made more compact by defining:
\begin{displaymath}
A=\frac{1}{\sigma+\delta_{x}+k\delta_{\pi}}
\begin{bmatrix}\sigma\ & 1-\beta\delta_{\pi}\ \\
k\sigma\ & k+\beta(\sigma+\delta_{x})\ \\
\end{bmatrix}
\end{displaymath}
\begin{displaymath}
K=\frac{1}{\sigma+\delta_{x}+k\delta_{\pi}}
\begin{bmatrix}1 \\
k \\
\end{bmatrix}
\end{displaymath}
\begin{displaymath}
\xi_{i}=\begin{bmatrix}x_{i} \\
p_{i}\ \\
\end{bmatrix}
\end{displaymath}
\begin{displaymath}
w_{i}=\left(\sigma u_{i}-v_{i}\right).
\end{displaymath}
 
Using the above definitions we  get  a more compact expression of the system:
\begin{equation}
\xi_{i}=A\mathbb{E}_{i}\xi_{i+1}+Kw_{i}.\label{System1}
\end{equation}
 We investigate the stability conditions, which is equivalent to ask what reaction function -- characterised by the parameters \(\delta_{\pi} \) and \(\delta_{x} \) -- keeps the economy on a stable path. For example, if the central bank only responds to inflation (i.e. \(\delta_{x}=0\)), we  ask whether \(\delta_{\pi} \) has to be greater or lower than one, i.e. if the central bank has to increase the short rate above its equilibrium level more or less than the realised inflation. Clarida, Gali \& Gertler \cite{Label51} show that \(\delta_{\pi}\)\(>1  \) is typical of the FED during the Volker tenure (in the early 1980s in the U.S.), which was characterized by lower inflation and output volatility.

The economic intuition is that a reaction parameter close to one means that the nominal rate is increased by the same  amount of inflation, thus keeping the real rate unchanged and not stimulating the economy. Bullard \& Mitra \cite{Label4} find that in general the system is stable if
and only if\begin{equation}
k(\delta_{\pi}-1) +(1-\beta)\delta_{x} >0\label{StabilityCondition}.
\end{equation}
 They obtain this rule by requiring that both eigenvalues of \(A\) lie inside the unit circle.
This  request is derived also by Blanchard \&\ Khan \cite{Label111} and used by Flashel \& Franke \cite{Label101} or  Walsh \cite{Label20}.
\section{Using the DSGE for pricing purposes}
\subsection{Arbitrage-free pricing}

The set-up introduced is standard (we have followed Walsh  \cite{Label20}), however it lets one to use a DSGE macroeconomic model  to price inflation derivatives in a no-arbitrage framework with a few minor changes. In general, the pricing kernel \(\psi_{i}\) properties discussed for example in  Constantinides \cite{Label7}   enable one to write the present value at time \(t_{i}\) of a derivative \(h_{i}\) paying the inflation-linked payoff \(H^{\pi}_{N} \) at time \(t_{N} \) in the form:
\begin{displaymath}
h_{i}=\mathbb{E}^{\mathbb{P}}[\psi _{N}H^{\pi}_{N}|\mathcal{F}_{i}]\frac{1}{\psi_{i}}.
\end{displaymath}

Here we  build a toy pricing model in discrete time that is based on the DSGE  model.

\subsubsection{Use of the macroeconomic model: inputs and outputs}

We  make a distinction between  input parameters  (the structural parameters of the DSGE model, equilibrium nominal rates, inflation expectations, output-gap expectations), and   calibrated parameters  (the volatilities and the market prices of risk, introduced later).

Calibrating  the market prices of risk
is not a usual procedure in derivatives pricing, because  the real-world drift is not an input   in the classic Black-Scholes framework   to price contingent claims: however, the DSGE model  takes expectations (under the \(\mathbb{P} \) measure) as an input. Since these expectations play the role of the drift in \eqref{System1} we need to take both inflation expectations and market implied levels (from the zero-coupon inflation swaps, for example) to calibrate the market prices of risk. The expectation of inflation is a kind of self-fulfilling prophecy: if there are   expectations of inflation, then  inflation will rise. This exercise is particularly useful for  inflation markets, since it is often observed that inflation forecasts and expectations can significantly differ from  levels of inflation calculated on a forward basis. Such differences can arise both because of risk aversion and market supply and demand factors: the market can be to a significant extent a \textquotedblleft one-way street",  overall \textquotedblleft short" inflation. In other words, market participants on the whole wish to hedge themselves against inflation. In particular, pension funds liabilities have to be covered.

 The idea of using market forecasts as model input, although not commonly used in standard derivatives pricing, lets one  use a theoretically consistent macroeconomic model for the   pricing of inflation derivatives.

 The algorithm we suggest  calibrates to both the nominal term structure and  the zero-coupon inflation index swaps (ZCIIS),  leaving much flexibility to calibrate to  market smiles.
To achieve this we explore the statistical properties of the main economic variables, as implied by the DSGE model presented above.
\subsubsection{Statistical properties of the inflation rate}
From equation \eqref{System1} we  write explicitly the dynamics of the   inflation rate:\begin{equation}
p_{i}=A_{2,1}\mathbb{E}_{i}x_{i+1}+A_{2,2}\mathbb{E}_{i}p_{i+1}+K_{2}w_{i}.
\end{equation}Here \(A_{i,j}\) is the \((i,j)\)-th element of the matrix \(A\). This equation states that the inflation dynamics depend on future expectations of output gap and inflation, plus a stochastic noise term introduced by the dynamics of the output gap and the central bank behaviour: we can safely assume that other factors, such as measurement errors,  price index basket rebalancing or any other idiosyncratic factor not directly modelled in this framework, may add noise to the inflation dynamics.\footnote{If one takes the view that this third source of randomness is not advisable to include, one  assumes that its value is always 0 with probability 1. As one notices in the following developments, this third source of randomness is mainly used in the calibration phase in order to have an additional degree of freedom and has no impact on the theoretical development of the model.} On the basis of these considerations, we add a further independent source of randomness, modelled with the adapted process
\(\{z_{i}\}_{i=0,1,...}  \): we require this pro\(\)cess to have zero mean, to be independent from its past realisations, to be independent from \(\{u_{i}\}_{i=0,1,...}  \) and \(\{v_{i}\}_{i=0,1,...}   \), and to have finite variance  \(\textrm{Var}(z_{i})  \), third and fourth moments (\(\textrm{Skew}(z_{i})  \) and \(\textrm{Kurt}(z_{i})   \) respectively).

The new expression for the inflation rate becomes:
\begin{equation}
p_{i}=A_{2,1}\mathbb{E}_{i}x_{i+1}+A_{2,2}\mathbb{E}_{i}p_{i+1}+K_{2}w_{i}+z_{i}.\label{inflationWithZeta}
\end{equation}

Its mean, variance and autocovariance are therefore:
\begin{displaymath}
\mathbb{E}[p_{i}]=A_{2,1}\mathbb{E}_{i}x_{i+1}+A_{2,2}\mathbb{E}_{i}p_{i+1}\label{EPinfl}
\end{displaymath}
\begin{displaymath}
\textrm{Var}(p_{i})=(K_{2})^{2}(\sigma^{2}\textrm{Var}(u_{i})+ \textrm{Var}(v_{i}))+ \textrm{Var}(z_{i}) \label{Var_pi}
\end{displaymath} 
\begin{displaymath}
\textrm{Cov}(p_{i},p_{i+j})=0 ,\qquad j\neq0.
\end{displaymath}

We note that the variance of the inflation process is a linear combination of the variances of the three processes  \(\{u_{i}\}_{i=0,1,...}  \), \(\{v_{i}\}_{i=0,1,...}  \)  and \(\{z_{i}\}_{i=0,1,...} \).
 
 Finally, we calculate the centered third and fourth moments: these may be needed in order to analyse their distribution in a more complete fashion:
\begin{equation}
\mathbb{E}\left[ \left(p_{i}-\mathbb{E}(p_{i})\right)^{3} \right]=(K_{2})^{3}\sigma^{3}\textrm{Skew}(u_{i})-(K_{2})^{3}\textrm{Skew}(v_{i})+\textrm{Skew}(z_{i})\label{SkewPi}
\end{equation}
\begin{displaymath}
\mathbb{E}\left[ \left(p_{i}-\mathbb{E}(p_{i})\right)^{4} \right]=(K_{2})^{4}\sigma^{4}\textrm{Kurt}(u_{i})+(K_{2})^{4}\textrm{Kurt}(v_{i})+\textrm{Kurt}(z_{i})+
\end{displaymath}
\begin{equation}
+6(K_{2})^{4}\sigma^{2}\textrm{Var}(u_{i})\textrm{Var}(v_{i})
+6(K_{2})^{2}\textrm{Var}(v_{i})\textrm{Var}(z_{i})+6(K_{2})^{2}\sigma^{2}\textrm{Var}(u_{i})\textrm{Var}(z_{i}).\label{KurtPI}
\end{equation}\subsubsection{Statistical properties of the short term nominal interest rate}
The nominal interest rate \(n_{i}\) is defined as\begin{equation}
n_{i}=\bar n(1+\hat n_{i})\label{ShortRate} 
\end{equation}
where \(\bar n\) is the equilibrium nominal interest rate, which is the short rate that would be chosen by the central bank if the  adjustment required by the Taylor rule  was zero---as  \(\hat n_{i}\) follows \eqref{TaylorRule}. This  follows by the definition of  \(\hat n\) as the log-linearised difference between the actual rate and equilibrium rate.

 We take the equilibrium nominal rate \(\bar n\) as a constant input that can be obtained from research and is therefore not calibrated to any traded asset.  We assume that the short rate is used to discount payments between different counterparties, i.e. it plays the role of the Libor rate : this assumption, albeit strong, simplifies the problem considerably.\footnote{We recall that in continuous time the short rate \(n(t)= f(t,t)= lim _{\Delta T\rightarrow0} F(t,t,t+\Delta T) \) where the forward rate is defined as \(F(t,S,T)=(P(t,S)/P(t,T)-1)/(T-S)\) with \(T>S. \) In discrete time we  define  \(n_{i}= f_{i,i}= F(t_{i},t_{i},t_{i+1})  \), therefore getting \(n_{i}=(1/P(t_{i},t_{i+1})-1)/\tau_{i+1}\). As a consequence,  the short rate can be used as the Libor rate, provided that there are no credit concerns in the interbank markets.} 

If we plug the Taylor rule  \eqref{TaylorRule} into \eqref{ShortRate} we  rewrite the nominal rate as
\begin{equation}
n_{i+1}=\bar n(1+\delta_{x}x_{i}+\delta_{\pi}p_{i}+v_{i}).
\end{equation}

We can compact the notation by introducing the vectors
\begin{displaymath}
\delta=\begin{bmatrix}\delta_{x}\ \\
\delta_{\pi}\ \\
\end{bmatrix};\;\;\;\;\xi_{i}=\begin{bmatrix}x_{i} \\
p_{i}\ \\
\end{bmatrix}.
\end{displaymath}

Therefore the interest rate can be written as \(n_{i+1}=\bar n(1+\delta^{T}\xi_{i})\)
where \((x)^{T}\) is the transpose of the vector \(x\). The source of randomness \(v_{i}\) is  included in the dynamics of \(\xi_{i}\), as can be deduced from \eqref{System1}.
Finally, by making use of  \eqref{System1} and \eqref{inflationWithZeta} we get:
\begin{displaymath}
n_{i+1}=\bar n(1+\delta^{T}A\mathbb{E}_{i}\xi_{i+1}+\delta^{T}Kw_{i}+\delta^{T}e_{2}z_{i})\label{ShortRateFull},
\end{displaymath}
where \(e_{2}^{T}=\begin{bmatrix}0 & 1 \\
\end{bmatrix}\).

By making use of this expression we   calculate the mean, variance  and the autocovariance of the nominal interest rate:
\begin{displaymath}
\mathbb{E}[n_{i+1}]=\bar n(1+\delta^{T}A\mathbb{E}\xi_{i+1})\label{EPn}
\end{displaymath}
\begin{displaymath}
\textrm{Var}(n_{i+1})=(\bar n)^{2}(\delta ^{T}K)^{2}\sigma ^{2}\textrm{Var}(u_{i})+ (\bar n)^{2}(\delta ^{T}K)^{2}\textrm{Var}(v_{i})+(\bar n)^{2}\delta^{2}_{\pi}\textrm{Var}(z_{i})\label{Var_n}
\end{displaymath}
\begin{displaymath}
\textrm{Cov}(n_{i},n_{i+j})=0, \;\;j\neq0.
\end{displaymath}

We note that the variance of the interest rate process is a linear combination of the variances of the three processes  \(\{u_{i}\}_{i=0,1,...}  \), \(\{v_{i}\}_{i=0,1,...}  \)  and \(\{z_{i}\}_{i=0,1,...}.    \) We take the equilibrium rates, output gap and inflation expectations as  inputs: they may be provided by macroeconomic research or can just be expression of the trader's views.

 Similarly to what could be done for the inflation rate, we can also calculate the centered third and fourth moments: these are needed in order to analyse their distribution in a more complete fashion:
\begin{equation}
\mathbb{E}\left[ \left(n_{i}-\mathbb{E}(n_{i})\right)^{3} \right]=[(\delta ^{T}K)^{3}\sigma^{3}\textrm{Skew}(u_{i})-(\delta ^{T}K)^{3}\textrm{Skew}(v_{i})+\textrm{\((\delta_{\pi})^{3}\)Skew}(z_{i})](\bar n)^{3}
\label{SkewN}\end{equation}
\begin{displaymath}
\mathbb{E}\left[ \left(n_{i}-\mathbb{E}(n_{i})\right)^{4} \right]=[(\delta ^{T}K)^{4}\sigma^{4}\textrm{Kurt}(u_{i})+(\delta ^{T}K)^{4}\textrm{Kurt}(v_{i})+\textrm{\((\delta_{\pi})^{4}\)Kurt}(z_{i})+
\end{displaymath}
\begin{equation}
+6(\delta ^{T}K)^{4}\sigma^{2}\textrm{Var}(u_{i})\textrm{Var}(v_{i})
+6(\delta ^{T}K\delta_{\pi})^{2}\textrm{Var}(v_{i})\textrm{Var}(z_{i})+6(\delta ^{T}K)^{2}(\sigma\delta_{\pi})^{2}\textrm{Var}(u_{i})\textrm{Var}(z_{i})](\bar n)^{4}.\label{KurtN}
\end{equation}

Finally, we calculate the covariance between the nominal rate \(n_{i+1}\) and the inflation \(p_{i}\), both \(\mathcal{F}_{i}\)-measurable:\begin{equation}
\textrm{Cov}(p_{i},n_{i+1})=\bar nK_{2}(\delta ^{T}K)\sigma ^{2}\textrm{Var}(u_{i})+ (\bar nK_{2}(\delta ^{T}K))\textrm{Var}(v_{i})+\bar n\delta_{\pi}\textrm{Var}(z_{i}).\label{CovPiN}
\end{equation}

The covariance depends  on the Taylor rule parameters vector \(\delta\), which makes explicit the philosophy of our modelling approach: any dependence between the nominal interest rate and inflation is not specified exogenously but is a consequence of the central bank reaction function. Furthermore,  
the correlation becomes one if there is no uncertainty in the Taylor rule, i.e.
\(\textrm{Var}(v_{i})=0  \) and if \(\textrm{Var}(z_{i})=0  \): in this case the central bank reacts deterministically to any change in the economy.

The other interesting limit case is when the output gap evolves deterministically, i.e.
\(\textrm{Var}(u_{i})=0 \), \(\delta ^{T}K<0\), and \(\textrm{Var}(z_{i})=0  \): rates evolve stochastically and correlation becomes \(-1\). As rates increase, the output gap decreases deterministically (because of the demand curve \eqref{Gap Dynamics}), bringing down the inflation according to the Phillips curve \eqref{Inflation Dynamics}. In this case the only source of randomness is the uncertainty in the short rate evolution due to the Taylor rule. 

The DSGE model augmented with the Taylor rule allows for this correlation to take values between -1 and 1, depending on the central bank reaction function and the specification of the sources of randomness: this can be arguably regarded as an interesting feature of the model, because it does not impose any constraint on the correlation range.\subsubsection{Calibrating to rates and inflation smiles: the normal case}
The prices of nominal rates and inflation caps/floors across different strikes and maturities are  available from brokers or investment banks (for example the Bloomberg pages VOLS or RILO): we can thus deduce the caplet/floorlet prices. Unlike options on other underlyings, inflation options are quoted in prices, not in implied volatilities.  By making some distributional assumptions on the nominal rates and inflation, we summarise the  distribution using only a few parameters.

For example, we can  assume a normal distribution and fit its volatility to the option prices for each maturity: this assumption is both convenient  from an analytical perspective (closed formulas for option prices are  obtained) and from a practitioner point of view: if rates are normally (and not lognormally) distributed, the distribution of their relative increments is skewed, (and not Gaussian as in the Black model). 
In this case we calibrate the variances of      \(\left\{u_{i}\right \}_{i=0,1,...}\), \(\left\{v_{i}\right \}_{i=0,1,...}\), and \(\left\{z_{i}\right \}_{i=0,1,...}\) to obtain the market implied variances for nominal rates and inflation for at-the-money trades.
We   calculate the market implied variances for      \(\left\{u_{i}\right \}_{i=0,1,...}\), \(\left\{v_{i}\right \}_{i=0,1,...}\), and \(\left\{z_{i}\right \}_{i=0,1,...}\) given the market implied variances of rates/inflation caplets/floorets. A word of caution should be issued, as 
there is no guarantee to obtain positive variances from this basic algorithm. Negative variances could be floored to zero or more sophisticated root-searching algorithms can be used.

\subsubsection{Measure change under normality assumptions} At this stage we make the measure change process \( \left\{\mu_{i}\right\}_{i=0,1,...}\)  explicit   to use the real-world  expectations to price derivatives in the risk-neutral measure. We define the measure change processes as  discretely sampled exponential Gaussian martingale: with this strategy  one  obtains  a positive martingale. A general introduction to  exponential L\'evy martingales can be found in Appelbaum \cite{Label113}. 

To simplify the notation, we rewrite equation \ref{DSGE_full_dyn} including the variable \(z_{i}\) in matrix format:
\begin{equation}
\xi_{i}=A\mathbb{E}_{i}\xi_{i+1}+Kw_{i}+e_{2}z_{i}=A\mathbb{E}_{i}\xi_{i+1}+K\sigma u_{_{i}}-Kv_{i}+e_{2}z_{i}.
\end{equation}

Defining the matrix \(C\) as follows:\begin{displaymath}
C=
\begin{bmatrix}\sigma K_{1} & -K_{1} & 0 \\
\sigma K_{2} & -K_{2} & 1 \\
\end{bmatrix}
\end{displaymath}
and compacting all three  sources of randomness in the three-dimensional vector \(\varepsilon_{i}\) defined as \(\varepsilon_{i}=
\begin{bmatrix}u_{i} & v_{i} & z_{i} \\
\end{bmatrix}\),
the notation
 is further simplified into \(\xi_{i}=A\mathbb{E}_{i}\xi_{i+1}+C\varepsilon^{T}_{i}\).

One notes that the variance-covariance matrix for the vector \(\varepsilon_{i}\)  is written as:\begin{displaymath}
\Sigma^{\varepsilon}_{i}=\begin{bmatrix}\textrm{Var}(u_{i}) & 0 & 0 \\
0 & \textrm{Var}(v_{i}) & 0 \\
0 & 0 & \textrm{Var}(z_{i}) \\
\end{bmatrix}
=\begin{bmatrix}\textrm{Var}(\varepsilon_{i}^{1}) & 0 & 0 \\
0 & \textrm{Var}(\varepsilon_{i}^{2}) & 0 \\
0 & 0 & \textrm{Var}(\varepsilon_{i}^{3}) \\
\end{bmatrix}.
\end{displaymath}

At this point
we introduce the three-dimensional deterministic vector process
\(\{\lambda_{i}\}_{i=0,1,...}\) defined as:
\begin{displaymath}
\lambda_{i}=\begin{bmatrix}\lambda_{i}^{u} \\
\lambda_{i}^{v} \\
\lambda_{i}^{z} \\
\end{bmatrix}.
\end{displaymath}
The quantities defined above are used to specify the measure change process \( \left\{\mu_{i}\right\}_{i=0,1,...}\), following and generalising Shreve \cite{Label00180a}. The measure change process is therefore defined as a multivariate Gaussian exponential martingale in the form: \(\frac{d\mathbb{Q}}{d\mathbb{P}}|_{\mathcal{F}_{i}}=\mu_{i}=e^{-\epsilon_{i}  \lambda_{i}-1/2\lambda_{i}^{T}\Sigma^{\varepsilon}_{i}\lambda_{i}}\).

One requires \(\mu_{0}=1\) and  the market price of risk vector process \( \left\{\lambda_{i}\right\}_{i=0,1,...}\)to be regular enough for the measure change process \( \left\{\mu_{i}\right\}_{i=0,1,...}\) to be a positive martingale (i.e. its expectation has always to be  finite) and square-integrable. Moving to the risk-neutral measure \(\mathbb{Q, }\) one obtains that the new process \(\nu_{i}=\varepsilon_{i}+\lambda_{i} \) is a zero-mean Gaussian process
with independent realisations  under \(\mathbb{Q}\).
In this measure we also write \(u^{*}_{i}=u_{i}+\lambda_{i}^{u}\), \(v^{*}_{i}=v_{i}+\lambda_{i}^{v}\), \(z^{*}_{i}=z_{i}+\lambda_{i}^{z}\), \(w^{*}_{i}=\sigma u_{i}^{*}-v^{*}_{i}\).

 We rewrite the expression for the macroeconomic variables (output gap and  inflation) once the measure change from \(\mathbb{P}\) to \(\mathbb{Q }\) has been performed: \( \xi_{i}=A\mathbb{E}_{i}\xi_{i+1}+C\nu_{i}=A\mathbb{E}_{i}\xi_{i+1}+C\lambda _{i}+C\varepsilon^{T}_{i}\).

Informally, one can think
 to the
linear function  of the market prices of risk  \(\lambda_{i}\) as a \textquotedblleft wedge" that is multiplied by some coefficients in the matrix \(C\) and then added to the deterministic linear function of the expectations \(A \mathbb{E}_{}\xi_{i+1}\) in order to calibrate the model to the traded prices of nominal bonds and inflation breakevens (through the relationship between nominal bonds, real bonds and inflation  zero-coupon swaps).

Finally, one finds a compact expression for the nominal short rate \(n_{i}\) and the inflation rate \(p_{i}\) under \(\mathbb{Q}\):
\begin{displaymath}
n_{i+1}=\bar n (1+\delta^{T}\xi_{i})= \bar n (1+\delta^{T}(A\mathbb{E}_{i}\xi_{i+1}+C\nu^{T}_{i}))=\bar n (1+\delta^{T}(A\mathbb{E}_{i}\xi_{i+1}+C\lambda _{i}+C\varepsilon^{T}_{i}))
\end{displaymath}
\begin{displaymath}
p_{i}=A_{2,1}\mathbb{E}_{i}x_{i+1}+A_{2,2}\mathbb{E}_{i}p_{i+1}+K_{2}w_{i}^{*}+z_{i}^{*}=A_{2,1}\mathbb{E}_{i}x_{i+1}+A_{2,2}\mathbb{E}_{i}p_{i+1}+\sigma K_{2}u_{i}^{*}-K_{2}v_{i}^{*}+z_{i}^{*}=
\end{displaymath}
\begin{displaymath}
=A_{2,1}\mathbb{E}_{i}x_{i+1}+A_{2,2}\mathbb{E}_{i}p_{i+1}+\sigma K_{2}(u_{i}+\lambda^{u}_{i})-K_{2}(v_{i}+\lambda^{v}_{i})+(z_{i}+\lambda^{z}_{i})=A_{2,1}\mathbb{E}_{i}x_{i+1}+A_{2,2}\mathbb{E}_{i}p_{i+1}+h\nu^{T}_{i}
\end{displaymath}
where the vector \(h\) has been defined as \(h=\begin{bmatrix}\sigma K_{2} & -K_{2} & 1 \\
\end{bmatrix} \).
\subsubsection{Calibrating to the nominal term structure}
We show how to calibrate the model to the nominal interest rates observed in the market by making some approximations. We use market prices of one period  discount factors to provide some expressions to be used in the calibration. We  write:
\begin{displaymath}
P(t_{0},t_{i+1})=\mathbb{E}_{0}^{\mathbb{P}}\left[ \psi_{i+1} \right] =\mathbb{E}_{0}^{\mathbb{Q}}\left[ \prod^{i}_{j=0}(1+n_{j+1}\tau_{j+1} )^{-1}\right] \cong \mathbb{E}_{0}^{\mathbb{Q}}\left[ e^{-\sum ^{i}_{j=0}n_{j+1}\tau_{j+1}} \right]
\end{displaymath}
 
where the last linearisation creates some error that can be reduced by calibrating the model on a finer time grid.
The term
\(\tau_{i+1}\) is the year fraction: \(\tau_{i+1}=t_{t+1}-t_{i}\).
In practice
one  does a bootstrapping over each time step, thanks to the fact that the interest rates level is independent from its previous levels.

The following step is to introduce the closed form expression for the nominal rate \(n_{i}\):
 \begin{displaymath}
\mathbb{E}_{0}^{\mathbb{Q}}\left[ e^{-n_{i+1}\tau_{i+1}} \right]=\mathbb{E}_{0}^{\mathbb{Q}}\left[ e^{- \bar n (1+\delta^{T}(A\mathbb{E}_{i}\xi_{i+1}+C\nu^{T}_{i}))\tau_{i+1}}\right]=e^{- \bar n\tau_{i+1} (1+\delta^{T}(A\mathbb{E}_{i}\xi_{i+1}))+\frac{\bar n^{2}\tau_{i+1}^{2}}{2}\delta^{T}C\Sigma^{\varepsilon}_{i}C^{T}\delta)} 
\end{displaymath}

 By taking the expectations  under the normality assumption for the vector \( \nu_{i}\), the one-period  discount factors approximated closed form is:\begin{equation}
\mathbb{E}_{0}^{\mathbb{Q}}\left[ e^{-n_{i+1}\tau_{i+1}} \right]\cong e^{c_{1}+c_{2}\textrm{Var}(u_{i})+c_{3}\textrm{Var}(v_{i})+c_{4}\textrm{Var}(z_{i})}\label{BondCF}
\end{equation}
where:
\begin{displaymath}
c_{1}=-\tau_{i+1} \bar n (1+\delta^{T}(A\mathbb{E}_{i}\xi_{i+1}))
\end{displaymath}

\begin{displaymath}
c_{2}=\frac{1}{2}\left(\tau_{i+1}\bar n\delta^{T}K\sigma \right)^{2}
\end{displaymath}

\begin{displaymath}
c_{3}=+\frac{1}{2}\left(\tau_{i+1}\bar n\delta^{T}K \right)^{2}
\end{displaymath}

\begin{displaymath}
c_{4}=\frac{1}{2}(\tau_{i+1}\bar n_{}\delta_{\pi})^{2}.
\end{displaymath}

\subsubsection{Calibrating to the ZCIIS\\}
As shown in Brigo \& Mercurio \cite{Label2}, the value of a zero-coupon inflation index swap (ZCIIS) can be regarded as the difference between the real and  nominal zero-coupon bond prices with the same maturity date. For the full definition of the real bond and term structure we refer to Hughston \cite{Label12}.

We exploit the model-independent relationship between real and nominal bond to write:
\begin{displaymath}
P^{R}(t_{0},t_{i+1})=P(t_{0},t_{i+1})+ZCIIS(t_{0},t_{i+1}).\label{ZCCIS}
\end{displaymath}

Since we  observe  the market prices of nominal bonds and ZCIIS for different maturities, we  deduce the value of a real bond, even if these instruments are not traded in the market.

We assume that the real bond  pays at maturity \(t_{i+1}\) the unit nominal multiplied by the underlying inflation index appreciation between times \(t_{0}\) and \(t_{i}\): this is to introduce the inflation publication lag in the formula, which becomes necessary since in reality the inflation rate is only published after a time lag. 

The approximated closed form is obtained as follows:
\begin{displaymath}
P^{R}(t_{0},t_{i+1})=\mathbb{E}_{0}^{\mathbb{P}}\left[ \frac{I_{i}}{I_{0}}\frac{\psi_{i+1}}{\psi_{0}} \right]=\mathbb{E}^{\mathbb{P}}\left[ \frac{I_{i}}{I_{0}}\psi_{i+1} \right]
=\mathbb{E}^{\mathbb{Q}}\left[ \frac{I_{i}}{I_{0}} \prod^{i+1}_{j=1}\frac{1}{1+\tau_{j}n_{j}}\right]
=\mathbb{E}^{\mathbb{Q}}\left[ \prod^{i+1}_{j=1}\frac{1+\tau_{j-1}p_{j-1}}{1+\tau_{j}n_{j}}\right]
\end{displaymath}
 By making some straightforward Taylor expansions the last expression can be rewritten as:
\begin{displaymath}
P^{R}(t_{0},t_{i+1})=\mathbb{E}^{\mathbb{Q}}\left[ \prod^{i+1}_{j=1}\frac{e^{log(1+\tau_{j-1}p_{j-1})}}{e^{log(1+\tau_{j}n_{j})}}\right]
\cong\mathbb{E}^{\mathbb{Q}}\left[ \prod^{i+1}_{j=1}\frac{e^{\tau_{j-1}p_{j-1}}}{e^{\tau_{j}n_{j}}}\right]
=\mathbb{E}^{\mathbb{Q}}\left[e^{ \sum^{i+1}_{j=1}(\tau_{j-1}p_{j-1}-\tau_{j}n_{j})}\right].
\end{displaymath}

We assume that \( p_{0}=0 \) and focus the attention on the one-period  real discount factor.

  By  the same Gaussianity assumptions used above, the following closed formula is obtained by plugging \eqref{inflationWithZeta} into the above expression:

\begin{displaymath}
 \mathbb{E}^{\mathbb{Q}}_{0}[e^{-\tau_{i+1}n_{i+1}+\tau_{i}p_{i}}]=\mathbb{E}^{\mathbb{Q}}_{0}[e^{-\tau_{i+1}\bar n (1+\delta^{T}(A\mathbb{E}_{i}\xi_{i+1}+C\nu^{T}_{i}))+\tau_{i}(A_{2,1}\mathbb{E}_{i}x_{i+1}+A_{2,2}\mathbb{E}_{i}p_{i+1}+h\nu^{T}_{i})}]= \\ \label{IF}
\end{displaymath}
\begin{equation}
 \\ \mathbb{=E}^{\mathbb{Q}}_{0}[e^{-\tau_{i+1} \bar n (1+\delta^{T}A\mathbb{E}_{i}\xi_{i+1})+\tau_{i}(A_{2,1}\mathbb{E}_{i}x_{i+1}+A_{2,2}\mathbb{E}_{i}p_{i+1})+\nu^{T}_{i}(\tau_{i}h-\tau_{i+1} \bar n\delta^{T}C)}]=e^{b_{1}+b_{2}\textrm{Var}(u_{i})+b_{3}\textrm{Var}(v_{i})+b_{4}\textrm{Var}(z_{i})}\label{IF}
\end{equation}
where
\begin{displaymath}
b_{1}=\tau_{i} A_{2,1}\mathbb{E}^{}_{i}x_{i+1}+\tau_{i} A_{2,2}\mathbb{E}^{}_{i}p_{i+1}-\tau_{i+1}\bar n(1+\delta^{T}A\mathbb{E}^{}_{i}\xi_{i+1})
\end{displaymath}

\begin{displaymath}
b_{2}=\frac{1}{2}\left( \tau_{i} K_{2}\sigma-\tau_{i+1}\bar n\delta^{T}K\sigma \right)^{2}
\end{displaymath}

\begin{displaymath}
b_{3}=\frac{1}{2}\left( \tau_{i}K_{2}-\tau_{i+1}\bar n(\delta^{T}K )  \right)^{2}
\end{displaymath}

\begin{displaymath}
b_{4}=\frac{1}{2}(\tau_{i} -\tau_{i+1}\bar n\delta_{\pi})^{2}.
\end{displaymath}

We stress that the variances calibrated from option prices are taken as an input in the above expression. The two approximated closed formulas for the nominal \eqref{BondCF} and the real bond \eqref{IF}  can be used to find the values of \(\lambda_{i}\)   that  calibrate the model to the market, given the variances of the distributions of the shock factors  \(u_{i}\),    \(v_{i}\),   and    \(z_{i}\).

To conclude this section, we observe that
the adaptation of the DSGE model to pricing proposed above is extremely respectful of the the original macroeconomic model, but for this reason it is also not straightforward to price derivatives. In fact, to obtain closed forms for the nominal and real bonds one has to resort to approximations and linearisations of exponentials, which are doable but not elegant. The model offers an insight of the macroeconomic forces operating behind the yield curve and the inflation dynamics, but all pricing has to happen using Monte Carlo simulations, which can be cumbersome and time consuming. Interestingly, the above section shows a first attempt to bridge the gap between two disciplines (monetary macroeconomics and financial mathematics) that are dealing with the same problem (inflation) in two different ways (DSGE modelling versus arbitrage pricing). This represent a step forward in the same direction indicated by Hughston \& Macrina \cite{Label131}, who derive some inflation dynamics from a macroeconomic model --- even if there is no concept of central bank policy in their work.  

With these ideas in mind, in the following section  we propose some continuous-time dynamics that, while retaining the most significant aspects of the DSGE model presented in the previous sections, are more tractable from a derivatives pricing perspective. It is important to stress that the new dynamics we propose are not a one-to-one translation of the discrete-time DSGE model, but rather they are inspired by it and take into account that in the post-Lehman environment the short rate is not the only policy tool used by the central bank.\footnote{When short rates are low or ineffective to stimulate the economy, the central bank can purchase assets to reduce long term interest rates and increase the money supply to spur growth.} To ensure that the proposed dynamics are meaningful, we bring some empirical evidence that shows that the proposed dynamics are realistic. Finally, we show that the discrete-time DSGE model and the continuous-time model proposed generate similar distributions for the main economic variables.
\section{Building  the continuous-time version}
Here we propose a strategy to loosely translate the DSGE model into  continuous time by making some  assumptions. Therefore we show that some continuous-time dynamics can be derived from  a widely-accepted macroeconomic model: they are used in the next section to build the inflation model. From this point, the notation for the variable \(y\) in continuous time is \(y(t)\). 

The following assumptions are made:\begin{enumerate}
\item 
There is no price flexibility for the firms, i.e. \( \omega = 1 \) and \( k = 0.  \) This assumption is reasonable as markets tend to be far from the perfect competition model, and therefore prices are sticky, especially over a shorter time step.
\item The one-period subjective discount factor is equal to the inverse of the inflation targeting parameter: \(\beta\delta_{\pi} =1.  \) This assumption is sensible because, when the central bank fights inflation aggressively ~(i.e. \(\delta_{\pi} \gg 1), \) interest rates  increase, pushing down the discount factor \(\beta\).
\item The  GDP growth rate is  modelled in the same way as the output gap. In fact, because the output gap is defined  as the difference between the actual and the potential GDP growth rate, and because the latter is an abstract concept (normally deemed to be constant over time),   this  means adding the constant potential growth rate to the output gap.

\item The GDP growth rate is defined as the percentage change of the GDP level from one period to the next one: \(x_{i}=(X_{i}-X_{i-1})/X_{i-1}\). One  changes the notation  and write: \(x_{t_{i}}=(X_{t_{i}}-X_{t_{i-1}})/X_{t_{i-1}}\). Furthermore, one  generalises the time step and write: \(\Delta t_{i}=t_{i}-t_{i-1}\), therefore obtaining:  \(x_{t_{i}}=(X_{t_{i}}-X_{t_{i-\Delta t_{i}}})/X_{t_{i-\Delta t_{i}}}=\Delta X_{t_{i}}/X_{t_{i-\Delta t_{i}}}\). Moving to continuous time one  writes \(x(t) \) as \(dX(t)/X(t)  \).  One needs to assume that the positive process \(\{X_{i}\}_{i=0,1,.. . }\) is regular enough for the limit to exist.

\item A similar line of thought can be followed to show how one moves from the discrete-time definition of inflation, as the percentage change in the price index level (\(p_{i}=(I_{i}-I_{i-1})/I_{i-1}\)), to the equivalent continuous-time definition (\(p(t)\) is written as \(dI(t)/I(t)\)).
Again we make an obvious request of positivity for the price index process \(\{I_{i}\}_{i=0,1,.. . }\).\item There are measurement errors and other sources of uncertainty for both inflation and growth rate, modelled by the \emph{m}-dimensional zero-mean random variable \(z_{i}\). The \(m\) components of this random variable (called \(z^{j}_{i},  \) with \(1,2,...,j, ...,m\)) are independent from each other.
The random variable \(z_{i}\) is also independent from  \(w_{i}\). The effects of the shock \(z^{j}_{i}\) on \(x_{i}\) and \(p_{i}\) are modelled by the \textit{m-}dimensional real-valued deterministic processes  \(\{a_{i}\}_{i=0,1, ...}\) and \(\{b_{i}\}_{i=0,1, ...}\), where their single components have notation \(a_{i,j}\) and \(b_{i,j}\).

\item 
The product of the expectation terms by some constants that appear in the DSGE model can be written as \(\sigma/(\sigma+\delta_{x}+k\delta_{\pi}) )\mathbb{E}_{i} x_{i+1}=m_{X}(t_{i})(t_{i+1}-t_{i})\) and  \((k+\beta(\sigma+\delta_{x}))/(\sigma+\delta_{x}+k\delta_{\pi}) \mathbb{E}_{i} p_{i+1}=m_{I}(t_{i})(t_{i+1}-t_{i})\) respectively. We assume that the quantities \(m_{X}(t_{i})\) and \(m_{I}(t_{i}) \)  are realisations of adapted stochastic processes. This means that these expectations are not dependent on the chosen time lag, and can be written as the product by a real function of time (\(m_{X}(t_{i})\) and \(m_{I}(t_{i})\) respectively) and the chosen time lag. One  generalises the time lag by writing  \(\sigma /(\sigma+\delta_{x}+k\delta_{\pi}) )\mathbb{E}_{t_{i}} x_{t_{i}+\Delta t_{i}}=m_{X}(t_{i})\Delta t_{i}\) and  \((k+\beta (\sigma+\delta_{x}))/(\sigma+\delta_{x}+k\delta_{\pi})\mathbb{E}_{t_{i}} p_{t_{i}+\Delta t_{i}}=m_{I}(t_{i})\Delta t_{i}\) respectively. When one moves to continuous time, \( \Delta t_{i}
\rightarrow dt\), and the real quantities \(m_{X}(t)\) and \(m_{I}(t) \)  do not change. Therefore one  writes the products of expectation terms and constants as a continuous time drift (\(m_{X}(t)d t\) and \(m_{I}(t)d t\) respectively).

\item The random variables \(u_{i}\) and \(z^{j}_{i}\) are independent and normally distributed, with mean 0 and unit variance. 
\item 
The random variables \(u_{i}\) and \(z^{j}_{i}\)  are independent from their previous levels. For example, taken \(u_{i}\), one  writes \(\text{Cov}(u_{i},u_{l})=\text{\(\delta_{i,l}\)}.\) In this context \(\delta_{i,l}\) is the Kronecker's delta sign, taking value 0 in all cases where \(i\neq l\) and 1 when \(i=l\).
\item Taken the random standard normal variable \(w_{i}\), one  introduces the random variable \(U_{i}\), defined as \(U_{i}=\sum^{i}_{k=1}w_{k}\), with \(U_{0}=0. \) Based on all the assumptions made, one  shows that \(U_{i}\sim N(0,i)\). By construction, the process \(\{U_{i}\}_{i=0,1,...}\) has zero mean, independent increments and \(U_{i}-U_{l}\sim N(0,i-l),\;i>l\). The increment \(U_{i+k}-U_{l+k}\) has the same distribution as the increment \(U_{i}-U_{l}\), for each \(k\).
\item By generalising the time lag, one finds that \(\Delta U_{t_{i}}=U_{t_{i}}-U_{t_{i}
-\Delta t_{i}}\sim N(0,\Delta t_{i}) \). Moving to continuous time one gets a  Brownian motion.
A similar discussion can be held for the \textit{m}-dimensional random variable \(z_{i}\), which becomes an \textit{m}-dimensional Brownian motion with independent components.
Shreve \cite{Label00180a} gives full details of this procedure to build the Brownian motion starting from a discrete-time Gaussian process.
\item
To compact notation, one introduces the \textit{m+1}-dimensional (or alternatively \textit{n}-dimensional) vectors, defined  as \(s^{X}_{t_{i}}=[ \sigma, a^{1}_{t_{i}},..., a^{M}_{t_{i}}]\), and \(s^{I}_{t_{i}}=[0, b^{1}_{t_{i}},..., b^{M}_{t_{i}}]\).
The idea is to compact all the random terms to express them using a lighter notation.\item 
To move to continuous time, one assumes that the processes \(\{s^{X}_{t_{i}}\}_{t_{i}=0,1,...}\) and \(\{s^{I}_{t_{i}}\}_{t_{i}=0,1,...}\)are regular enough for the limits  \(s^{X}_{t_{i}} \rightarrow\ s_{X}(t) \) and \(s^{I}_{t_{i}}\rightarrow\ s_{I}(t)\) to exist and for the total variance to be the same. \end{enumerate} 
The  system \ref{DSGE_full_dyn} can be rewritten in discrete time using a generic time step \(\Delta t_{i}\) as:
\begin{equation}
\begin{bmatrix}x_{t_{i}} \\
p_{t_{i}}\ \\
\end{bmatrix}
=
\left(
\begin{bmatrix}m_{X}(t_{i})\ \\
 m_{I}(t_{i})\ \\
\end{bmatrix}\Delta t_{i}+\begin{bmatrix}\sigma\ \\
0 \\
\end{bmatrix}
\left(w_{t_{i}}\right)\Delta t_{i}^{1/2}+\sum^{M}_{j=1}\begin{bmatrix}a_{i,j} \\
b_{i,j} \\
\end{bmatrix}
\left( z^{j}_{t_{i}} \right)\Delta t_{i}^{1/2}
\right)
\end{equation}From the assumptions made above, the two above equations can be  translated in continuous time 
as follows:
\begin{equation}
dX(t)/X(t)=m_{X}(t)dt\:+s_{X}(t)\cdot\:dW(t)
\end{equation} 
\begin{equation}
dI(t)/I(t)=m_{I}(t)dt\:+s_{I}(t)\cdot\:dW(t),
\end{equation} 
where \(\{W(t)\}_{t\geqslant0}\) is an \textit{n}-dimensional Brownian motion.
The notation \(\cdot\)  is used to refer to the vector product.\\ 
At this stage one  complements this model with some dynamics for the expectations of the drift: in fact, as shown in the following section, empirical evidence suggests that expectations themselves are subject to frequent revisions (as the economic agents process new information and data) and therefore are themselves stochastic. A possible expression for the dynamics of the expectations is the following:
\begin{equation}
dm_{X}(t)=a_{X}(t)dt+b_{X}(t)\cdot\:dW^{\mathbb{}}(t)
\end{equation}
\begin{equation}
dm_{I}(t)=a_{I}(t)dt+b_{I}(t)\cdot\:dW^{\mathbb{}}(t).
\end{equation}
where the scalar processes \(\{a_{X}(t)\}_{t\geq0}\) and \(\{a_{I}(t)\}_{t\geq0}\), and the \textit{m+1}-dimensional processes \(\{b_{X}(t)\}_{t\geq0}\) and \(\{b_{I}(t)\}_{t\geq0}\) are  deterministic processes regular enough for the SDEs to be integrated and to have a unique strong solution. To conclude, the  above stochastic differential equations are derived from a well-established macroeconomic model. They are consistent with empirical evidence (as shown in the next section) and are used in the following section as a part of a wider setup to build a structural continuous-time pricing model for inflation derivatives,  based on macroeconomic assumptions.
\subsection{Testing the dynamics against empirical evidence}In this section we show some economic time series to confirm that, over time, the growth rate of real GDP and of the price index are stationary processes that show some randomness. This paper is not about econometrics, so the evidence is presented in a somewhat intuitive and loosely defined fashion. The actual levels of real GDP and price index are growing in an exponential fashion over time: these two
observations confirm that the choice of a  Brownian motion with time-changing coefficients and stochastic drift is  sensible.

Evidence is shown for the US and the UK economy, however similar results hold for all economies.
All
data are sourced from Bloomberg.
We
do not perform any statistical test, because for our purposes it would suffice to gain intuition on the behaviour of the economic variables simply by looking at the proposed graphs.
 
\textbf{Fact 1} - Over time both price indexes and  GDP have grown steadily, as shown by the first four figures.

\textbf{Fact 2} - Over time their growth rate has been subject to some randomness, as shown by the fifth to the eighth figure of this section.

Further, we show some evidence of expectations (or forecast) of UK GDP growth rate (compiled by Bloomberg) and of the US inflation rate (compiled by the University of Michigan): both series show that the expectations themselves are stochastic, which suggests that the assumption of assuming the processes for the expectations is sensible and consistent with empirical evidence.

\textbf{Fact 3} - Growth rate and inflation expectations are subject to randomness: this is shown by the last two figures of this section.

\begin{figure}[H]
\centering
\begin{minipage}{.5\textwidth}
  \centering
  \includegraphics[width=.8\linewidth]{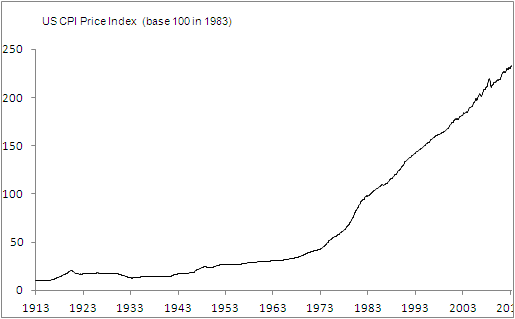} 
    \caption
   {Time series of US CPI Price Index}
  
\end{minipage}%
\begin{minipage}{.5\textwidth}
  \centering
  \includegraphics[width=.8\linewidth]{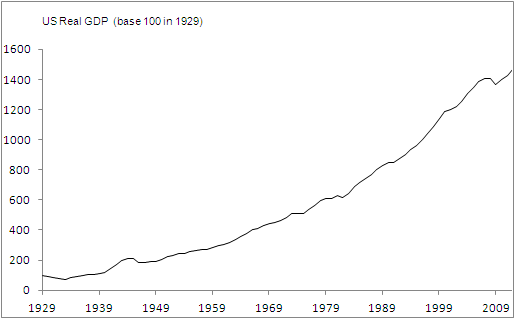}
  \caption
  {Time series of US real GDP.}
 
\end{minipage}
\end{figure}

\begin{figure}[H]
\centering
\begin{minipage}{.5\textwidth}
  \centering
  \includegraphics[width=.8\linewidth]{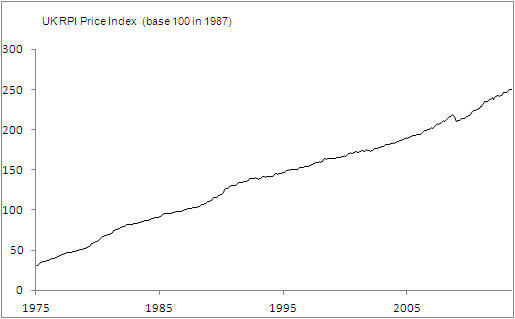}
    \caption
   {Time series of UK RPI price index.}
  
\end{minipage}%
\begin{minipage}{.5\textwidth}
  \centering
  \includegraphics[width=.8\linewidth]{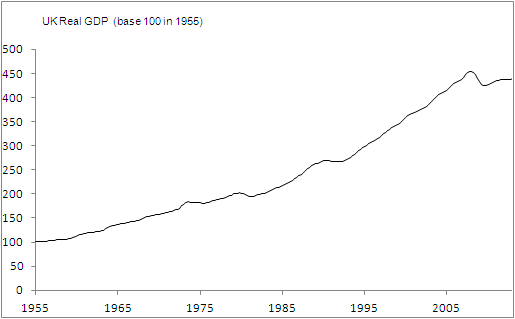}
  \caption
  {Time series of UK real GDP.}
 
\end{minipage}
\end{figure}

\begin{figure}[H]
\centering
\begin{minipage}{.5\textwidth}
  \centering
  \includegraphics[width=.8\linewidth]{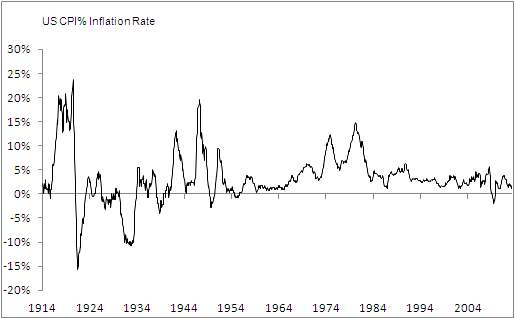}
    \caption
  {Time series of US CPI inflation.}
  
\end{minipage}%
\begin{minipage}{.5\textwidth}
  \centering
  \includegraphics[width=.8\linewidth]{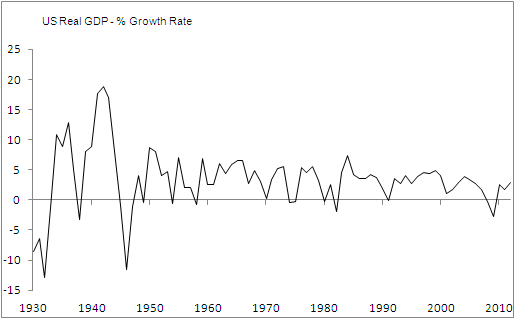}
  \caption
 {Time series of US real GDP growth rate.}
 
\end{minipage}
\end{figure}

\begin{figure}[H]
\centering
\begin{minipage}{.5\textwidth}
  \centering
  \includegraphics[width=.8\linewidth]{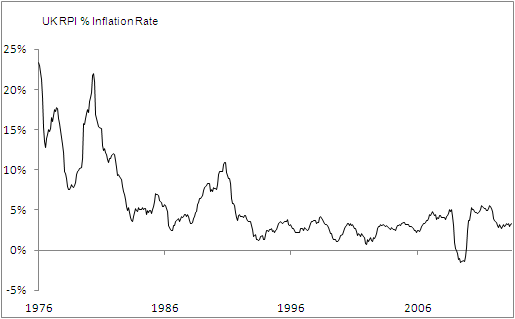}
    \caption
 {Time series of UK RPI inflation.} 
\end{minipage}%
\begin{minipage}{.5\textwidth}
  \centering
  \includegraphics[width=.8\linewidth]{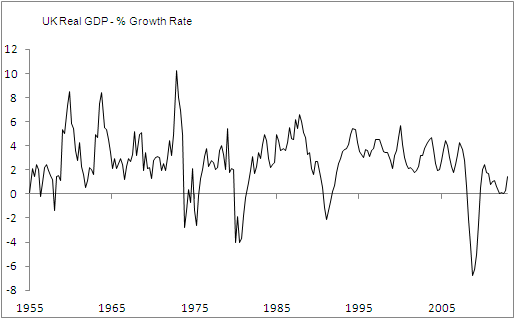}
  \caption
  {Time series of UK real GDP growth rate.}
 
\end{minipage}
\end{figure}

\begin{figure}[H]
\centering
\begin{minipage}{.5\textwidth}
  \centering
  \includegraphics[width=.8\linewidth]{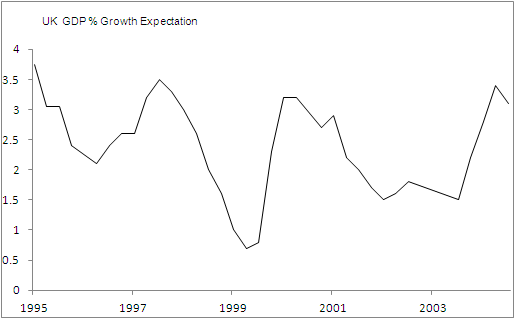}
    \caption
 {{Time series of UK real GDP growth expectations (survey by Bloomberg).}} 
\end{minipage}%
\begin{minipage}{.5\textwidth}
  \centering
  \includegraphics[width=.8\linewidth]{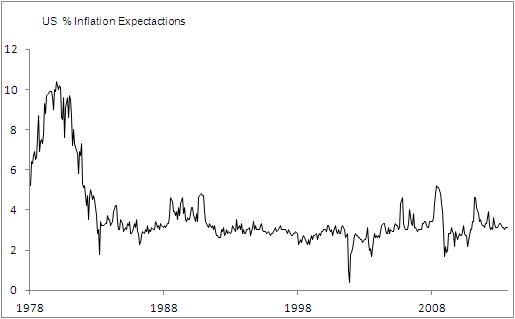}
  \caption
{Time series of US inflation expectations (survey by University of Michigan).}
 
\end{minipage}
\end{figure}

 \subsection{Comparing the DSGE model with the continuous-time model }This section shows that the discrete-time DSGE model and the continuous-time model we propose
can deliver similar distributions for the main economic variables if  parametrized in a consistent way. Therefore in the following section we  choose the continuous-time model to develop the theory  as it is superior compared to the DSGE model as far as its analytical tractability is concerned.

Further, in the following section we  show that one  finds closed form expressions in  the continuous-time model to both the nominal and inflation term structure,  to both nominal rates and inflation options, and to year-on-year inflation forward, without having to resort to the linearisations and approximations proposed earlier in this section when dealing with the discrete-time DSGE model.

In order to obtain similar distributions for the most relevant financial quantities in both cases, one applies a moment-matching technique to the two models.
We assume that all parameters in the continuous-time model are expressed as right-continuous step functions
and that the dimensionality of the Brownian motion is 3.
We focus our attention on second order moments, as the first order moments are straightforward to match.

\textbf{Inflation rate.} In the discrete-time DSGE model, we showed that the variance of the inflation rate  is:
\begin{displaymath}
\textrm{Var}(p_{i})=(K_{2})^{2}(\sigma^{2}\textrm{Var}(u_{i})+ \textrm{Var}(v_{i}))+ \textrm{Var}(z_{i}).
\end{displaymath}
In the next section (see \vref{InflationDiff}) we show that the diffusion term
of the inflation rate (approximated by the ratio \(dI(t)/I(t) \)) is \([b_{I}(t)(T-t)+s_{I}(t)].\)

This implies that the total variance over the first year (\(t=0\) and \(T=1\)) is \(\sum^{3}_{i=1}[b_{I}(0)+s_{I}(0)]^{2}\).

In the discrete time grid, we have \(t_{i-1}=0\) and  \(t_{i}=1\). Therefore it makes sense to
match both 
conditions by requesting that:
\begin{equation}
\textrm{Var}(p_{i})/(t_{i}-t_{i-1})=[(K_{2})^{2}(\sigma^{2}\textrm{Var}(u_{i})+ \textrm{Var}(v_{i}))+ \textrm{Var}(z_{i})]/(t_{i}-t_{i-1})=\sum^{3}_{i=1}[b_{I}(0)+s_{I}(0)]^{2}.
\end{equation}\textbf{Short rate.} A similar method can be applied to the variance of the nominal short rate, that in the DSGE set-up is calculated as:
\begin{displaymath}
\textrm{Var}(n_{i+1})=(\bar n)^{2}(\delta ^{T}K)^{2}\sigma ^{2}\textrm{Var}(u_{i})+ (\bar n)^{2}(-\delta ^{T}K)^{2}\textrm{Var}(v_{i})+(\bar n)^{2}\delta^{2}_{\pi}\textrm{Var}(z_{i}).
\end{displaymath}
Because
the short term nominal rate level at time \(t_{i+1}\) is independent
from its level at the previous time \(t_{i}\) (this follows because the nominal rate is a linear combination of the output gap and inflation, both of which are driven by  Gaussian processes that are independent from their own realisations over time), one  writes the variance of the change in the nominal rate as: \begin{displaymath}
\textrm{Var}(n_{i+1}-n_{i})=\textrm{Var}(n_{i+1})+\textrm{Var}(n_{i})=
\end{displaymath}
\begin{displaymath}
(\bar n)^{2}[(\delta ^{T}K)^{2}\sigma ^{2}\textrm{(Var}(u_{i})_{}+\textrm{Var}(u_{i-1}))+ (-\delta ^{T}K)^{2}\textrm{(Var}(v_{i})+\textrm{Var}(v_{i-1}))+\delta^{2}_{\pi}\textrm{(Var}(z_{i})+\textrm{Var}(z_{i-1}))].
\end{displaymath}
In the next section   (see \vref{ShortRateVol}) we  show that the diffusion term of the nominal short rate differential \(dn(t)\) is:

\begin{displaymath}
-\frac{h_{x}b_{X}(t)+h_{p}b_{I}(t)}{\zeta(t)}.
\end{displaymath}
The matching condition is therefore: \begin{equation}
[(\bar n)^{2}[(\delta ^{T}K)^{2}\sigma ^{2}\textrm{(Var}(u_{i})_{}+\textrm{Var}(u_{i-1}))+ (-\delta ^{T}K)^{2}\textrm{(Var}(v_{i})+\textrm{Var}(v_{i-1}))+\delta^{2}_{\pi}\textrm{(Var}(z_{i})+\textrm{Var}(z_{i-1}))]]/(t_{i}-t_{i-1})=
\end{equation}\begin{displaymath}
=-\frac{h_{x}b_{X}(0)+h_{p}b_{I}(0)}{\zeta(0)}.
\end{displaymath}
We assume that \(\zeta(t)\) is a mere positive calibration function, and that the real positive parameters \(h_{x}\) and \(h_{p}\) are taken exogenously.
In fact, as shown in the next section, they have a precise financial meaning. This said, the purpose of this exercise at
this stage is simply to show that some statistical properties in two different models can be matched.

\textbf{Covariance between nominal short rate and inflation. }The covariance in the DSGE model is:
\begin{displaymath}
\textrm{Cov}(p_{i},n_{i+1})=\bar nK_{2}(\delta ^{T}K)\sigma ^{2}\textrm{Var}(u_{i})+ (\bar nK_{2}(\delta ^{T}K))\textrm{Var}(v_{i})+\bar n\delta_{\pi}\textrm{Var}(z_{i}).
\end{displaymath}

Because
the short term nominal rate level at time \(t_{i}\) is independent
from the inflation level at the same time \(t_{i}\) (as discussed above), this covariance can be interpreted also as \begin{displaymath}
\textrm{Cov}(p_{i},n_{i+1})=\textrm{Cov}(p_{i},n_{i+1}-n_{i}).
\end{displaymath}
The correlation is calculated as follows:

\begin{displaymath}
\textrm{Corr}(p_{i},n_{i+1}-n_{i})=\frac{\bar nK_{2}(\delta ^{T}K)\sigma ^{2}\textrm{Var}(u_{i})+ (\bar nK_{2}(\delta ^{T}K))\textrm{Var}(v_{i})+\bar n\delta_{\pi}\textrm{Var}(z_{i})}{(K_{2})^{2}(\sigma^{2}\textrm{Var}(u_{i})+ \textrm{Var}(v_{i}))+ \textrm{Var}(z_{i}) )^{1/2}\textrm{(Var}(n_{i+1}) +\textrm{Var}(n_{i}))^{1/2}}.
\label{DTIrInflCorr}\end{displaymath}By doing some basic calculations, and by taking into account results \vref{ShortRateVol} and \vref{InflationDiff}, one  calculates the instantaneous covariance
of the nominal short rate change and the inflation rate between times \(t\) and \(T\) :\begin{displaymath}
-[b_{I}(t)(T-t)+s_{I}(t)][\zeta(t)^{-1}(h_{x}b_{X}(t)+h_{p}b_{I}(t))].
\end{displaymath}

 Therefore the matching condition is:
 \begin{equation}
\textrm{Cov}(p_{i},n_{i+1}-n_{i})/(1-0)=-[b_{I}(0)(1-0)+s_{I}(0)][\zeta(0)^{-1}(h_{x}b_{X}(0)+h_{p}b_{I}(0))].
\end{equation}
\textbf{Example}. To show the application of the above methodology, we simulate over the first year the GDP growth rate, the inflation rate, and the short nominal interest rate over 5,000 Monte Carlo trials. We assume that the shocks in the DSGE model are normally distributed with some variances that  are  calibrated below.
The parametrisation we propose has no specific economic meaning, as we only want to show how the moment-matching works in practice.
The assumptions made on the economy and the financial market are the following:

\begin{enumerate}
\item Market agents are risk-neutral. This translates in a \(\sigma\) parameter of 0 in the DSGE model, and in zero market prices of risk, both in the discrete-time and in the continuous-time Gaussian processes.
\item 
The inflation rate is expected to be 3\% in the first year, with a standard deviation of 1.1\%.
This standard deviation can be either an empirical estimate or can be inferred from traded derivatives markets. The source of this standard deviation is not relevant in this exercise.\item 
The output gap is expected to be -2\% in the first year.
\item The potential growth rate of the economy is 2\%.  \item The equilibrium level of the short term nominal rate is 4\%.
   \item  The standard deviation of the nominal short rate is 0.45\%.
The short rate is currently at 2.1\%.\   \item  The central bank is attaching three times more importance to fighting inflation than to stimulating growth.
\item The correlation between the nominal rate change and the inflation rate is positive (given that the central bank is targeting inflation in a very aggressive way), and is 65\%.
\end{enumerate}

The time index at 1 means that the parameter is relative to the first year, that is the point in time that we are simulating. The parametrisation we choose for the DSGE model is the following:
\\

\begin{tabular}{|c|c|c|}\hline
\textbf{Parameter} & \textbf{Value} & \textbf{Comment} \\\hline

\(\sigma\) & 0 & Agents are risk-neutral \\\hline
\(k\) & 0.01 & Prices are sticky \\\hline
\(\delta_{\pi}\) & 3 & The central bank fights inflation aggressively\\\hline
\(\delta_{x}\) & 1 & The central bank is not targeting growth \\\hline
\(\beta\) & 0.95 & Standard subjective discount factor \\\hline
Var(\(u_{1}\)) & 0.01 &  \\\hline
Var(\(v_{1}\)) & 0.01 &  \\\hline
Var(\(z_{1}\)) & 0.0001 &  \\\hline
\end{tabular}

\begin{displaymath}
\end{displaymath}

Moving to the continuous time model, we propose the following parametrisation based on the same economic assumptions proposed above, applying the moment matching conditions stated above. The time indexes are either 0 (initial condition) or 1 (final condition). Between these two times one can think to the parametric functions like \(a_{X}(t)\), \(a_{I}(t)\), \( b_{X}(t)\), \(b_{I}(t)\), \(s_{I}(t)\), and \(s_{X}(t)\) as right-continuous step functions: because the dimensionality of the Brownian motion is 3, there are three values for the volatility vectors specified below. The Monte Carlo simulation has been run in one time step equal to one year. Finally, some model parameters that appear in the  table below have not been explained in the continuous-time construction presented above, but are introduced in the following section. The aim of this section is simply to show that the two models provide results that are broadly in line, however the full explanation of the continuous-time model and of its parameters is given in the following section.\begin{displaymath}
\end{displaymath}
\begin{tabular}{|c|c|c|c|c|}\hline
\textbf{Parameter} & \textbf{Value}  & \textbf{Value (2)} & \textbf{Value (3)} & \textbf{Comment} \\\hline
\(h_{x}\) & 3 &  &  & Central bank focus on growth\\\hline
\(h_{p}\) & 1 &  &  & Central bank focus on inflation\\\hline
\(\bar x\) & 2\%\ &  &  & Central bank target growth\\\hline
\(\bar p\) & 2\%\ &  &  & Central bank target growth\\\hline
\(\zeta(0)\) & 2.015 &  &  & Please refer to next section for this parameter\\\hline

\(a_{X}(0)\) & 0.5\%\ &  &  &  \\\hline
\(a_{I}(0)\) & -1.5\%\ &  &  &  \\\hline
\(\mu_{X}(0)\) & -0.5\%\ &  &  &  \\\hline
\(\mu_{I}(0)\)& 4.5\%\ &  &  &  \\\hline
\(s_{X}(0)\) & 0.01 & 0.01 & 0.01 &  \\\hline
\(s_{I}(0)\) & 0.01 & 0.01 & 0.01 &  \\\hline
\(b_{X}(0)\) & 0.004 & -0.01 & -0.00005 &  \\\hline
\(b_{I}(0)\) & -0.001 & 0.0005 & -0.0005 &  \\\hline

\end{tabular}
\\

\textbf{Results} Here we
show a table comparing the target levels, the  results in the DSGE model, and the results in the continuous-time model for the short rate change and the inflation rate in the first year.
\\

\begin{tabular}{|c|c|c|c|}\hline
\textbf{Statistic} & \textbf{Target} & \textbf{DSGE simulation} & \textbf{Continuous-time simulation} \\\hline
\(\mathbb{E}[n_{i+1}-n_{i}]\) & 2\%\ & 2.01\%\ & 1.98\%\ \\\hline
\(\mathbb{E}[p_{i}]\) & 3\%\ & 3.01\%\ & 2.97\%\ \\\hline
StDev\([n_{i+1}-n_{i}]\) & 0.45\%\ & 0.42\%\ & 0.44\%\ \\\hline
StDev\([p_{i}]\) & 1.1\%\ & 1.08\%\ & 1.07\%\ \\\hline
Corr\([n_{i+1}-n_{i},p_{i}] \)& 65\%\ & 64.16\%\ & 69.07\%\ \\\hline
\end{tabular}
\begin{displaymath}
\end{displaymath}
Here we
show the scatter-plot of the two variables together. Both the marginal  and the joint distributions are matched well.

\begin{figure}[H]
\centering
\begin{minipage}{.5\textwidth} 
  \centering
  \includegraphics[width=.8\linewidth]{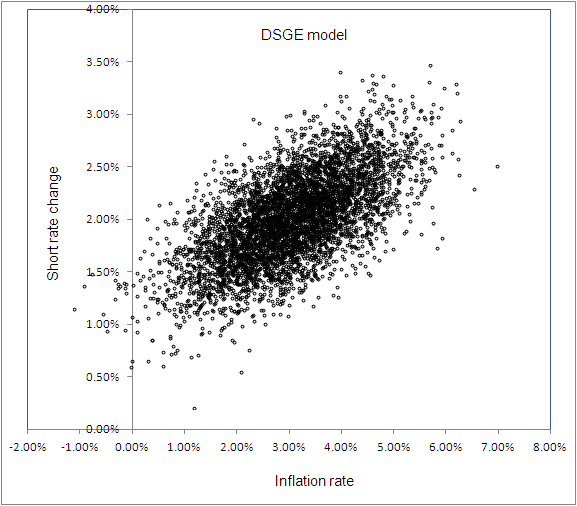}
  \caption
  {Scatter plot of 5,000 Monte Carlo simulations of the DSGE model. }
  \label{fig:test1}
\end{minipage}%
\begin{minipage}{.5\textwidth}
  \centering
  \includegraphics[width=.8\linewidth]{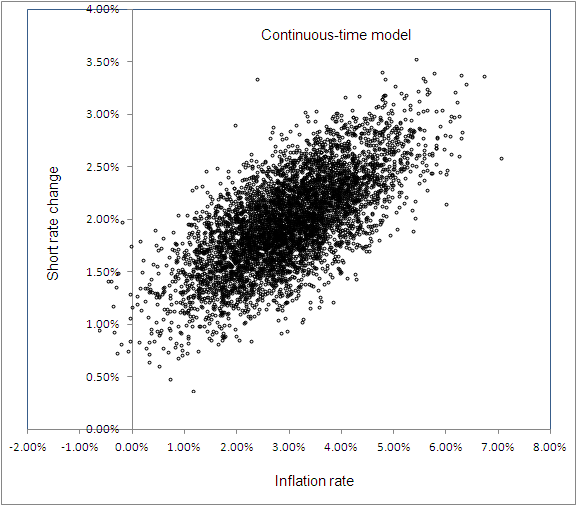}
  \caption
  {Scatter plot of 5,000 Monte Carlo simulations of the continuous-time model. }
  \label{fig:test2}
\end{minipage}

\end{figure}

\section{Inflation derivatives pricing with a central bank reaction function: the CTCB model}

In this section  we propose a continuous-time model to price inflation-linked derivatives  by use of a model that explicitly takes into account the economic dynamics and the central bank behaviour.
Therefore, as it happened in the discrete-time DSGE model analysed in the previous section, the co-movement of interest rates and inflation is not specified exogenously but rather is  the result of  central bank policy.

To achieve this, we make some standard assumptions regarding the structure of the financial market (absence of arbitrage)
and model the relative changes of  both real GDP (Gross Domestic Product) and Price Index  using  Brownian motions with stochastic drifts. Furthermore  the central bank trades nominal bonds to change the money supply in the economy, to keep the growth rate and inflation 
around some pre-specified targets (see for example Walsh \cite{Label20}).
These bond trades have an impact on the nominal bond prices, and therefore on the term structure of interest rates.
Normally inflation-linked pricing models model the co-movement of inflation and nominal interest rates  exogenously, without specifying the economic rationale behind this: we think that bridging the gap between economics and finance is beneficial for both disciplines.\footnote{Another example of inflation-linked pricing model based on sound economic assumptions can be found in Hughston \& Macrina \cite{Label13}, Hughston \& Macrina \cite{Label131}, and Alexander \cite{Label1}: the spirit of these papers has been a source of inspiration for the current model, as they use a microeconomic approach based on Sidrauski \cite{Label155}  to determine the continuous-time dynamics of the price index (however they do not model the central bank reaction function).}

 The advantages of this approach are manifold. Firstly, the dynamics assumed in this model appear to be consistent with the behaviour of central banks in  recent years, when significant purchases of bonds (the so-called \textquotedblleft quantitative easing") have been made since short term interest rates have  reached  and sometimes crossed the zero  bound in many developed economies. One can ask why it is important that a pricing model generates asset prices using  realistic dynamics: after all,  this could be irrelevant once the model has calibrated to a set of market observables. The answer is that such model would fail to minimise hedging profit and loss volatility if a dynamic hedging simulation was run and would make realistic stress testing difficult.\footnote{ Dynamic hedging simulations can be  used to assess the quality of a mode. The idea is to generate some  \textquotedblleft real world" dynamics and assess how the delta hedging done through a model performs, in terms of reducing hedging profit and loss volatility (\textquotedblleft slippages").
Examples of these techniques can be found, for example, in Rebonato \cite{Label001801}.}

Secondly,  this approach  does not rely on the so-called \textquotedblleft Forex Analogy", which assumes the existence of the \textquotedblleft real" economy (see Hughston \cite{Label12}). The Forex analogy has roots in the economic theory (see Fisher \cite{Label1012}). We  use this only as a calculation device, however the quantities we model are all market observables: this makes this model different from the Jarrow-Yildirim model (see Jarrow \& Yildirim \cite{Label15} or Brody, Crosby \& Li \cite{Label222}). The main advantage   is that the model parameters are calibrated in a transparent way to liquid market observables (nominal bonds, inflation swaps, nominal and inflation caps and floors), as opposed to using and  estimating a real rate volatility which is hardly observable in the market.
In practice, one avoids taking costly uncertain-parameters reserves or valuation adjustments, as required by accounting principles.
Examples of  models that do not rely on the Forex Analogy can be found in Dodgson \& Kainth\ \cite{Label1811}, in Mercurio \cite{Label017}
or in Brigo \& Mercurio \cite{Label2}: however none of these models is based on macroeconomic foundations.

Thirdly,
although the model is extremely complex and takes into account many market features (dynamics for the price index and the real economy,  their expectations,  the  no-arbitrage principle, central bank policy, liquidity effects), we show that the dynamics of the short term nominal rate can be reconciled with a well-established short  interest rate model (the generalised Hull-White model, which is a time-varying parameters version of the Ornstein-Uhlenbeck process). This provides both an elegant link with the established theory and some closed forms for interest rates derivatives that are useful for the model calibration.

Fourthly,
zero-coupon and year-on-year forward and options prices  are derived in closed forms:  the model remains  tractable even if is based on  realistic assumptions.

Fifthly, the extension of this model to the open economy is straightforward.

Sixthly, the calibration of this model is computationally not intensive, which allows fast pricing of all kind of trades, from inflation options, to year-on-year caps and floors, to more complex inflation structures as LPI (Limited Price Index).
The main reason for this computational simplicity is that we propose a separable calibration. This point is  fully developed in the following section.

The reader  interested in the inflation derivatives market can refer to some marketing notes edited by investment banks, like Barclays \cite{Label112} or Lehman Brothers \cite{Label15000}, or alternatively Deacon, Derry \& Mirfendereski\ \cite{Label8}, Benaben \cite{Label112221a}, Campbell, Shiller \& Viceira \cite{Label4000}, McGrath \& Windle \cite{Label11411}, \ and J\"ackel \& Bonneton \cite{Label1555}.\ 
\section{Inflation model assumptions}
\subsection{Probabilistic set-up }
\begin{enumerate}
\item The model is set in continuous time \(t\).
Time is a positive real number and is expressed in years. From this point the notation for the continuous-time variable \(y\) at time \(t\) is \(y(t)\).\item  Randomness is modelled via a \textit{n}-dimensional \(\mathbb{P}\)-Brownian motion \(\{W(t)\}_{t\geq0}\).
The probability measure \(\mathbb{P}\) is fully defined in the next point. The \textit{n} components of the Brownian motion \(W(t)\) are independent.\item  We work with the  probability triplet \(\{\Omega, \mathcal{F}, \mathbb{P}\} \) equipped with the natural filtration \(\{\mathcal{F}_{t}\}_{t\geq0}\)  generated by the Brownian motion \(\{W(t)\}_{t\geq0}\).  All filtration-related concepts  are defined with respect to this  filtration. In particular \(\mathbb{P}\) is the real-world (\textquotedblleft physical") probability measure. 
If no probability measure is specified, the expectation is taken with respect to the real-world measure (\(\mathbb{P})\). 
\item Derivatives pricing is carried out in the \(\mathbb{P} \) measure via the pricing kernel (defined below), or in the risk-neutral measure \(\mathbb{Q}\) (defined as the pricing measure which is  characterised by having    the money market account \(B(t)  \) as numeraire), or finally in the \(T-\)forward measure \(\mathbb{Q}^{T}\), defined as the pricing measure using the bond price \(P(t,T)\) as numeraire. The    money market account \(B(t)  \)  and the bonds  \(P(t,T)\) are defined in detail in the next section.  Expectations of a payoff \textit{\(\Pi\)} taken under the generic measure \(\mathbb{M}\) given the information available at time \textit{t} are denoted as \(\mathbb{E}^{\mathbb{M}}[\textit{\(\Pi\)} |\mathcal{F}_{t}]\) or alternatively \(\mathbb{E}^{\mathbb{M}}_{t}[\textit{\(\Pi\)} ].\)
\end{enumerate}
\subsection{Financial instruments}
All  instruments listed below and their related quantities are modelled as \(\{\mathcal{F}_{t}\}_{t\geq0}\)-adapted stochastic processes and   are regular enough to ensure the existence of the expectations  introduced later. The list of instruments is not exhaustive but only contains the ones  needed to build the model.
\begin{enumerate}
\item \textbf{Nominal zero-coupon bonds}, that pay with certainty (i.e. they are risk-free) one unit of currency at time \(T\), have  price \(P(t,T) \) at time \(t\). There exists a continuum of bond prices, i.e. \(T\in[t,+\infty) \).\footnote{When calibrating the model to market observables, this assumption is relaxed because only a finite amount of bond maturities are quoted on the market.}
From the nominal bond prices one  derives all kind of rates, for example instantaneous forward rates \(f(t,T)= - \partial\log(P(t,T))/\partial T\) and the short rate \(n(t)=f(t,t)\). See Brigo  \&  Mercurio \cite{Label2} for further details.
\item \textbf{Money market account} \(B(t)\), with dynamics \(dB(t)=n(t)B(t)dt\), \(B(0)=1\).
\item \textbf{Price Index }\(I(t)\), which is a positive stochastic process that reflects the price level of the economy. Its dynamics are specified later in the economy set-up.
\item \textbf{\textbf{Zero-Coupon Inflation  Index Swap}s} \(ZCIIS(t,T)\): the inflation leg pays \(I(T)/I(t) -1\), while the fixed leg pays \((\,1\,+ \,K(t,T))^{T-t}-1\). Both payments happen at maturity \textit{T} and there is no time lag. The inflation breakeven \(K(t,T)\) is agreed at time \textit{t}: there exists a continuum of inflation breakevens, i.e. \(T\in[t,+\infty)\footnote{When calibrating the model to market observables, this assumption is relaxed because only a finite amount of inflation swaps maturities are quoted on the market.}\). In a zero-coupon inflation swap the strike
\(K(t,T) \) is such that at inception the expected value at maturity of the swap is zero:  \( \mathbb{E}^{\mathbb{Q}}_{t}[((I(T)/I(t)-(1+K(t,T))^{T-t})B(t)/B(T))]=0 \). 
\item \textbf{Inflation  bonds}
\(P^{I}(t,T)\), which pay at time \textit{T }the level of the price
index \(I(T) \), with no time lag. There exists a continuum of inflation bond prices, i.e. \(T\in[t,+\infty)\). They are also known as \textquotedblleft linkers". Because  we are working in a market without liquidity concerns, the inflation dynamics implied by the inflation bond prices are the same as the ones implied by the inflation swap market. The zero-coupon linker price is \(P^{I}(t,T)=\mathbb{E}^{\mathbb{Q}}_{t}[I(T)B(t)/B(T)] \).
Normally these bonds have an implicit deflation floor: the bond holder will not pay the issuer in case of deflation. This feature is ignored in this work.\end{enumerate}
\subsection{Financial market}
The assumptions  regarding the financial market are standard:\begin{enumerate}
\item There is no credit risk in the economy.
\item The financial market is arbitrage-free. A thorough treatment of absence of arbitrage and its implications can be found in Bj\"ork \cite{Label3}, Cochrane \cite{Label6}, or Duffie \cite{Label10}.
\item Assuming that we use the money market account \(B(t)\) as numeraire, we are working in the risk-neutral measure \(\mathbb{Q}\). This implies that the bond price \(\mathbb{Q}\)-dynamics are given by
\begin{equation}
dP(t,T)/P(t,T)=n(t)dt \ +\sigma_{P}(t,T)\cdot dW^{\mathbb{Q}}(t)\label{Q_BondDyn}
\end{equation} 
 where the bond volatility \(\sigma_{P}(t,T)  \) is an \textit{n}-dimensional deterministic  process.\footnote{Given two \textit{n}-dimensional vectors \(a,b\), with components \(a_{1},...,a_{n}\) and  \(b_{1},...,b_{n}\) respectively, the notation \(a\cdot b\) is equivalent to \(\sum^{n}_{i=1}a_{i}b_{i}\). This notation is used extensively in this work. Under no circumstances this notation has to be confused with a Stratonovich integral.}  These volatilities are referred to as \textquotedblleft model" volatilities in the calibration section, as opposed to volatilities implied by market prices of options. The form of these bond volatilities is left general at this point, however at a later stage we  characterise them fully in terms of model parameters.\item 
The Radon-Nikodym derivative \(L(t)=d\mathbb{Q}/d\mathbb{P} |_{\mathcal{F}_{t}}\) has the  dynamics: \(dL(t) =-\,\ L(t)\lambda(t)\cdot dW^{\mathbb{P}}(t)\), where \(\lambda(t)\) is an \textit{n}-dimensional deterministic  process. 
The process \(\left\{\lambda(t)\right\}_{t\geq0}\) is called \textquotedblleft market price of risk". \item 
The pricing kernel \(\psi(t)  \) defined as \(\psi(t)=L(t)/B(t) \) has dynamics: \(d\psi(t)/\psi(t)=-n(t)dt\,-\,\lambda(t)\cdot dW^{\mathbb{P}}(t)\).
The pricing kernel has many useful properties, however for these purposes we remember that \(P(t,T)=\mathbb{E}^{\mathbb{P}}_{t}[\psi(T)/\psi(t)]\).
Further analysis of the pricing kernel properties can be found in Constantinides \cite{Label7}, Hughston \cite{Label11}, Leippold \& Wu \cite{Label16}, Shefrin \cite{Label00180}, and Rogers \cite{Label00181}.\item 
The real bond,
which is not an  asset traded on the market, is defined as the ratio between the inflation bond and the current price index level: \(P^{R}(t,T)=P^{I}(t,T)/I(t)\).\footnote{It is worth stressing that, although the model proposed in this paper does not require the concept of real bond and real rates, these  are often found in the literature:  therefore it is useful to show how these quantities can be recovered in the present set-up. We  use it as a calculation device to obtain the year-on-year forward.} Both \(P^{I}(t,T)\) and \(I(t)\) have been defined previously. Using the same logic  above, from the real bond prices one  extracts  a real term structure of interest rates: in particular one  defines the real short rate \(r(t)=f^{R}(t,t)\), where  \(f^{R}(t,T)= - \partial\log(P^{R}(t,T))/\partial T\).
The process \(\left\{r(t)\right\}_{t\geq 0}\) is \(\{\mathcal{F}_{t}\}_{t\geq0}\)-adapted and can be used to define the real money market account \(B^{R}(t),\) with dynamics \(dB^{R}(t)=r(t)B^{R}(t)dt\):  \(B^{R}(t)\) is   locally riskless in the real risk-neutral measure \(\mathbb{Q}^{R}\)   (as  \(B(t)\) is in the \(\mathbb{Q}\) measure).
One  also defines the real pricing kernel \(\psi^{R}(t)=I(t)\psi(t) \): similarly one  shows that \(P^{R}(t,T)=\mathbb{E}^{\mathbb{P}}_{t}[\psi^{R}(T)/\psi^{R}(t)]\). Furthermore, one  introduces the \(\{\mathcal{F}_{t}\}_{t\geq0}\)-adapted process \(\left\{\lambda^{R}(t)\right\}_{t\geq0}\), called \textquotedblleft real market price of risk" and obtain the dynamics: \(d\psi^{R}(t)=-r(t)\psi^{R}(t)dt\,-\,\psi^{R}(t)\lambda^{R}(t)\cdot dW^{\mathbb{P}}(t)\). 
  \item 
The definition of the real bond, real rates and real pricing kernel is sufficient to define another economy, labelled \textquotedblleft real" economy. This is the cornerstone of the so-called \textquotedblleft Forex Analogy" (see Hughston \cite{Label12}, Hughston \cite{Label11}, or Brigo \& Mercurio \cite{Label2}): because one  writes \(I(t)=\psi^{R}(t)/\psi(t)\), one  regards the price index \(I(t)\) as the exchange rate between the real and the nominal economy. With this  one 
obtains, in analogy with the FX spot rate drift, the risk-neutral drift for the price index: \(\mathbb{E}^{\mathbb{Q}}_{t}[I(t+dt)-I(t)]=(n(t)-r(t))I(t)dt\).
\end{enumerate}

\subsection{Economy dynamics and central bank role}
We make some assumptions regarding the economy. With these assumptions  one  makes a realistic description of the economy based on the continuous-time model, that  is based on the widely-used\  discrete-time DSGE macroeconomic model (presented and discussed in the previous section).\begin{enumerate}
\item 
At time \textit{t}, the closed economy is described by three positive quantities \(X(t), I(t), \) and \(M(t),\) that represent the real output of the economy, the price level in the economy, and the money supply respectively.
The real output is an alternative expression for the real Gross Domestic Product, also referred to as real GDP. Money supply is defined as the total amount of cash available in the economy: in this paper we do not distinguish between the different monetary aggregates like M1, M2, ... . The processes \(\left\{X(t)\right\}_{t\geq0}\) , \(\left\{ I(t)\right\}_{t\geq0}\), and  \(\left\{M(t)\right\}_{t\geq0}\) are positive \(\{\mathcal{F}_{t}\}_{t\geq0}\)-adapted processes.
\item 
The \(\mathbb{P} \)-dynamics of instantaneous output and price index are defined as follows:
\begin{equation}
dX(t)=X(t)[m_{X}(t)dt\:+s_{X}(t)\cdot\:dW^{\mathbb{P}}(t)]
\end{equation} 
\begin{equation}
dI(t)=I(t)[m_{I}(t)dt\:+s_{I}(t)\cdot\:dW^{\mathbb{P}}(t)]
\end{equation} 
where
\(m_{X}(t)\) and \(m_{I}(t)\) are one-dimensional stochastic \(\{\mathcal{F}_{t}\}_{t\geq0}\)-adapted processes whose dynamics are to be defined below, and \(s_{X}(t)\) and \(s_{I}(t)\) are \textit{n}-dimensional  deterministic processes. 
These volatilities are referred to as \textquotedblleft model" volatilities in the calibration section, as opposed to volatilities implied by market prices of options. The choice of  Brownian motions for the real output and price index relative change processes is reasonable as these quantities are always positive and have historically shown an upward trend with some noise\footnote{In order to ensure that this assumption is reasonable, we have run a normality test (Jarque-Bera) to the time series of the Eurozone GDP (quarterly readings) and Consumer Price Index (monthly readings) for the last 30 years. In both cases the normality assumptions is accepted.}. In particular we showed in the previous section that the above two equations for the growth and inflation rate can be derived from a well-specified macroeconomic model.
\item
The dynamics of the expectations  are modelled using the SDEs:\begin{equation}
dm_{X}(t)=a_{X}(t)dt+b_{X}(t)\cdot\:dW^{\mathbb{P}}(t)\label{dmx}
\end{equation}
\begin{equation}
dm_{I}(t)=a_{I}(t)dt+b_{I}(t)\cdot\:dW^{\mathbb{P}}(t)\label{dmi}
\end{equation} where the processes  \(a_{X}(t)\) and \(a_{I}(t)\) are one-dimensional  deterministic processes  and \(b_{X}(t)\) and \(b_{I}(t)\) are \textit{n}-dimensional  deterministic processes.
\item
We assume that the central bank
is the only institution responsible for  money supply. The central bank uses the money supply as a policy tool and tries to keep the economy close to a target annual growth rate \(\bar x\) and to a target  annual inflation rate  \(\bar p\). The targets \( \bar x\) and \( \bar p\) are constant real numbers. According to standard macroeconomic theory, an increase in money supply can increase both the price level and the output: the central bank can attach more importance to the growth target or to the price stability.
The
relative importance of these two goals is modelled with the two real positive constants  \(h_{x}\) and \(h_{p}\). 
 To summarise the above assumptions,  we assume that the central bank policy is explained by the newly proposed \(\mathbb{P}\)-dynamics
\begin{equation}
dM(t)/M(t)=-h_{p}(dI(t)/I(t)-  \bar pdt)-h_{x}(dX(t)/X(t)-  \bar xdt)+s_{M}(t)\cdot\:dW^{\mathbb{P}}(t).
 \label{React} 
 \end{equation}
This curve is also known as \textquotedblleft central bank reaction function".\footnote{The above expression for the reaction function attaches more importance to intuition than to formal correctness: if one wanted to write an expression containing only stochastic differentials and not involving ratios of stochastic differentials (like \(dM(t)/M(t)\), \(dI(t)/I(t)\), or \(dX(t)/X(t)\)) one can define the reaction function in logarithmic differential terms and adjust the equilibrium levels from \(\bar p\) and \(\bar x\) to \(\bar p^{*}\) and \(\bar x^{*}\) respectively for the change in drifts:\begin{equation}
d\log M(t)=[-h_{p}(d\log I(t)-  \bar p^{*}dt)-h_{x}(d\log X(t)-  \bar x^{*}dt)+s_{M}(t)\cdot\:dW^{\mathbb{P}}(t)].
 \label{ReactLog} 
 \end{equation}In the rest of the paper we will not be using the above expression and will develop our theory using \ref{React}.}  Here \(s_{M}(t)\) is an \textit{n}-dimensional deterministic process that measures the uncertainty around the central bank policy.
These volatilities are referred to as \textquotedblleft model" volatilities in the calibration section, as opposed to volatilities implied by market prices of options. The above equation can be read as follows: modulo some uncertainty (modelled by the term \(s_{M}(t)\cdot\:dW(t)\)), the central bank  reduces the money supply (both \( -h_{p}\) and \(-h_{x}\) are negative real numbers\footnote{Because the so-called \textquotedblleft quantitative easing" has been implemented only in the last few years by some central banks, it is not possible to provide  data-based estimates of these parameters for the moment.}) when inflation or output growth are above their targets. It should be noted that the above specification for the central bank policy is similar to the Taylor rule (see Walsh \cite{Label20}, Woodford \cite{Label21}, Taylor \cite{Label18},  Clarida, Dali \& Gertler\ \cite{Label51}, or Clarida, Dali \& Gertler \cite{Label5}): however, because the Taylor rule assumes that the short term interest rate is the monetary policy tool (as opposed to the money supply), the Taylor rule can lead to negative policy rates, while in a low rates environment central banks tend to use open market operations as policy tools.\footnote{This consideration is even more relevant in the current low rates environment, where the main option left to the central banks in the USA, UK, Japan and the Euro area is to purchase bonds to stimulate and reflate the economy (so-called \textquotedblleft quantitative easing").}\item 
The central bank changes the money supply by trading in the secondary bond market, which has some feedback effects on bond prices. These effects are known by  market participants and discounted into the market prices. The central bank can also target some specific sectors of the yield curve, for example it can decide to sell short maturity bonds and buy longer maturities bonds to make the curve flatter while not inflating its balance sheet.\footnote{The  \textquotedblleft operation twist" implemented by the FED in 2011 is a good example.}
We assume that the relative increase in the money supply has a linear relationship with the relative increase in the bond prices, weighted for each maturity \textit{T} by a weight function \(Z(T)\). These effects are priced by the market and  are modelled by the  \(\mathbb{Q}\)-dynamics for the money supply:
\begin{equation}
dM(t)/M(t) = \gamma dt+\int^{t+\Omega}_{t}Z(T)[dP(t,T)/P(t ,T)]dT+s_{L}(t)\cdot dW^{\mathbb{Q}}(t).\label{MonetaryPolicyBondImpact2}
\end{equation}
Here \( \gamma  \) is a real constant that models the natural growth of the money supply;  \(Z(T) \) is a real, positive and increasing deterministic scalar function of  the bond  maturity \textit{T}.
The request that the function \(Z(T) \) is always positive comes from economic considerations: if bond prices increase, nominal rates decrease, which is equivalent to saying that the money supply goes up; in this framework the interest rate itself is not the policy tool, because all monetary policy is modelled via the money supply \(M(t)\). The integral in the above expression is  a deterministic one, given that at time \(t\) the quantity \(dP(t,T)/P(t ,T)\) is known for each maturity \(T\) and therefore deterministic: the integral in the above expression has to be regarded as a way to weight the impact of relative changes in the bond prices across the different maturities \(T\in[ t,t+\Omega]
 \) of the term structure.\footnote{An  observation similar to the comment made on the reaction function  can be done at this stage: the liquidity relationship can be rewritten as: \begin{equation}
d\log M(t)=[\gamma dt+\int^{t+\Omega}_{t}Z(T)[d\log P(t,T)+1/2[\sigma _{P}(t,T)\cdot \sigma _{P}(t,T)]^{}]dT+s_{L}(t)\cdot dW^{\mathbb{Q}}(t)].\label{MonetaryPolicyLogBondImpact}
\end{equation}We will not be using the above expression in the model theory development, but rather \vref{MonetaryPolicyBondImpact2}.\\ } The real positive constant \(\Omega>0\) represents the time horizon used by the central bank to purchase or sell nominal bonds in order to influence the money supply \(M(t)\). For example, if the central bank is trading bonds up to the 30 years maturity, one sets the parameter \(\Omega\) to 30.
\\ Uncertainty around this relationship is captured by the stochastic differential, multiplied by a liquidity volatility deterministic  \textit{n}-dimensional process 
\( \{ s_{L}(t)\}_{t\geq0}\) .
\item We finally require the following relationship to hold:
\begin{equation}
h_{p} s_{I}(t)+h_{x} s_{X}(t)- s_{M}(t)=0.  \label{LiqLocRiskless}
\end{equation}
It should be noted that the reaction function is still stochastic as the drifts are stochastic, and that the liquidity relationship is stochastic as the short rate is stochastic. We also note that the above condition ensures that the diffusion term for the \(\mathbb{P}\)-dynamics \ref{React}  is the same as the diffusion term for the \(\mathbb{Q}\)-dynamics \ref{MonetaryPolicyBondImpact2}, therefore satisfying Girsanov's theorem.

\end{enumerate}

\section{CTCB Model construction}

We  build the pricing model, which is referred to as \textquotedblleft Continuous-Time Central Bank" (CTCB) model in the following sections. The assumptions made so far can be regarded as standard no-arbitrage assumptions in the financial market, in conjunction with some reasonable assumptions on growth and inflation rate (modelled as  Brownian motions with some stochastic drifts --- historically GDP and price levels have shown an upward trend with some noise). Furthermore,   the central bank trades nominal bonds to keep the economy around some target levels, and this has some (wanted) effects on the bond prices, and hence on the yield curve.

The model construction that follows  puts together the
financial market and macroeconomic assumptions to obtain a pricing framework that is consistent both with the economic theory and the no-arbitrage principle.

\textbf{Step 1} - The \(\mathbb{Q}\)-dynamics of the economic variables
and  expectations are
obtained with Girsanov theorem:
\begin{equation}
dm_{X}(t)=(a_{X}(t)-\lambda(t)\cdot b_{X}(t))dt+b_{X}(t)\cdot\:dW^{\mathbb{Q}}(t)\label{MX_diff}
\end{equation}
\begin{equation}
dm_{I}(t)=(a_{I}(t)-\lambda(t)\cdot b_{I}(t))dt+b_{I}(t)\cdot\:dW^{\mathbb{Q}}(t)\label{MI_diff}
\end{equation}
\begin{equation}
dX(t)/X(t)=(m_{X}(t)-\lambda(t)\cdot s_{X}(t))dt\:+s_{X}(t)\cdot\:dW^{\mathbb{Q}}(t) \label{Q_GDP}
\end{equation} 
\begin{equation}
dI(t)/I(t)=(m_{I}(t)-\lambda(t)\cdot s_{I}(t))dt\:+s_{I}(t)\cdot\:dW^{\mathbb{Q}}(t)\label{Q_PriceIndex}.
\end{equation} 

\textbf{Step 2} - Similarly, the \(\mathbb{Q}\)-dynamics for the central bank policy
are obtained using Girsanov theorem: \begin{equation}
dM(t)/M(t)=-h_{p}(dI(t)/I(t)-  \bar pdt)-h_{x}(dX(t)/X(t)-  \bar xdt)-\lambda(t)\cdot s_{M}(t)dt+s_{M}(t)\cdot\:dW^{\mathbb{Q}}(t).\label{Q_CentralBankPolicy}
\end{equation} 

\textbf{Step 3} - Putting together the central bank policy equation \ref{Q_CentralBankPolicy} and the economy dynamics (equations \ref{Q_GDP} and \ref{Q_PriceIndex}) in the risk-neutral measure we get:
\begin{displaymath}
dM(t)/M(t)=-h_{p}((m_{I}(t)-\lambda(t)\cdot s_{I}(t))dt\:+s_{I}(t)\cdot\:dW^{\mathbb{Q}}(t)-  \bar pdt)
\end{displaymath}
\begin{equation}
-h_{x}((m_{X}(t)-\lambda(t)\cdot s_{X}(t))dt\\ \\ \:+s_{X}(t)\cdot\:dW^{\mathbb{Q}}(t)-\bar xdt)-\lambda(t)\cdot s_{M}(t)dt+s_{M}(t)\cdot\:dW^{\mathbb{Q}}(t).\label{Q_CentralBankPolicyWithEco}
\end{equation}

\textbf{Step 4} - Equating the central bank policy equation \ref{Q_CentralBankPolicyWithEco} and  equation \ref{MonetaryPolicyBondImpact2}, which models the impact of central bank policy on bond prices,    we obtain:
\begin{displaymath}
\gamma dt+\int^{t+\Omega}_{t}Z(T)[dP(t,T)/P(t ,T)]dT+s_{L}(t)\cdot dW^{\mathbb{Q}}(t)=-h_{p}((m_{I}(t)-\lambda(t)\cdot s_{I}(t))dt\:+s_{I}(t)\cdot\:dW^{\mathbb{Q}}(t)-  \bar pdt)
\end{displaymath}
\begin{equation}
-h_{x}((m_{X}(t)-\lambda(t)\cdot s_{X}(t))dt\\ \\ \:+s_{X}(t)\cdot\:dW^{\mathbb{Q}}(t)-\bar xdt)-\lambda(t)\cdot s_{M}(t)dt+s_{M}(t)\cdot\:dW^{\mathbb{Q}}(t).\label{Q_CentralBankPolicyWithEcoandLiq}
\end{equation} 
\textbf{Step 5} - Combining the above  and the no arbitrage
condition for the bond price dynamics (\ref{Q_BondDyn}),  we obtain:
\begin{displaymath}
\gamma dt+\int^{t+\Omega}_{t}Z(T)[n(t)dt \ +\sigma_{P}(t,T)\cdot dW^{\mathbb{Q}}(t)]dT+s_{L}(t)\cdot dW^{\mathbb{Q}}(t)=-h_{p}((m_{I}(t)-\lambda(t)\cdot s_{I}(t))dt\:+s_{I}(t)\cdot\:dW^{\mathbb{Q}}(t)-  \bar pdt)
\end{displaymath}
\begin{equation}
-h_{x}((m_{X}(t)-\lambda(t)\cdot s_{X}(t))dt\\ \\ \:+s_{X}(t)\cdot\:dW^{\mathbb{Q}}(t)-\bar xdt)-\lambda(t)\cdot s_{M}(t)dt+s_{M}(t)\cdot\:dW^{\mathbb{Q}}(t).\label{Q_CentralBankPolicyWithEcoandLiqNoArb}
\end{equation}
We compact the notation by introducing the following quantities:
\begin{displaymath}
\zeta(t)=\int^{t+\Omega}_{t}{{Z(T)}}dT
\end{displaymath}
\begin{displaymath}
\Sigma_{P}(t)=\int^{t+\Omega}_{t}{{Z(T)}}\sigma_{P}(t,T)dT.
\end{displaymath} \\ We note that the quantity \(\zeta(t)\) is always strictly positive because the function \(Z(T)\) and the constant \(\Omega\) are requested to be strictly positive. With the above definitions the model equation becomes: \begin{displaymath}
\zeta (t)n(t)dt  +\Sigma_{P}(t)\cdot dW^{\mathbb{Q}}(t)=
-h_{p}[(m^{\mathbb{}}_{I}(t)-\lambda(t)\cdot s_{I}(t))dt\:+s_{I}(t)\cdot\:dW^{\mathbb{Q}}(t)-  \bar pdt]
\end{displaymath}
\begin{displaymath}
  -h_{x}[(m_{X}(t)-\lambda(t)\cdot s_{X}(t))dt+s_{X}(t)\cdot\:dW^{\mathbb{Q}}(t)-\bar xdt]
\:-\lambda(t)\cdot s_{M}(t)dt+s_{M}(t)\cdot\:dW^{\mathbb{Q}}(t) -\gamma dt-{s_{L}(t)\cdot\:dW^{\mathbb{Q}}(t). }
\end{displaymath}

\textbf{Step 6} - The no-arbitrage conditions
are obtained from the above equation by collecting  the terms multiplied by \(dt\) and  \(dW(t)\) in the following way: 
\begin{equation}
 \gamma=h_{p}  \bar p
  +h_{x}\bar x
  \end{equation}
\begin{equation}
 \zeta (t)n(t) \ =-h_{p}[m_{I}(t)-\lambda(t)\cdot s_{I}(t)]
  -h_{x}[m_{X}(t)-\lambda(t)\cdot s_{X}(t)]-\lambda(t)\cdot s_{M}(t) \label{IrCalibCOND}
\end{equation}

\begin{equation}
 \Sigma_{P}(t)=\int^{t+\Omega}_{t}{{z(T)}}\sigma_{P}(t,T)dT=-h_{p}s_{I}(t)
  -h_{x}s_{X}(t)+s_{M}(t)\textcolor[rgb]{0,0,0}{{-s_{L}(t)}}.\label{NoArbCond_dW}
\end{equation}
We note that we can equate the equation above to \(s_{L}(t)\) thanks to condition \ref{MonetaryPolicyBondImpact2}:\\ \begin{equation}
 \Sigma_{P}(t)=\int^{t+\Omega}_{t}{{z(T)}}\sigma_{P}(t,T)dT=-h_{p}s_{I}(t)
  -h_{x}s_{X}(t)+s_{M}(t)\textcolor[rgb]{0,0,0}{{-s_{L}(t)}}=\textcolor[rgb]{0,0,0}{{-s_{L}(t)}}.
\end{equation}Some 
observations can be made regarding these  conditions:
\begin{enumerate}
\item The first condition can be regarded as the risk-neutral drift for the money supply assuming that there is no uncertainty and no monetary policy.
Here we group all deterministic terms multiplied by \(dt\). Therefore we can refer to the constant \(\gamma\) as the natural money supply growth rate. We also note that the constant \(\gamma\) is likely to be positive, given that the central bank reaction function parameters \(h_{_{X}}\) and \(h_{P}\)  are positive by construction and that the target levels \(\bar x\) and \(\bar p\) are normally positive numbers. This matches the intuition that over time the money supply tends to grow, unless the central bank tries to reduce it. \item The second calibration conditions gives us a closed-form expression for the short rate that is used in the following section to get the short rate dynamics. If  remembers the  condition \ref{LiqLocRiskless} 
\begin{displaymath}
h_{p} s_{I}(t)+h_{x} s_{X}(t)- s_{M}(t)=0 
\end{displaymath} the calibration condition simplifies into\begin{equation}
\zeta (t)n(t) \ =-h_{p}m_{I}(t)
  -h_{x}m_{X}(t)\label{IRCalibNoArb} 
\end{equation}which shows that the second calibration condition contains all stochastic terms multiplied by \(dt\), given that \(m_{X}(t)\) and \(m_{I}(t)\) are stochastic.
\item 
The bond volatilities  \(\sigma_{P}(t,T)\) are determined  by other model parameters and can be regarded as a combination  of the economic factor volatilities. This is clear from the third calibration condition, which contains all terms multiplied by \(dW(t)\). There are constraints on  \(s_{M}(t)\) (see condition \ref{LiqLocRiskless}) for which  one needs to write  \(\Sigma_{P}(t)=-s_{L}(t)\).  \end{enumerate}

 \textbf{Step 7} - To price inflation derivatives, it can be useful to work with the \(T^{*}\)-forward measure. By using the techniques detailed for example in Brigo \& Mercurio \cite{Label2}, one   obtains the inflation and bond price dynamics under this measure:
\begin{equation}
dI(t)/I(t)=(m_{I}(t)-\lambda(t)\cdot s_{I}(t) \,+\sigma_{P}(t,T^{*})\cdot s_{I}(t)\,)dt\:+s_{I}(t)\cdot\:dW^{T^{*}}(t)\label{QTInflDyn}
\end{equation}
\begin{equation}
dP(t,T)/P(t,T)=(n(t) \,+s_{P}(t,T)\cdot \sigma_{P}(t,T^{*})\,)dt\:+s_{P}(t,T)\cdot\:dW^{T^{*}}(t).\label{QTBondDyn}
\end{equation}

 We close this section with an observation on instantaneous correlations. In this model the same \(n\)-dimensional Brownian motion is the  source of randomness for all variables. One remembers that, given two stochastic differential equations \(dX(t)=a\,dW^{1}(t)+b\,dW^{2}(t) \) and \(dY(t)=c\,dW^{1}(t)+f \,dW^{2}(t) \) (where  \(\{W^{1}(t)\}_{t\geq0}\) and \(\{W^{2}(t)\}_{t\geq0}\) are two independent one-dimensional Brownian motions and \(a,b,c, \) and \(f\) are deterministic real constants), one obtains \(dX(t)dY(t)=(ac+bf)dt\). Perhaps we can write a general formula for the instantaneous correlation \(\rho_{t}\) using quadratic variations and covariation \(\rho_{t}=\frac{d\left\langle X, Y\right\rangle_{t}}{\sqrt{d\left\langle X, X\right\rangle_{t}d\left\langle Y, Y\right\rangle_{t}}}=\frac{(ac+bf)}{\sqrt{a^{2}+b^{2}}\sqrt{c^{2}+d^{2}}}.\)
Therefore one can in principle use the model volatilities to calibrate also market-implied instantaneous correlations between the macroeconomic variables: this idea is further developed in the following sections. 
\section{Equivalent interest rates model}Here we show that the model presented in the previous section, although is a completely new model and is derived from macroeconomic assumptions, yields some dynamics  for the short rate that follow a mean-reverting Hull-White model. The Hull-White model, presented in Hull \& White \cite{Label14a} and further analysed in Brigo \& Mercurio \cite{Label2}, is a widely-used model for the short rate \(n(t)\) that has the properties to be mean-reverting and to calibrate to any given term structure of interest rates. Here we show how this model is derived within the macroeconomic framework and study its mean-reverting property as a function of the economy parameters.
 \\ The derivation is carried out as follows. The second calibration condition \ref{IrCalibCOND}
gives an expression containing the short term interest rate \(n(t)\):
\begin{displaymath}
\zeta (t)n(t) \ =-h_{p}[m_{I}(t)-\lambda(t)\cdot s_{I}(t)]
  -h_{x}[m_{X}(t)-\lambda(t)\cdot s_{X}(t)]-\lambda(t)\cdot s_{M}(t).
\end{displaymath}

If one differentiates this condition and remembers the  condition \ref{LiqLocRiskless} 
 one gets:

\begin{displaymath}
d\zeta (t)n(t)+\zeta (t)dn(t)=-h_{p}dm_{I}(t)-h_{x}dm_{X}(t).
\end{displaymath}

 There is no covariance term in the left-hand side of the above differential given that \(\zeta (t)\) is a deterministic quantity. One  remembers the expressions for the drift differentials \ref{MX_diff} and   \ref{MI_diff}  and substitute them in the above expression, obtaining:
\begin{displaymath}
d\zeta (t)n(t)+\zeta (t)dn(t)=-h_{p}[[a_{I}(t)-\lambda(t)\cdot b_{I}(t)]dt+b_{I}(t)\cdot dW^{\mathbb{Q}}(t)]
-h_{x}[[a_{X}(t)-\lambda(t)\cdot b_{X}(t)]dt+b_{X}(t)\cdot dW^{\mathbb{Q}}(t)].
\end{displaymath}

 Further, one needs to calculate the differential of \(\zeta (t)\):
\begin{displaymath}
d\zeta (t)=\left(\frac{\partial\int^{t+\Omega}_{t}{{Z(T)dT}}}{\partial t}\right)dt=[Z(t+\Omega)-Z(t)]dt.
\end{displaymath}
One substitutes the above in the differential, and after rearranging one obtains:
\begin{displaymath}
\zeta (t)dn(t)=-[Z(t+\Omega)-Z(t)]n(t)dt-h_{p}[[a_{I}(t)-\lambda(t)\cdot b_{I}(t)]dt+b_{I}(t)\cdot dW^{\mathbb{Q}}(t)]+
\end{displaymath}
\begin{displaymath}
-h_{x}[[a_{X}(t)-\lambda(t)\cdot b_{X}(t)]dt+b_{X}(t)\cdot dW^{\mathbb{Q}}(t)].
\end{displaymath}
We can  express the differential of the short rate \(n(t)\):
\begin{displaymath}
dn(t)=-[Z(t+\Omega)-Z(t)]/\zeta (t)n(t)dt-h_{p}/\zeta (t)[[a_{I}(t)-\lambda(t)\cdot b_{I}(t)]dt
\end{displaymath}
\begin{displaymath}
+b_{I}(t)\cdot dW^{\mathbb{Q}}(t)]-h_{x}/\zeta (t)[[a_{X}(t)-\lambda(t)\cdot b_{X}(t)]dt+b_{X}(t)\cdot dW^{\mathbb{Q}}(t)].
\end{displaymath}

 To compact notation one defines the following terms:
\begin{equation}
f_{2}(t)=[Z(t+\Omega)-Z(t)]/\zeta (t)   \label{MeanRevSpeed}
\end{equation}
\begin{equation}
f_{1}(t)=[-h_{p}a_{I}(t)-h_{x}a_{X}(t)]/\zeta (t) \label{MeanRevLevel}
\end{equation}
\begin{equation}
\sigma_{n}(t)=[-h_{x}b_{X}(t)-h_{p}b_{I}(t)]/\zeta (t). \label{ShortRateVol}
\end{equation}
This shows that the model implies some short nominal interest rates dynamics that are similar to the ones assumed by the generalised Vasicek model:\begin{displaymath}
dn(t)=[f_{1}(t)-f_{2}(t)n(t)-\lambda(t)\cdot\sigma_{n}(t)]dt+\sigma_{n}(t)\cdot dW^{\mathbb{Q}}(t).
\end{displaymath}It is important to notice that the requests made on the function \(Z(T)\) (to be an increasing and positive function) imply that \(f_{2}(t)\) is always positive, i.e. that the  nominal short rate is mean-reverting: therefore this condition makes sense both from a financial perspective (it guarantees mean reversion) and from a monetary macroeconomic perspective, given that these request implies that higher bond prices, and therefore lower interest rates, are equivalent to a higher money supply.
Before doing some further analysis, we notice that the source of randomness in the CTCB model is \(n\)-dimensional, and the volatility function \(\sigma_{n}(t)\) is \(n\)-dimensional accordingly. To stress the difference against the original Hull-White model, where the driving Brownian motion is scalar, we write the scalar Hull-White volatility as  \(\sigma_{n}^{*}(t)\): one  links the two processes by asking that the total  variance of the source of randomness of the CTCB model is the same as the total variance of the Hull-White model. The relationship is:\begin{equation}
[\sigma_{n}^{*}(t)]^{2}=\sum^{n}_{i=1}[\sigma_{n}^{i}(t)]^{2}\label{HWScalarVectorVol}
\end{equation} where \(\sigma_{n}^{i}(t)\) is the \textit{i}-th component of the \textit{n}-dimensional model volatility function \(\sigma_{n}(t)\).\\ Consistently with the Hull-White model, the distribution of the short rate is Gaussian and can generate negative short nominal rates: in the current low rates environment, when central banks are explicitly setting  negative deposit rates, we don't think this is a theoretical problem, rather we think that this model is probably better suited than other positive rates models to deal with the current market conditions. In Denmark the central bank set the short interest rate to  -0.2\%, the European Central Bank to -0.1\%: in practice central banks can set a negative short rate to stimulate commercial banks to lend to consumers and firms in times of economic distress. \\ The only differences w.r.t. to the original extended Vasicek model is that the volatility
is a multidimensional function of time, and that the driving source of randomness is a multidimensional Brownian motion: as explained in detail in point 4 below, one  finds the volatility  vector components to target a certain level of total volatility, and therefore the marginal distribution. However these differences
do not prevent us from reaching the following conclusions:
\begin{enumerate}
\item  If we only want to use this model to price interest rates derivatives, one calibrates the function \(f_{1}(t)\) to the nominal forward rates observed in the market, as suggested in the original Hull-White paper, modulo some changes. Alternatively, one  uses the calibration condition \(\mathbb{E}^{T^{*}}_{t}[n(T)]=f(t,T).\) \item   The nominal bond price is such that the volatility of the relative moves is a deterministic volatility function. This is no surprise given the original assumptions. This  is important because it can simplify the calculation of the year-on-year convexity adjustment, as shown in the following sections.
\item  Thanks to the above fact one  uses Black-Scholes formulas to price European bond options.
In the relevant forward measure  one  carries out discounting by simply multiplying by the market bond prices. Because it is trivial to price bond options in this model,  one  writes Black-type formulas for bond options. These yield closed forms for nominal cap/floors and swaptions  (using the method presented in Jamshidian \cite{Label15a}): this intuition is developed in the following sections.
 \item Because the process for the short rate is normal, trees can be easily constructed. In fact the \textit{n}-dimensional Brownian motion can be treated as a one-dimensional process for this purpose (this technique is also called \textquotedblleft flattening", where the independent components of the Brownian motion are \textquotedblleft summed" and considered as a single Brownian motion with the appropriate diffusion term). \end{enumerate}
\section{Further analysis on the mean-reversion property}
\subsection{General case}  Here we analyse the mean-reversion coefficient found in the previous section (result \ref{MeanRevSpeed}) and we link it to the general theory of mean-reverting Gaussian models developed by Hull \& White \cite{Label14a}. Here we refer to the original formulation of the model, where the driving Brownian motion \(\sigma_{n}^{*}(t)\) is a scalar process: one  translates it into vector by using \ref{HWScalarVectorVol}. This is not a major problem, as one can \textquotedblleft flatten" the vector volatility into an equivalent scalar volatility that leaves the total variance unchanged.\\ In particular, in that paper the authors present a version of the model with time-dependent coefficients, where the dynamics of the short rate are governed by the SDE: \begin{equation}
dn(t)=[\theta(t)-a(t)n(t)-\lambda(t)\sigma_{n}(t)]dt+\sigma_{n}^{*}(t)dW^{\mathbb{Q}}(t).
\end{equation} For simplicity we pose the parameter \(b=0\), using the notation of the Hull-White original paper. They suggest a calibration strategy that yields the model parameters as functions of two functions used to fit the term structure of interest rates using the Ansatz \(P(t,T)=A(t,T)e^{-n(t)B(t,T)}\).
At the initial time \(t=0 \) the positive functions \(A(0,T)\) and \(B(0,T)\) are numerically calibrated to  the market term structure \(P(0,T)\). \\ In particular, the Hull \& White\ find that the mean reversion speed \(a(t)\) has to satisfy the condition:\begin{equation}
a(t)=-\frac{\frac{\partial B^{2}(0,t)}{\partial t^{2}}}{\frac{\partial B^{}(0,t)}{\partial t^{}}}=-\frac{\beta ^{''}(t)}{\beta ^{'}(t)}   \label{MeanRevSpeedHW}
\end{equation} \\ where we have made the notation lighter by defining: \(\beta(t)=B(0,t).\\  \) Further,
the authors prove a calibration condition for the mean reversion level parameter \(\theta(t)\):\begin{equation}
\theta(t)=\lambda(t)\sigma^{*}_{n}(t)-a(t)\frac {\partial \log A(0,t)}{\partial t}-\frac {\partial ^{2}\log A(0,t)}{\partial t^{2}}+\left[ \frac {\partial  B(0,t)}{\partial t} \right]^{2}\int^{t}_{0}\left[\frac{\sigma^{*}_{n}(s)}{\frac {\partial  B(0,s)}{\partial s}}  \right]^{2}ds.
\end{equation}\\  
  At this stage we observe that the time-dependent version of the Hull-White model does not necessarily imply mean reversion: in fact, the mean reversion speed coefficient \(a(t)\) is positive (i.e. there is mean reversion) only if the sign of the first derivative \(\beta ^{'}(t)\) is different from the sign of the second derivative \(\beta ^{''}(t)\). \\ For example, if one takes \(A(0,t)=1  \) (which is a legitimate choice) and if \(n(0)>0\),  the function \(B(0,t) \) is increasing where the term structure is upward-sloping. Let us introduce the compound spot rate for maturity \(T\) observed at time \(t\) and denote it by \(Y(t,T)  \): it is defined as the flat interest rate such that: \(P(t,T)=e^{-Y(t,T)(T-t)}\). If \(A(0,t)=1, \)  this expression has to be equal to \(P(0,T)=e^{-n(0)B(0,T)}\). Equalling the two terms one gets \(Y(0,T)=n(0)\frac{B(0,T)}{T}\). 

 We focus our attention on the mean-reversion speed, and equate the result from  Hull \& White \cite{Label14a} to the expression found in the previous section. One  draws some conclusions on the function \(Z(T)\). By equating \ref{MeanRevSpeed} and \ref{MeanRevSpeedHW} one gets:
\begin{displaymath}
-\frac{\beta ^{''}(t)}{\beta ^{'}(t)}=[Z(t+\Omega)-Z(t)]/\zeta (t)
\end{displaymath}

\begin{displaymath}
\int^{t+\Omega}_{t}{{Z(T)}}dT=-[Z(t+\Omega)-Z(t)]
\frac{\beta ^{'}(t)}{\beta ^{''}(t)}.
\end{displaymath}
We take a derivative of the above expression w.r.t. \(t\):\begin{displaymath}
[Z(t+\Omega)-Z(t)]=-[Z^{'}(t+\Omega)-Z^{'}(t)]
\frac{\beta ^{'}(t)}{\beta ^{''}(t)}-[Z(t+\Omega)-Z(t)]
\frac{(\beta ^{''}(t))^{2}-\beta ^{'}(t)\beta ^{'''}(t))}{(\beta ^{''}(t))^{2}}.
\end{displaymath}One  rearranges the above expression as:
\begin{displaymath}
\frac{[Z^{}(t+\Omega)-Z^{}(t)]}{[Z'(t+\Omega)-Z'(t)]}=-\frac{\frac{\beta ^{'}(t)}{\beta ^{''}(t)}}{\left( \frac{2(\beta ^{''}(t))^{2}-\beta ^{'}(t)\beta ^{'''}(t)}{(\beta ^{''}(t))^{2}} \right)}=-\frac{\beta ^{'}(t)\beta ^{''}(t)}{2(\beta ^{''}(t))^{2}-\beta ^{'}(t)\beta ^{'''}(t)}.
\end{displaymath}By defining \(u(t)=Z(t+\Omega)-Z(t)\)
and \(\alpha(t)=-\frac{\beta ^{'}(t)\beta ^{''}(t)}{2(\beta ^{''}(t))^{2}-\beta ^{'}(t)\beta ^{'''}(t)}\) 
and requiring that \(\beta ^{'}(t)\neq0\), \(\beta ^{''}(t)\neq0\), one rewrites the above expression as:\begin{displaymath}
\frac{u'(t)}{u(t)}=\frac{d\log\ u(t)}{dt}=\alpha^{-1}(t).
\end{displaymath}
One recalls that the first two conditions \(\beta ^{'}(t)\neq0\), \(\beta ^{''}(t)\neq0\) have already been used in the above examples and are equivalent to requiring that the term structure is not flat (a trivial case that never happens in practice) and that the term structure is not explosive, so that mean reversion can be ensured.\\   This linear ODE is solved in \(t>t_{0}\) to yield \(u(t)=u(t_{0})e^{\int^{t}_{t_{0}}\alpha^{-1}(s)ds}\).

The meaning of the above result is a relationship between the functions \(Z(t)  \) and \(\beta(t)\):\begin{equation}
[Z(t+\Omega)-Z(t)]=[Z(t_{0}+\Omega)-Z(t_{0})]e^{\int^{t}_{t_{0}}-\frac {2(\beta ^{''}(s))^{2}-\beta ^{'}(s)\beta ^{'''}(s)}{\beta ^{'}(s)\beta ^{''}(s)}   ds}
\label{zetabeta}.\end{equation}
We make two observations from this expression:\begin{enumerate}
\item 
To ensure mean reversion, one has to require that the function \(Z(t)  \)  is positive and increasing: this request was made in the previous section and is  confirmed by looking at the properties of the Hull-White model, in particular \ref{zetabeta}.
In fact, if  \(Z(t)  \)  is positive and increasing, \(Z(t_{0}+\Omega)>Z(t_{0})\) (an exponential is always positive), being consistent with the request \(Z(t+\Omega)>Z(t)\).\item One 
uses the relationship above to calibrate the function  \(Z(t)  \)   as  a function of \(\beta(t)\) and \(A(0,t) \) (i.e. the term structure of interest rates): alternatively one  sets the functions \(A(0,t)  \) and  \(Z(t)   \) using some functional forms, and obtain the function  \(\beta(t)\) by calibrating to the term structure. The latter seems more appropriate, as we may want to make some assumptions on the market liquidity function   \(Z(t)  \). \end{enumerate} One also  impose a further calibration constraint on the model bond volatilities. In the previous section we have introduced the bond volatilities \(\sigma_{P}(t,T) \) without specifying more details: taking in consideration result \ref{ShortRateVol} and the bond volatility formula in the Hull-White model, one  imposes a further calibration condition:\begin{equation}
\sigma_{P}(t,T)=[h_{x}b_{X}(t)+h_{p}b_{I}(t)]/\zeta (t)\left[ \frac{\beta(T)-\beta(t)}{\beta'(t)} \right].
\end{equation}\subsection{Constant mean reversion speed  }We conclude this section with an observation regarding the constant mean reversion speed of the Hull-White model, that is used in many applications. The main result we find is that if one imposes that the function \(Z(T)\) is an exponential in the form \(Z(T)=e^{\delta T}\) (with \(\delta >0\) to ensure that \(Z(t)\) is increasing), one immediately shows that the mean reversion speed has the property: \begin{equation}
[Z(t+\Omega)-Z(t)]/\zeta (t)=\frac{Z(t+\Omega)-Z(t)}{\int^{t+\Omega}_{t}{{Z(T)}}dT}=\delta.
\end{equation}This result is interesting from a theoretical perspective, because a higher mean reversion speed decreases the intra-curve rates correlation, as it is well known in literature, and this has a similar meaning as increasing the parameter \(\delta\): a higher parameter \(\delta\) means that the longer maturities of the curve react more strongly to monetary policy compared to the short end of the curve, therefore increasing the intra-curve decorrelation.\\   
However this result has a very practical implication too: the relationship \ref{zetabeta}
can be difficult to implement numerically, as one wants to impose the liquidity function \(Z(t)\) and imply \(\beta(t)\) (the converse would be trivial): this could reduce the flexibility of the model. \\ However, we remember result \ref{MeanRevSpeedHW} and obtain the differential equation \(\delta=-\frac{\beta ^{''}(t)}{\beta ^{'}(t)}\), which  is solved by \(\beta(t)=e^{-\delta t}\). Therefore, known \(Z(t)\), one has \(B(0,t)=\beta(t)\). By observing from the market the short rate \(n(0)\) and the term structure \(P(0,t)\), one uses the Ansatz \(P(0,T)=A(0,T)e^{-n(0)B(0,T)}\) to obtain \(A(0,t)=\frac{P(0,t)}{e^{-n(0)e^{-\delta t}}}\), which fully calibrates the model to the nominal yield curve. This result is exploited in the calibration process in the following section. From this point we  assume that \(Z(T)=e^{\delta T}\).
\\ By doing some calculations one sees how  the form \(Z(T)=e^{\delta T}\) compares with the relationship \ref{zetabeta}:\begin{displaymath}
[Z(t+\Omega)-Z(t)]=[Z(t_{0}+\Omega)-Z(t_{0})]e^{\int^{t}_{t_{0}}-\frac{2(\beta ^{''}(s))^{2}-\beta ^{'}(s)\beta ^{'''}(s)}{\beta ^{'}(s)\beta ^{''}(s)} ds}
\end{displaymath}

\begin{displaymath}
e^{\delta (t+\Omega)}-e^{\delta t}=[e^{\delta (t_{0}+\Omega)}-e^{\delta t_{0}}]e^{\int^{t}_{t_{0}}\frac{2\delta^{4}-\delta^{4}}{\delta^{3}}ds}=[e^{\delta (t_{0}+\Omega)}-e^{\delta t_{0}}](e^{\delta (t-t_{0})}).
\end{displaymath}
Finally it is useful to show how the choice of  \(Z(T)=e^{\delta T}\) is consistent with  the classical integration of the Ohrstein-Uhlenbeck process: in fact, given the SDE \(dn(t)=[\theta(t)-an(t)]dt+\sigma^{*}_{n}(t)dW^{\mathbb{}}(t)\) with known initial condition \(n(s)\), one writes the differential for \(n(t)e^{at}\), integrates and obtains the standard result:\begin{displaymath}
n(t)=n(s)e^{-a(t-s)}+\int^{t}_{s}e^{-a(t-u)}\theta (u)du+\int^{t}_{s}e^{-a(t-u)}\sigma^{*}_{n} (u)dW(u).
\end{displaymath}In the above formula \(\sigma^{*}_{n} (u)\) refers to the Hull-White scalar short rate volatility. If one  remembers the calibration condition \ref{IrCalibCOND} one  substitutes the expression for \(\zeta(t)\) inside it and confirm that one gets the same result stated in the above formula. In fact:\begin{displaymath}
\zeta(t)=\int^{t+\Omega}_{t}{{Z(T)}}dT=\int^{t+\Omega}_{t}{{e^{\delta T}}}dT=\frac{e^{\delta (t+\Omega)}-e^{\delta t}}{\delta}=\frac{e^{\delta t}(e^{\delta\Omega }-1)}{\delta}.
\end{displaymath}
If one recalls the calibration condition \ref{IrCalibCOND} and the  condition \ref{LiqLocRiskless} one writes:\begin{displaymath}
\zeta (t)n(t) \ =-h_{p}[m_{I}(t)-\lambda(t)\cdot s_{I}(t)]
  -h_{x}[m_{X}(t)-\lambda(t)\cdot s_{X}(t)]-\lambda(t)\cdot s_{M}(t)=-h_{p}m_{I}(t)-h_{x}m_{X}(t).
\end{displaymath}One  integrates the SDEs \vref{dmi} and \vref{dmx} and obtains:
 \begin{displaymath}
\frac{e^{\delta t}(e^{\delta\Omega }-1)}{\delta}n(t) \ =-[h_{p}m_{I}(s)+h_{x}m_{X}(s)]-\int^{t}_{s}[h_{p}a_{I}(u)+h_{x}a_{X}(u)]du-\int^{t}_{s}[h_{p}b_{I}(u)+h_{x}b_{X}(u)] \cdot dW(u)
\end{displaymath}
\begin{displaymath}
n(t)=n(s)e^{-\delta(t-s)}-\int^{t}_{s}\frac{[h_{p}a_{I}(u)+h_{x}a_{X}(u)]\delta e^{-\delta( t-u)}}{(e^{\delta\Omega }-1)}\frac{1}{e^{\delta u}}du-\int^{t}_{s}\frac{[h_{p}b_{I}(u)+h_{x}b_{X}(u)]\delta e^{-\delta( t-u)}}{(e^{\delta\Omega }-1)}\frac{1}{e^{\delta u}}\cdot dW(u)
\end{displaymath}where we recall the calibration condition \vref{IRCalibNoArb}  written as \(n(s)=\frac{-[h_{p}m_{I}(s)+h_{x}m_{X}(s)]}{\zeta(s)}\). 

Further calculations yield:
\begin{displaymath}
n(t)=n(s)e^{-\delta(t-s)}-\int^{t}_{s}\frac{[h_{p}a_{I}(u)+h_{x}a_{X}(u)] e^{-\delta( t-u)}}{\zeta(u)}du-\int^{t}_{s}\frac{[h_{p}b_{I}(u)+h_{x}b_{X}(u)]\delta e^{-\delta( t-u)}}{\zeta(u)}\cdot dW(u).
\end{displaymath}
If one remembers the definitions \ref{MeanRevLevel} and \ref{ShortRateVol}, one finally obtains the desired result:\begin{displaymath}
n(t)=n(s)e^{-\delta(t-s)}+\int^{t}_{s}e^{-\delta(t-u)}\theta (u)du+\int^{t}_{s}e^{-\delta(t-u)}\sigma_{n} (u)\cdot dW(u).
\end{displaymath}In the above formula we stress that  the function \(\sigma_{n} (u)\) is the CTCB vector short rate volatility.
\section{Pricing of vanilla interest rates derivatives}Finding that our macro-based CTCB model yields a short rate model that is a Hull-White model makes the pricing of interest rates derivatives much simpler.
The main result is that, because bond prices are lognormally distributed, one  uses Black-type formulae to price bond options. Bond options are used also to find the prices of caplets and floorlets, to price caps and floors: this is explained for example in Brigo \& Mercurio \cite{Label2}. For swaptions, the method suggested by Jamishidian \cite{Label15a} can be followed.
Closed forms allow faster pricing of vanilla interest rates derivatives, therefore speeding up the calibration. We quote some results that are useful and that can be found for example in Hull \& White \cite{Label14a} and\ Brigo \& Mercurio \cite{Label2}. In the following calculations we  use the  quantity  \(P(t,T_{1},T_{2}),   \) defined as a portfolio of a long bond \(P(t,T_{2})\) and a short bond \(P(t,T_{1}) \).\begin{lemma} The undiscounted price of a European vanilla option on a lognormally distributed asset X(T) with strike K, whose logarithm  has    expectation  \(\mathbb{E}[\log X(T)]=M \) and variance \(Var[\log X(T)]=V^{2}\), is:  \begin{equation}
\mathbb{E}[\omega(X-K)^{+}]=\omega e^{M+1/2V^{2}}N(\omega(M-\log(K)+V^{2})/V)-\omega KN(\omega(M-\log(K))/V)\label{LogNormalOptionPrice}
\end{equation} where the function N(x) is the cumulative standard Gaussian distribution, i.e. \(N(x)=\int^{x}_{-\infty}(2\pi)^{-1}e^{-\frac{t^{2}}{2}}dt \) and \(\omega\in \{-1,1 \} \) for puts and calls respectively.\end{lemma} 
 \begin{lemma} The variance between times \(t\) and \(T_{1}\) of the  quantity \(P(t,T_{1},T_{2})  \), with \(t<T_{1}<T_{2}\), in the Hull-White model is \begin{equation}
V_{P}
(t,T_{1},T_{2})=[\beta(T_{2})-\beta(T_{1})]^{2}\int^{T_{1}}_{t}\left[ \frac{\sigma^{*}_{n}(u)}{\beta^{'}(u)} \right]^{2}du.
 \end{equation}
\end{lemma} 

\begin{lemma} The price at time \(t\)  of  an option  on the  quantity  \(P(t,T_{1},T_{2}) \) with option maturity \(T_{1}\) and option strike \(K\) in the Hull-White model with \(t<T_{1}<T_{2}\) is \begin{equation}
ZBO(call, t,T_{1},T_{2},K)=P(t,T_{2})N(h)-KP(t,T_{1})N(h-(V_{P}
(t,T_{1},T_{2}))^{\frac{1}{2}})
\end{equation}\begin{equation}
ZBO(put, t,T_{1},T_{2},K)=-P(t,T_{2})N(-h)+KP(t,T_{1})N((V_{P}
(t,T_{1},T_{2}))^{\frac{1}{2}}-h)
\end{equation}where \(h=\frac{1}{V_{P}
(t,T_{1},T_{2}))^{\frac{1}{2}}}\log\left[\frac{P(t,T_{2})}{P(t,T_{1})K}\right]+\frac{V_{P}
(t,T_{1},T_{2}))^{\frac{1}{2}}}{2}\).
\end{lemma} 

\ \begin{lemma} The  price at time \(t\) of a caplet (floorlet) with maturity \(T_{1}\), strike K, and notional M, on the forward rate between times \(T_{1}\) and \(T_{2}\), denoted as \(F(t,T_{1},T_{2})\), is the  price of a put (call) option with strike \((1+K(T_{2}-T_{1}))^{-1}\), notional \(M(1+K(T_{2}-T_{1}))\),  maturity  \(T_{1}\) on the quantity  \(P(t,T_{1},T_{2})\).
The result is model independent and we assume  \(t<T_{1}<T_{2}\). \begin{equation}
Caplet(t,T_{1},T_{2},K,N)=M(1+K(T_{2}-T_{1}))ZBO(put, t,T_{1},T_{2},(1+K(T_{2}-T_{1}))^{-1})
\end{equation}\begin{equation}
Floorlet(t,T_{1},T_{2},K,N)=M(1+K(T_{2}-T_{1}))ZBO(call, t,T_{1},T_{2},(1+K(T_{2}-T_{1}))^{-1}).
\end{equation}\end{lemma} 

\begin{lemma} The price of   a coupon-bearing bond option with maturity T in the Hull-White model  is equivalent to pricing a portfolio of zero-coupon bond options  using the special nominal rate \(n^{*}\).
Coupons paid at times \(T_{i}>T\) are denoted by \(c_{i}\). The last coupon includes the payment of the notional.\begin{equation}
CBO(call,t,T,T_{1 }...T_{M},c_{1}...c_{M},K)=\sum ^{M}_{i=1}c_{i}ZBO(call, t,T_{},T_{i},P(t,T_{i},n^{*}))
\end{equation}
\begin{equation}
CBO(put,t,T,T_{1 }...T_{M},c_{1}...c_{M},K)=\sum ^{M}_{i=1}c_{i}ZBO(put, t,T_{},T_{i},P(t,T_{i},n^{*})).
\end{equation}\end{lemma} 
 \begin{lemma} The price at time \(t\)  of   a payer (P) swaption (that gives the right to enter at time T into a payer swaption with fixed rate K and payment dates \(T_{i}>T\)) is equivalent to the price of a coupon-bearing bond option.
The result is model-independent.\begin{equation}
Swtpn(P,t,T,T_{1}
...T_{M},K)=CBO(put,t,T,T_{1 }...T_{M},c_{1}...c_{M},K)=\sum ^{M}_{i=1}c_{i}ZBO(put, t,T,T_{i},X_{i})\end{equation}
\begin{equation}
Swtpn(R,t,T,T_{1}
...T_{M},K)=CBO(call,t,T,T_{1 }...T_{M},c_{1}...c_{M},K)=\sum ^{M}_{i=1}c_{i}ZBO(call, t,T_{},T_{i},X_{i})\end{equation}
\begin{displaymath}
X_{i}=A(T,T_{i})e^{-n^{*}B(T,T_{i})}.
\end{displaymath}\end{lemma} 
 \begin{lemma} The price of   a  swaption with strike \(K \), maturity \(T\) and payment dates \(T_{i}>T\) in the Hull-White model is:
\begin{equation}
S(P,t,T,T_{1}
...T_{M},K)=\sum ^{M}_{i=1}c_{i}ZBO(put, t,T,T_{i},X_{i})=\sum^{M}_{i=1}c_{i}[-P(t,T_{i})N(-h_{i})+X_{i}P(t,T_{i-1})N((V_{P}
(t,T_{i-1},T_{i}))^{\frac{1}{2}}-h_{i})]
\end{equation}
\begin{equation}
S(R,t,T,T_{1}
...T_{M},K)=\sum ^{M}_{i=1}c_{i}ZBO(call, t,T_{},T_{i},X_{i})=\sum^{M}_{i=1}c_{i}P(t,T_{i})N(h_{i})-X_{i}P(t,T_{i-1})N(h_{i}-(V_{P}
(t,T_{i-1},T_{i}))^{\frac{1}{2}})
\end{equation}
\begin{displaymath}
h_{i}=\frac{1}{V_{P}
(t,T_{i},T_{i-1}))^{\frac{1}{2}}}\log\left[\frac{P(t,T_{i})}{P(t,T_{i-1})X_{i}}\right]+\frac{V_{P}
(t,T_{i-1},T_{i}))^{\frac{1}{2}}}{2}
\end{displaymath}
\begin{displaymath}
X_{i}=A(T,T_{i})e^{-n^{*}B(T,T_{i})}.
\end{displaymath}
\end{lemma}

 We can  start pricing derivatives based on the above results using the macroeconomic model defined in the previous sections and leveraging on the equivalent short rate model.  

 \begin{lemma} The price of   bond options, caplets and floorlets, and swaptions in the CTCB model  follow the formulas proposed above with the following parametrisation\begin{displaymath}
[\sigma^{*}_{n} (t)]^{2}= \sum^{n}_{i=1}\{[-h_{x}b^{i}_{X}(t)-h_{p}b^{i}_{I}(t)]/\zeta (t)\}^{2}\ 
\end{displaymath}where n
is the dimensionality of the Brownian motion W(t). Here \(b^{i}_{X}(t) \) is the i-th component of the volatility vector \( b_{X}(t)\), and \(  b^{i}_{I}(t)\)  is the i-th component of the volatility vector \( b_{I}(t). \)\end{lemma}    \section{Pricing zero-coupon inflation swaps and options}
In this section we calculate the full expression for the price index \(I(t) \): its conditional lognormality  translates into closed forms (\textquotedblleft Black type") for  zero-coupon Inflation options.
This
makes the model calibration much faster. The price index dynamics in the forward measure can be used to simplify the problem by discounting via multiplication by the zero-coupon bond.

To do these analyses, we  calculate the closed form dynamics of \(I(t)\) taking into account the stochastic dynamics of its drift \(m_{I}(t). \)\\ We start by obtaining their \(T^{*}\)-forward dynamics:
\begin{displaymath}
dI(t)/I(t)=(m_{I}(t)-\lambda(t)\cdot s_{I}(t) \,+\sigma_{P}(t,T^{*})\cdot s_{I}(t)\,)dt\:+s_{I}(t)\cdot\:dW^{T^{*}}(t)
\end{displaymath}
\begin{displaymath}
dm_{I}(t)=[a_{I}(t)-\lambda(t)\cdot b_{I}(t)+\sigma_{P}(t,T^{*})\cdot b_{I}(t)]dt+b_{I}(t)\cdot dW^{{T^{*}}}(t).
\end{displaymath}
We compact the notation by defining \(g_{1}(t)=-s_{I}(t)\cdot(\lambda(t)-\sigma_{P}(t,T^{*}))\) and \(g_{2}(t)=a_{I}(t)-b_{I}(t)\cdot(\lambda(t)-\sigma_{P}(t,T^{*}))\).

We notice that \(g_{2}(t)\) is deterministic as all the quantities used to build it are deterministic. At this stage we recall that in the CTCB model the bond option volatilities are expressed as: \begin{displaymath}\sigma_{P}(t,T^{})=
\frac{[h_{x}b_{X}(t)+h_{p}b_{I}(t)]}{(e^{\delta (t+\Omega)}-e^{\delta t})}(1-e^{\delta^{(T-t)}}).
\end{displaymath}\\ The dynamics are therefore rewritten in a more compact form as:
\begin{displaymath}
dI(t)/I(t)=(m_{I}(t)+g_{1}(t))dt+s_{I}(t)\cdot\:dW^{T^{*}}(t)
\end{displaymath}\begin{displaymath}
dm_{I}(t)=g_{2}(t)dt+b_{I}(t)\cdot dW^{{T^{*}}}(t).
\end{displaymath}

We  integrate the
expression for  \(m_{I}(s)\) between times \(t\) and \(T\):
\begin{displaymath}
\int^{T}_{t}m_{I}(s)ds=m_{I}(t)(T-t)+\int^{T}_{t}\int^{s}_{t}g_{2}(u)duds+\int^{T}_{t}\int^{s}_{t}b_{I}(u)\cdot dW^{{T^{*}}}(u)ds. 
\end{displaymath}
Applying Fubini's theorem, we  write the integral of the price index drift in a simpler form:\begin{displaymath}
\int^{T}_{t}m_{I}(s)ds=m_{I}(t)(T-t)+\int^{T}_{t}(T-s)g_{2}(s)ds+\int^{T}_{t}(T-s)b_{I}(s)\cdot dW^{{T^{*}}}(s).
\end{displaymath}We   write the normal distribution of the  integral of the drift:\begin{displaymath}
\int^{T}_{t}m_{I}(s)ds\sim\mathcal{N}\left(m_{I}(t)(T-t)+\int^{T}_{t}(T-s)g_{2}(s)ds,\int^{T}_{t}(T-s)^{2}b_{I}(s)\cdot b_{I}(s)ds\right).
\end{displaymath}

With
the above results in mind
we derive the expression for the price index level \(I(t)\):
\begin{displaymath}
I(T)=I(t)e^{\int^{T}_{t}(m_{I}(t)+(T-s)g_{2}(s)+g_{1}(s)-\frac{1}{2}s_{I}(s)\cdot\:s_{I}(s)\:)ds+\int^{T}_{t}((T-s)b_{I}(s)+s_{I}(s))\cdot\:dW^{T^{*}}(s)}.
\end{displaymath}
To achieve a lighter notation, we define:
\begin{equation}
g_{3}(s)=m_{I}(t)+(T-s)g_{2}(s)+g_{1}(s)-\frac{1}{2}s_{I}(s)\cdot\:s_{I}(s)\:
\end{equation}
\begin{equation}
g_{4}(s)=(T-s)b_{I}(s)+s_{I}(s). \label{InflationDiff}
\end{equation}
We note that \(g_{4}(t)\) and \(g_{3}(t)\) are deterministic functions. Based on the above, we obtain the following expression for the  \(T^{*}\)-dynamics and the terminal distribution of \(I(t)\):
\begin{equation}
d\log I(t)=g_{3}(t)dt+g_{4}(t)\cdot dW^{{T^{*}}}(t)
\end{equation}
\begin{equation}
dI(t)/I(t)=[g_{3}(t)+\frac{1}{2}g_{4}(t)\cdot g_{4}(t)]dt+g_{4}(t)\cdot dW^{{T^{*}}}(t)
\end{equation}
\begin{equation}
\log\ \frac{I(T)}{I(t)}={\int^{T}_{t}g_{3}(s)ds+\int^{T}_{t}g_{4}(s)\cdot dW^{{T^{*}}}(s)}\sim\mathcal{N}\left(\int^{T}_{t}g_{3}(s)ds,\int^{T}_{t}g_{4}(s)\cdot g_{4}(s)ds\right).\label{logpriceindexdistr}
\end{equation}A similar analysis can be done for \(X(t)\). We are  in a position to price zero-coupon inflation options.

\begin{lemma} The undiscounted price of  a zero-coupon inflation option priced at time t with maturity T and strike K in the CTCB model is
\begin{equation}
\omega e^{M+1/2V^{2}}N(\omega(M-(1+K)^{T-t}+V^{2})/V)-\omega KN(\omega(M-(1+K)^{T-t})/V)
\end{equation}where \(N(x)=\int^{x}_{-\infty}(2\pi)^{-1}e^{-\frac{s^{2}}{2}}ds \) and \(\omega\in \{-1,1 \} \) for puts and calls respectively. Further, \begin{displaymath}
M=\int^{T}_{t}g_{3}(s)ds
\end{displaymath}
\begin{displaymath}
V^{2}=\int^{T}_{t}g_{4}(s)\cdot g_{4}(s)ds.
\end{displaymath}
\end{lemma} 
\textit{Proof}. Using result \ref{LogNormalOptionPrice} and the distribution of the logarithm of \(I(t)\)  shown in \ref{logpriceindexdistr} one obtains the above.

\section{Pricing year-on-year inflation swaps and options}Here we focus our attention on year-on-year payoffs, that are model dependent. In fact, a convexity adjustment has to be introduced to take into account the co-movement of the nominal interest rate (used for discounting between times \(t\) and \(T_{i}\)) and the price index.\\ 

\textbf{Step 1} -- We work with the real economy as a calculation device, and obtain the dynamics of the real bond, defined as:
\begin{displaymath}
P^{r}(t,T)=\mathbb{E}^{\mathbb{Q}}_{t}[I(T)/I(t)e^{-\int^{T}_{t}n(s)ds}]=P(t,T)\mathbb{E}^{\mathbb{Q}^{T}}_{t}[I(T)/I(t)]=\\ \end{displaymath}
\begin{displaymath}
=P(t,T)e^{\int^{T}_{t}g_{3}(s)\,+\frac{1}{2}g_{4}(s)\cdot g_{4}(s)ds}=P(t,T)e^{\int^{T}_{t}[m_{I}(s)+g_{5}(s)\,]ds}=P(t,T)e^{m_{I}(t)(T-t)+\int^{T}_{t}g_{5}(s)\,ds}
\end{displaymath} 
where we  define \(g_{5}(s)=g_{3}(s)+\frac{1}{2}g_{4}(s)\cdot g_{4}(s)-m_{I}(t)(T-t)=g_{1}(s)+(T-s)g_{2}(s)+\frac{1}{2}g_{4}(s)\cdot g_{4}(s)-^{1}_{2}s_{I}(s)\cdot\ s_{I}(s).\\ \\ \)By applying Ito's lemma to \(P^{r}(t,T)\),  and taking into account the dynamics of  \(P(t,T)\) and \(m_{I}(t)\), we obtain:
 \begin{displaymath}
dP^{r}(t,T)=P^{r}(t,T)[(...)dt \ +\sigma_{P^{r}}(t,T)\cdot dW^{\mathbb{Q}^{T}}(t)]
\end{displaymath} where \(\sigma_{P^{r}}(t,T)=\sigma_{P^{}}(t,T)+b_{I}(t)(T-t)\). 
We are not interested in the drift component, but only in the diffusion term. This result is obtained by explicitly calculating the diffusion term:\begin{displaymath}
\left(\frac{\partial P^{r}(t,T)}{\partial P(t,T)}P(t,T)\sigma_{P^{}}(t,T)+\frac{\partial P^{r}(t,T)}{\partial m_{I}(t)}b_{I}(t)\right)\cdot dW^{\mathbb{Q}^{T}}(t)=
\end{displaymath}
 
\begin{displaymath}
\left(e^{m_{I}(t)(T-t)+\int^{T}_{t}g_{5}(s)\,ds}P(t,T)\sigma_{P^{}}(t,T)+(T-t)P(t,T)e^{m_{I}(t)(T-t)+\int^{T}_{t}g_{5}(s)\,ds}b_{I}(t)\right)\cdot dW^{\mathbb{Q}^{T}}(t)=
\end{displaymath}
\begin{displaymath}
P^{r}(t,T)(\sigma_{P^{}}(t,T)+b_{I}(t)(T-t))\cdot dW^{\mathbb{Q}^{T}}(t).
\end{displaymath}
\textbf{Step 2 }-- We build a \textit{T}-forward martingale by building a portfolio with a zero-coupon inflation swap with notional\emph{ I(t)} and maturity \(T\) and divide  by the numeraire, i.e. the nominal bond \(P(t,T)\). We recall a model-independent result that states that the present value (PV) of    a zero-coupon inflation swap is the difference between the real and nominal bond of the same maturity. We get:\begin{displaymath}
I(t)(P^{r}(t,T)-P(t,T))/P(t,T)=I(t)(P^{r}(t,T)/P(t,T) -1).
\end{displaymath}
We consider the quantity \(I(t)P^{r}(t,T)/P(t,T)\),  known as the forward price index:  \(\hat I(t,T)=I(t)P^{r}(t,T)/P(t,T).\)

The reason why this is called forward price index is clear if one makes the following observation:\begin{displaymath}
P^{r}(t,T)=\mathbb{E}^{\mathbb{Q}}_{t}[I(T)/I(t)e^{-\int^{T}_{t}n(s)ds}]=P(t,T)/I(t)\mathbb{E}^{\mathbb{Q}^{T}}_{t}[I(T)].
\end{displaymath}
Therefore one obtains:
\begin{displaymath}
\hat I(t,T)=I(t)P^{r}(t,T)/P(t,T)=\mathbb{E}^{\mathbb{Q}^{T}}_{t}[I(T)].
\end{displaymath}

By using Ito's Lemma on  \(\hat I(t,T)=I(t)P^{r}(t,T)/P(t,T), \) we obtain its risk-neutral dynamics. Again, we confirm that in the \textit{T}-forward dynamics the forward price index has to be a positive martingale:\begin{displaymath}
d\hat  I(t,T)=\hat  I(t,T)s_{\hat  I}(t,T)\cdot dW^{\mathbb{Q}^{T}}(t)
\end{displaymath}
where \(s_{\hat  I}(t,T)^{} \) is determined from the other model volatilities via Ito's Lemma in the way showed below. 

In particular one obtains:
\begin{displaymath}
s_{\hat  I}(t,T_{})=s_{I}(t)+b_{I}(t)(T-t).
\end{displaymath}
 To show this, one applies Ito's lemma for the diffusion part of the forward price index   \(\hat I(t,T)=I(t)P^{r}(t,T)/P(t,T)\):
\begin{displaymath}
\left(\frac{\partial \hat I(t,T_{})}{\partial I(t)} I(t)s_{I}(t)+\frac{\partial \hat I(t,T_{})}{\partial P^{r}(t,T)}P^{r}(t,T)\sigma_{P^{r}}(t,T)+\frac{\partial \hat I(t,T_{})}{\partial P^{}(t,T)}P(t,T)\sigma_{P^{}}(t,T)\right)\cdot dW^{\mathbb{Q}^{T_{}}}(t)=
\end{displaymath}

\begin{displaymath}
\left(\hat I(t,T_{})s_{I}(t)+\hat I(t,T_{})\sigma_{P^{r}}(t,T)-\hat I(t,T_{})\sigma_{P^{}}(t,T)\right)\cdot dW^{\mathbb{Q}^{T_{}}}(t)=s_{I}(t)+b_{I}(t)(T-t).
\end{displaymath}
This
final step was possible thanks to the expression of the diffusion term of the real bond found in step 1.\\  
\textbf{Step 3 }-- 
By a simple application of Ito's lemma one  shows that, taken some deterministic and regular functions \(a\), \(b\), and \(s\), (here \(a\) is a scalar function, \(b\) and \(s\) are vectorial functions with the same dimension of the driving Brownian motion \(W(t)\)) if one has two SDEs  defined as \(dX(t)=X(t)s\cdot \,dW(t)\ \)and \(dY(t)=Y(t)[a\,dt+b\cdot dW(t)]\), the ratio \(Z(t)=X(t)/Y(t)\) has dynamics \(dZ(t)=Z(t)[(-a+b\cdot b-s\cdot b)dt+(s-b)\cdot dW(t)]\).
This result is used in Step 5.\\
\textbf{Step 4 }-- 
Similarly to what is done for the BGM model, one chooses a reference tenor \(T^{*}\) and changes the dynamics of the inflation forwards to the same forward measure: an example of this technique is available in Belgrade \& Benhamou \cite{Label112221}. The dynamics of the inflation forwards were found in step 2: therefore we know explicitly the dynamics of \(\hat I(t,T_{i})\) and \(\hat I(t,T_{j}).  \) For example, if the reference tenor is \(T_{i}\) one obtains the following dynamics for the inflation forwards at tenors  \(T_{i}\) and \(T_{j}\) (\(T_{i}>T_{j}\)):
\begin{displaymath}
d\hat  I(t,T_{i})=\hat  I(t,T_{i})s_{\hat  I}(t,T_{i})\cdot dW^{\mathbb{Q}^{T_{i}}}(t)
\end{displaymath}
\begin{displaymath}
d\hat  I(t,T_{j})=\hat  I(t,T_{j})(-(\sigma _{P}(t,T_{i})-\sigma _{P}(t,T_{j})) \cdot s_{\hat  I}(t,T_{j})dt+s_{\hat  I}(t,T_{j})\cdot dW^{\mathbb{Q}^{T_{i}}}(t)).
\end{displaymath}
\textbf{Step 5 }-- 
One  introduces the  price index ratio process \(\mathcal{I}(t,T_{j},T_{i})=I(t,T_{i})/I(t,T_{j})\): using Ito's lemma and the results found at step 3 above one obtains its dynamics.
\begin{displaymath}
d\mathcal{I}(t,T_{j},T_{i})=\mathcal{I}(t,T_{j},T_{i})[((\sigma _{P}(t,T_{i})-\sigma _{P}(t,T_{j}))\cdot  s_{\hat I}(t,T_{j})
+s_{\hat  I}(t,T_{j})\cdot s_{\hat  I}(t,T_{j})-s_{\hat  I}(t,T_{i})\cdot s_{\hat  I}(t,T_{j}))dt
\end{displaymath}
\begin{displaymath}
+(s_{\hat  I}(t,T_{i})-s_{\hat  I}(t,T_{j}))\cdot dW^{\mathbb{Q}^{T_{i}}}(t)].
\end{displaymath}
Assuming \(t<t_{h}<T_{j}<T_{i}\), we can write the expectation of the ratio as:
 \begin{displaymath}
\mathbb{E}^{\mathbb{Q}^{T_{i}}}_{t}[\mathcal{I}(t_{h},T_{j},T_{i}]=\mathcal{I}(t,T_{j},T_{i})e^{\int
_{t}^{T_{j}}((\sigma _{P}(u,T_{i})-\sigma _{P}(u,T_{j}))\cdot  s_{\hat I}(u,T_{j})+s_{\hat  I}(u,T_{j})\cdot s_{\hat  I}(u,T_{j})-s_{\hat  I}(u,T_{i})\cdot s_{\hat  I}(u,T_{j}))du}.
\end{displaymath}
\textbf{Step 6 }-- 
We link the price index  ratio to the year-on-year payoff: the year-on-year forward can be expressed as an expectation of \(\mathcal{I}\):\begin{displaymath}
\mathbb{E}^{\mathbb{Q}^{T_{i}}}_{t}[I(T_{i}) / I(T_{j}) ]=\mathbb{E}^{\mathbb{Q}^{T_{i}}}_{t}[\hat I(T_{i},T_{i})/\hat I(T_{j},T_{j})]  =
\mathbb{E}^{\mathbb{Q}^{T_{i}}}_{t}[\mathbb{E}^{\mathbb{Q}^{T_{i}}}_{T_{j}}[\hat I(T_{i},T_{i})/\hat I(T_{j},T_{j})]]=\mathbb{E}^{\mathbb{Q}^{T_{i}}}_{t}[\hat I(T_{j},T_{i})/\hat I(T_{j},T_{j})]=
\end{displaymath}

\begin{displaymath}
\mathbb{E}^{\mathbb{Q}^{T_{i}}}_{t}[\mathcal{I}(T_{j},T_{j},T_{i}]=\mathcal{I}(t,T_{j},T_{i})e^{\int^{T_{j}}_{t}((\sigma _{P}(u,T_{j})-\sigma _{P}(u,T_{i}))\cdot s_{\hat  I}(u,T_{j})+s_{\hat  I}(u,T_{j})\cdot s_{\hat  I}(u,T_{j})-s_{\hat  I}(u,T_{i})\cdot s_{\hat  I}(u,T_{j}))du}=\end{displaymath}
\begin{displaymath}=\frac{ \hat I(t,T_{i})}{ \hat I(t,T_{j})}e^{\int^{T_{j}}_{t}((\sigma _{P}(u,T_{i})-\sigma _{P}(u,T_{j}))\cdot  s_{\hat I}(u,T_{j})+s_{\hat  I}(u,T_{j})\cdot s_{\hat  I}(u,T_{j})-s_{\hat  I}(u,T_{i})\cdot s_{\hat  I}(u,T_{j}))du}.
\end{displaymath}We  give the following lemma, thanks to the analysis done above.
 \begin{lemma} The undiscounted price of  a year-on-year inflation caplet/floorlet priced at time t with maturity T and strike K in the CTCB model is
\begin{equation}
\omega e^{M+1/2V^{2}}N(\omega(M-(1+K)+V^{2})/V)-\omega KN(\omega(M-(1+K))/V)
\end{equation}where \(N(x)=\int^{x}_{-\infty}(2\pi)^{-1}e^{-\frac{t^{2}}{2}}dt \) and \(\omega\in \{-1,1 \} \) for floorlets and caplets respectively. The year-on-year inflation is calculated between times \(T_{j}\) and \(T_{i}\). Further, \begin{displaymath}
M=\int^{^{T_{j}}}_{t}((\sigma _{P}(u,T_{i})-\sigma _{P}(u,T_{j}))\cdot  s_{ \hat I}(u,T_{j})+s_{\hat  I}(u,T_{j})\cdot s_{\hat  I}(u,T_{j})-s_{\hat  I}(u,T_{i})\cdot s_{\hat  I}(u,T_{j}))du+\int^{^{T_{i}}}_{T_{j}}g_{3}(u)du
\end{displaymath}
\begin{displaymath}
V^{2}=\int
_{T_{j}}^{^{T_{i}}}((\sigma _{\hat I}(u,T_{i})-\sigma_{\hat I}(u,T_{j}))\cdot((\sigma _{\hat I}(u,T_{i})-\sigma_{\hat I}(u,T_{j}))du.
\end{displaymath}
\end{lemma} 
\textit{Proof}. Using result \ref{LogNormalOptionPrice}, the  result in step 6, and the distribution of the logarithm of \(I(t)\) shown above one obtains the result.

\section{Single currency derivatives pricing simulation}To test the results found in the previous sections, we  implemented a Monte Carlo simulation and the closed forms for zero-coupon and year-on-year inflation options. We have run 20,000 simulations over 10 years, and here we show the results, the standard error and the closed form results.  We price caps with strikes 0, 1, 2, 3, 4, 5
percent with maturities from 1 to 10 years. We assume that the dimensionality of the driving Brownian motion is 3.\\ For this simulation, we assume the following set of  parameters, that are constant over time: \(a_{X}(t)=0\%\), \(a_{I}(t)=0.5\%\), \(b_{X}(t)=0\%\), \(b_{I}(t)=0.3\%\),    \(s_{X}(t)=0\%\), \(s_{I}(t)=0.3\%\),  \(\sigma_{P}(t,T)=1\%\ \), \(\lambda(t)=0\)\%,\ \(\mu_{I}(0)=0\%\); in case of vector functions, like the volatilities, we assume that the value is the same for all the 3 components. For this analysis we have only presented the parameters that are directly relevant for the pricing of inflation derivatives: a full calibration exercise  is presented in the following section.\\ The data show that there is good agreement between the Monte Carlo simulation (denoted as MC PV in the following tables) and the closed forms (denoted as PV forms) and that the number of simulations is high enough to control the numerical error.\\ The results for zero-coupon options are the following (strikes in columns, maturities in rows):
 \\

\begin{tabular}{|c|c|c|c|c|c|c|}\hline MC PV  & 0\%\  & 1\%\  & 2\%\  & 3\%\  & 4\%\  & 5\%\ \\\hline 1  & 0.00209  & 0.00006  & 0  & 0  & 0  & 0 \\\hline 2  & 0.00772  & 0.00058  & 0  & 0  & 0  & 0 \\\hline 3  & 0.01804  & 0.00268  & 0.00009  & 0  & 0  & 0 \\\hline 4  & 0.03359  & 0.00784  & 0.00056  & 0.00001  & 0  & 0 \\\hline 5  & 0.05477  & 0.01731  & 0.00219  & 0.00009  & 0  & 0 \\\hline 6  & 0.08208  & 0.03225  & 0.00625  & 0.00046  & 0.00002  & 0 \\\hline 7  & 0.11599  & 0.05352  & 0.01421  & 0.00167  & 0.00009  & 0 \\\hline 8  & 0.15721  & 0.08208  & 0.02753  & 0.00475  & 0.00038  & 0.00002 \\\hline 9  & 0.20641  & 0.1188  & 0.04775  & 0.01112  & 0.00129  & 0.00008 \\\hline 10  & 0.26444  & 0.16451  & 0.07619  & 0.02256  & 0.00363  & 0.00032 \\\hline \end{tabular}
\begin{displaymath}
\end{displaymath} 

\begin{tabular}{|c|c|c|c|c|c|c|}\hline MC Error  & 0\%\  & 1\%\  & 2\%\  & 3\%\  & 4\%\  & 5\%\\\hline 1  & 0.00002  & 0  & 0  & 0  & 0  & 0 \\\hline 2  & 0.00006  & 0.00002  & 0  & 0  & 0  & 0 \\\hline 3  & 0.00012  & 0.00005  & 0.00001  & 0  & 0  & 0 \\\hline 4  & 0.00018  & 0.0001  & 0.00003  & 0  & 0  & 0 \\\hline 5  & 0.00027  & 0.00018  & 0.00006  & 0.00001  & 0  & 0 \\\hline 6  & 0.00036  & 0.00027  & 0.00012  & 0.00003  & 0  & 0 \\\hline 7  & 0.00047  & 0.00039  & 0.00021  & 0.00007  & 0.00002  & 0 \\\hline 8  & 0.0006  & 0.00052  & 0.00033  & 0.00014  & 0.00004  & 0.00001 \\\hline 9  & 0.00074  & 0.00068  & 0.00049  & 0.00024  & 0.00008  & 0.00002 \\\hline 10  & 0.00091  & 0.00086  & 0.00067  & 0.00038  & 0.00015  & 0.00004 \\\hline \end{tabular}

\begin{displaymath}
\end{displaymath} 

\begin{tabular}{|c|c|c|c|c|c|c|}\hline PV - form  & 0\%\  & 1\%\  & 2\%\  & 3\%  & 4\%\  & 5\%\ \\\hline 1  & 0.00212  & 0.00006  & 0  & 0  & 0  & 0 \\\hline 2  & 0.00779  & 0.00059  & 0.00001  & 0  & 0  & 0 \\\hline 3  & 0.01814  & 0.00271  & 0.00009  & 0  & 0  & 0 \\\hline 4  & 0.03374  & 0.00786  & 0.00057  & 0.00001  & 0  & 0 \\\hline 5  & 0.05502  & 0.0174  & 0.00224  & 0.00009  & 0  & 0 \\\hline 6  & 0.08244  & 0.03244  & 0.00631  & 0.00047  & 0.00001  & 0 \\\hline 7  & 0.11649  & 0.05391  & 0.01427  & 0.00172  & 0.00008  & 0 \\\hline 8  & 0.15781  & 0.08265  & 0.02766  & 0.00483  & 0.00038  & 0.00001 \\\hline 9  & 0.20712  & 0.11952  & 0.04799  & 0.01121  & 0.00133  & 0.00007 \\\hline 10  & 0.26534  & 0.16545  & 0.07674  & 0.02262  & 0.00371  & 0.00031 \\\hline \end{tabular}

\begin{displaymath}
\end{displaymath}
\begin{tabular}{|c|c|c|c|c|c|c|}\hline Difference: PV - form, MC sim.   & 0  & 1\%\  & 2\%\  & 3\%  & 4\%\  & 5\%\ \\\hline 1  & -0.00003  & 0  & 0  & 0  & 0  & 0 \\\hline 2  & -0.00006  & -0.00001  & 0  & 0  & 0  & 0 \\\hline 3  & -0.0001  & -0.00003  & 0  & 0  & 0  & 0 \\\hline 4  & -0.00015  & -0.00003  & -0.00001  & 0  & 0  & 0 \\\hline 5  & -0.00025  & -0.00009  & -0.00005  & 0  & 0  & 0 \\\hline 6  & -0.00036  & -0.00019  & -0.00007  & -0.00002  & 0  & 0 \\\hline 7  & -0.0005  & -0.00038  & -0.00006  & -0.00005  & 0.00001  & 0 \\\hline 8  & -0.0006  & -0.00057  & -0.00013  & -0.00008  & 0  & 0 \\\hline 9  & -0.00071  & -0.00073  & -0.00025  & -0.00009  & -0.00004  & 0.00001 \\\hline 10  & -0.0009  & -0.00095  & -0.00055  & -0.00006  & -0.00008  & 0.00001 \\\hline \end{tabular}
\\ \\ The results for year-on-year options are the following:
\\

 \begin{tabular}{|c|c|c|c|c|c|c|}\hline MC PV  & 0\%\  & 1\%\ & 2\%\ & 3\% & 4\%\  & 5\%\ \\\hline 1  & 0.0021  & 0.00006  & 0  & 0  & 0  & 0 \\\hline 2  & 0.00621  & 0.00114  & 0.00007  & 0  & 0  & 0 \\\hline 3  & 0.01081  & 0.00375  & 0.00066  & 0.00005  & 0  & 0 \\\hline 4  & 0.01562  & 0.00736  & 0.00229  & 0.00041  & 0.00004  & 0 \\\hline 5  & 0.02055  & 0.01159  & 0.0049  & 0.00141  & 0.00025  & 0.00003 \\\hline 6  & 0.02549  & 0.01611  & 0.0083  & 0.00322  & 0.00087  & 0.00015 \\\hline 7  & 0.03051  & 0.0209  & 0.01233  & 0.00588  & 0.00212  & 0.00055 \\\hline 8  & 0.03558  & 0.02581  & 0.01672  & 0.00922  & 0.0041  & 0.00142 \\\hline 9  & 0.04066  & 0.0308  & 0.02139  & 0.01311  & 0.00681  & 0.00288 \\\hline 10  & 0.04566  & 0.03575  & 0.02613  & 0.01731  & 0.01005  & 0.00493 \\\hline \end{tabular}
  \begin{displaymath}
 \,
\end{displaymath}
 \begin{tabular}{|c|c|c|c|c|c|c|}\hline MC Error  & 0\%\  & 1\%\  & 2\%\ & 3\%\  & 4\%\  & 5\%\\\hline 1  & 0.00002  & 0  & 0  & 0  & 0  & 0 \\\hline 2  & 0.00004  & 0.00002  & 0  & 0  & 0  & 0 \\\hline 3  & 0.00006  & 0.00004  & 0.00002  & 0  & 0  & 0 \\\hline 4  & 0.00007  & 0.00006  & 0.00003  & 0.00001  & 0  & 0 \\\hline 5  & 0.00008  & 0.00007  & 0.00005  & 0.00003  & 0.00001  & 0 \\\hline 6  & 0.00009  & 0.00008  & 0.00007  & 0.00004  & 0.00002  & 0.00001 \\\hline 7  & 0.0001  & 0.00009  & 0.00008  & 0.00006  & 0.00004  & 0.00002 \\\hline 8  & 0.00011  & 0.0001  & 0.00009  & 0.00008  & 0.00005  & 0.00003 \\\hline 9  & 0.00011  & 0.00011  & 0.00011  & 0.00009  & 0.00007  & 0.00004 \\\hline 10  & 0.00012  & 0.00012  & 0.00012  & 0.0001  & 0.00008  & 0.00006 \\\hline \end{tabular}
 \begin{displaymath}
 \,
\end{displaymath}
\begin{tabular}{|c|c|c|c|c|c|c|}\hline PV - form  & 0\%\  & 1\%\  & 2\%\  & 3\%  & 4\%\  & 5\%\ \\\hline 1  & 0.00002  & 0  & 0  & 0  & 0  & 0 \\\hline 2  & 0.00004  & 0.00002  & 0  & 0  & 0  & 0 \\\hline 3  & 0.00006  & 0.00004  & 0.00002  & 0  & 0  & 0 \\\hline 4  & 0.00007  & 0.00006  & 0.00003  & 0.00001  & 0  & 0 \\\hline 5  & 0.00008  & 0.00007  & 0.00005  & 0.00003  & 0.00001  & 0 \\\hline 6  & 0.00009  & 0.00008  & 0.00007  & 0.00004  & 0.00002  & 0.00001 \\\hline 7  & 0.0001  & 0.00009  & 0.00008  & 0.00006  & 0.00004  & 0.00002 \\\hline 8  & 0.00011  & 0.0001  & 0.00009  & 0.00008  & 0.00005  & 0.00003 \\\hline 9  & 0.00011  & 0.00011  & 0.00011  & 0.00009  & 0.00007  & 0.00004 \\\hline 10  & 0.00012  & 0.00012  & 0.00012  & 0.0001  & 0.00008  & 0.00006 \\\hline \end{tabular}
\\
\\

 \begin{tabular}{|c|c|c|c|c|c|c|}\hline Difference: PV - form, MC sim.  & 0\%\  & 1\%\  & 2\%\  & 3\%\  & 4\%\  & 5\%\ \\\hline 1  & 0.00003  & 0  & 0  & 0  & 0  & 0 \\\hline 2  & 0.00068  & 0.00065  & 0.00006  & 0  & 0  & 0 \\\hline 3  & 0.00036  & 0.0007  & 0.00034  & 0.00004  & 0  & 0 \\\hline 4  & 0.00019  & 0.00051  & 0.00053  & 0.0002  & 0.00003  & 0 \\\hline 5  & 0.00011  & 0.00034  & 0.0005  & 0.00035  & 0.00011  & 0.00002 \\\hline 6  & 0  & 0.00016  & 0.00036  & 0.00041  & 0.00022  & 0.00006 \\\hline 7  & -0.00002  & 0.00008  & 0.00028  & 0.0004  & 0.00031  & 0.00014 \\\hline 8  & -0.00002  & 0.00005  & 0.00018  & 0.00033  & 0.00035  & 0.00024 \\\hline 9  & 0  & 0.00004  & 0.00014  & 0.00029  & 0.00038  & 0.00032 \\\hline 10  & -0.00006  & -0.00004  & 0.00004  & 0.00017  & 0.0003  & 0.00033 \\\hline \end{tabular}
\section{Extension to the open economy}In this setting one also prices inflation derivatives  struck in a different currency. To do this, one simply defines the quantities defined in the previous sections also for the foreign economy and then introduces a domestic risk-neutral process for the FX rate \(\{Y(t)\}_{t\geq0}\), expressed in the FORDOM convention. One assumes that the foreign economy works in a similar way, that there is a foreign central bank and that there is a  liquidity relationship in the foreign bond market between foreign bond prices and foreign money supply. In particular, all parameters for the foreign economy variables are the same used in the domestic one, with an index \(f\).\\ The dynamics of the foreign assets and other  quantities are: \\ \begin{displaymath}
dX^{f}(t)/X^{f}(t)=(m_{X^{f}}(t)-\lambda^{f}(t)\cdot s_{X^{f}}(t) \,\,)dt\:+s_{I}(t)\cdot\:dW^{\mathbb{Q}^{f}}(t)\end{displaymath}
\begin{displaymath}
dI^{f}(t)/I^{f}(t)=(m_{I^{f}}(t)-\lambda^{f}(t)\cdot s_{I^{f}}(t) \,\,)dt\:+s_{I}(t)\cdot\:dW^{\mathbb{Q}^{f}}(t)
\end{displaymath}
\begin{displaymath}
dm_{X^{f}}(t)=(a_{X^{f}}(t)-\lambda^{f}(t)\cdot b_{X^{f}}(t))dt+b_{X^{f}}(t)\cdot\:dW^{\mathbb{Q}^{f}}(t)
\end{displaymath}
\begin{displaymath}
dm_{I^{f}}(t)=(a_{I^{f}}(t)-\lambda^{f}(t)\cdot b_{I^{f}}(t))dt+b_{I^{f}}(t)\cdot\:dW^{\mathbb{Q}^{f}}(t)
\end{displaymath}
\begin{displaymath}
dP^{f}(t,T)/P^{f}(t,T)=n^{f}(t)dt\:+\sigma_{P^{f}}(t,T)\cdot\:dW^{\mathbb{Q}^{f}}(t)
\end{displaymath}  
\begin{displaymath}
dn^{f}(t)=[f^{f}_{1}(t)-f^{f}_{2}(t)n^{f}(t)-\lambda^{f}(t)\cdot\sigma_{n^{f}}(t)]dt+\sigma_{n^{f}}(t)\cdot dW^{\mathbb{Q}^{f}}(t)
\end{displaymath}
\begin{displaymath}
dY(t)/Y(t)=(n^{}(t)-n^{f}(t))dt\:+s_{Y}(t)\cdot\:dW^{\mathbb{Q}}(t).
\end{displaymath}  
We are assuming that the same Brownian motion drives both the domestic and the foreign economy. In particular the parameters for the foreign short rate dynamics are defined in the same way  as the domestic ones:
\begin{equation}
f^{f}_{2}(t)=[Z^{f}(t+\Omega)-Z^{f}(t)]/\zeta ^{f}(t)
\end{equation}
\begin{equation}
f^{f}_{1}(t)=[-h^{f}_{p}a_{I^{f}}(t)-h^{f}_{x}a_{X^{f}}(t)]/\zeta ^{f}(t)
\end{equation}
\begin{equation}
\sigma^{f}_{n}(t)=[-h^{f}_{x}b_{X^{f}}(t)-h^{f}_{p}b_{I^{f}}(t)]/\zeta^{f} (t). \label{ShortRateVolForeign}
\end{equation}
By changing the numeraire in the foreign economy from \(B^{f}(t)\) to
 \(Y(t)B^{f}(t)\), one achieves the domestic risk-neutral dynamics for the foreign economic variables. This translates into a change of drift of \(s_{(\cdot)}(t)\cdot s_{Y}(t)\), where \(s_{(\cdot)}(t)\) is the Brownian volatility for a generic model variable.
 \section{Uncertain-parameters extension: modelling the inflation smile}The model presented  gets its randomness from an \textit{n}-dimensional Brownian motion \(W(t). \) In principle, one  extends the theory proposed  to the Merton jump-diffusion case, which could add  flexibility especially to model inflation options skew. Here we show that the Merton equation can be obtained in the framework proposed above if one assumes that the model has uncertain parameters. An uncertain-parameters model is a model whose parameters can take random values that are known at inception. Normally one assumes that there is a finite number of possible levels for the parameters and that the parameter set is determined loosely speaking one instant before the process starts. The distributions of the state variables are mixtures of distributions.
For an introduction to these models
one can see Brigo \& Mercurio \cite{Label2}.

In the Merton Model the source of randomness  is the  process
\(\Lambda(t)=s_{I}(t)W(t)+\sum^{N(t)}_{i=1}(J_{i}-1)\), where:\begin{enumerate}
\item 
\(N(t)\) is a Poisson process with intensity \(h\)  independent from the Brownian motion \(W(t)\) and the jump sizes \(J_{1},J_{2},...\)
\item 
The logarithm of the jump size \(J\) has a normal distribution with constant mean \(\mu_{J}\) and variance \((\delta_{J})^{2}\).
Therefore the expected jump size is \(\mathbb{E}[J-1]=e^{\mu_{J}+\frac{1}{2}(\delta_{J})^{2}}-1=k. \)  The logarithm of the jump size is also independent from the Brownian motion \(W(t)\).\item 
The drift of the process has been adjusted to take into account the compensator: \(\mu^{}_{I^{}}(t)-hk.\)\end{enumerate}
For simplicity, here we consider a one-dimensional source of randomness. The equation governing the evolution of the price index would read:\begin{displaymath}
dI(t)=I(t)[(\mu^{}_{I^{}}(t)-hk)dt+d\Lambda^{\mathbb{Q}}(t)]  =I(t)[(\mu^{}_{I^{}}-hk)dt+s_{I}(t)dW^{\mathbb{Q}}(t)+(J-1)dN^{\mathbb{Q}}(t)].
\end{displaymath}        As Merton has showed, because the distribution of \textit{J} is lognormal, the distribution of \(\log [I(T)/I(t_{0})]\) is still normal conditional to the event \(\{N(T)=n\}\).

Therefore one  regards such model as an uncertain-parameters model, where, with probability \(\mathbb{Q}(N(T)=n)=e^{-hT}(hT)^{n}/n!\), the SDE for \(I(t) \) is:\begin{displaymath}
dI(t)=I(t)[(\mu^{}_{I^{}}(t)-hk+n(\mu_{j}+\frac{1}{2}\delta^{2}_{J})/(T-t_{0}))dt+((s_{I}(t))^{2}+n(\delta_{J})^{2}/(T-t_{0}))^{\frac{1}{2}}dW^{\mathbb{Q}}(t)].
\end{displaymath}
Therefore the theory developed so far for the Brownian case is extended to the Merton case by making some assumptions regarding uncertain model parameters.

\section{Model calibration and applications }

  Here we propose a strategy to calibrate the CTCB model to market  observables by finding suitable model parameters,  and show some practical applications. Two main advantages become apparent: firstly, the CTCB  calibration process is separable and  model is analytically tractable and secondly, because it is based on economic theory, one  runs some economic stress scenarios and obtains the answers directly from the model itself, without  making assumptions on how an economic shock would impact on financial quantities such as inflation and rates volatilities. 

\subsection{At-the-money calibration strategy }
We are calibrating the model at time \(t_{0}=0\): we are still making the assumption that the market observables are continuous functions of the maturity, to keep the notation light (this assumption is removed in the next sections). When calibrating vector parameters, like \(b_{I}(t)\) or \(s_{I}(t)\) to name a few, we do not make any assumptions regarding how the total quantity needed to calibrate the model is split across the single components to model instantaneous correlations: this topic is analysed later. \subsection{Calibration steps: a first strategy}
\begin{enumerate}
\item 
One  makes explicit the assumptions on the structural parameters of the model, namely the reaction function parameters \(h_{x}\) and \(h_{p}\), the reaction function targets \(\bar x\) and \(\bar p\), the liquidity horizon of the central bank \(\Omega\), and  the function \(Z(T)=e^{\delta T}\) by choosing the parameter \(\delta>0\). In practice, these parameters are to be regarded not as a target for the calibration, but as an input from economic research that is expected to stay  constant over time.
The GDP volatility \(s_{X}(t)\) can be regarded as an input of the model, and therefore estimated based on some historic data.\item If one knows the parameter \(\delta >0\), one  writes \(Z(T)=e^{\delta T}\) and \(\zeta(t)=\int^{t+ \Omega}_{t}Z(T)dt=(\delta^{-1})(e^{\delta (t+\Omega)}-e^{\delta t})\). Further one finds the function \(\beta(T)=B(0,T)=e^{-\delta T}\): we remind the reader that this function is the one used in the Ansatz  \(P(t,T)=A(t,T)e^{-n(t)B(t,T)}\) that characterises the nominal bond prices \(P(t,T)\) as a function of the short rate \(n(t)\).
Once \(B(0,t)\) is found, from the market bond  prices \(P(0,T)\)  and the market quote for \(n(0)\) one  deduces the function \(A(0,T)=P(0,T)/e^{-n(0)e^{-\delta T}}\).
One should remember that the functions \(A(t,T)\) and \(B(t,T)\) are not core functions of the CTCB model, but are only relevant to its dual Hull-White  model. Finally, if needed one  gets the function \(B(t,T)=[B(0,T)-B(0,t)]/(\partial B(0,t)/\partial t)=(e^{-\delta T}-e^{-\delta t})/(-\delta e^{-\delta t})=(\delta^{-1})(1-e^{-\delta^{(T-t)}})  \): this is a standard result in the Hull-White model.
\item By exploiting the fact that a CTCB model implies an equivalent (dual) Hull-White\ model for the short rate \(n(t)\), one   calculates the mean reversion speed \(a(t)\): as proved in the previous section,  the parametrisation 
 \(Z(T)=e^{\delta T}\) implies that the mean reversion speed is constant and equivalent to \(\delta\); in fact we recall that \begin{equation}
a(t)=[Z(t+\Omega)-Z(t)]/\zeta (t)=\frac{Z(t+\Omega)-Z(t)}{\int^{t+\Omega}_{t}{{Z(T)}}dT}=\delta.
\end{equation}We notice that we are still missing the short rate volatility \(\sigma_{n}^{*}(t)\) and the market price of risk \(\lambda(t)\) to get the mean reversion level function \(\theta(t) \). This function is found in the following steps.
 \item 
 One takes  the market quotes of at-the-money  caps and floors: from these it is straightforward to get the single at-the-money caplets and floorlets. These are sensitive to the interest rate volatility, and  can be used to calibrate some CTCB model volatilities: we recall that at time \(t\) the price of a caplet with strike \(K\) on the Libor between times \(T_{i-1}\) and \(T_{i}\) is equivalent to  the price of a put option with expiry \(T_{i-1}\) on a zero-coupon bond with maturity \(T_{i}>T_{i-1}\). The price of such option can be obtained in closed form via a Black-type formula in the Hull-White model,   where the total variance used for pricing is: \begin{equation}
V^{2}(t,T_{i-1},T_{i})=[\beta(T_{i})-\beta(T_{i-1})]^{2}\int^{T_{i-1}}_{t}\left[ \frac{\sigma^{*}_{n}(u)}{\beta^{'}(u)} \right]^{2}du.
\end{equation}
Because  \(\beta(T)=B(0,T)=e^{-\delta T}\) and  \(\beta'(T)=\partial B(0,T)/\partial T=-\delta e^{-\delta T}\) , the above  is written as:\begin{equation}
V^{2}(t,T_{i-1},T_{i})=[e^{-\delta T_{i}}-e^{-\delta T_{i-1}}]^{2}\int^{T_{i-1}}_{t}\left[ \frac{\sigma^{*}_{n}(u)}{-\delta e^{-\delta u}} \right]^{2}du.\label{ExampleHWInterval}
\end{equation} 

Finally,
we recall
that in the equivalent Hull-White model the short rate volatility was expressed as: \begin{equation}
\sigma_{n}(t)=[-h_{x}b_{X}(t)-h_{p}b_{I}(t)]/\zeta (t)=-\delta[h_{x}b_{X}(t)+h_{p}b_{I}(t)]/(e^{\delta (t+\Omega)}-e^{\delta t}).
\end{equation}

One should refer to \ref{HWScalarVectorVol} to show how one moves from the scalar original Hull-White volatility \(\sigma^{*}_{n}(t)\) to the short rate \textit{n}-dimensional  model volatility in the CTCB model, denoted as \(\sigma_{n}(t)\).

 Therefore, at the end of this step we have fully calibrated the CTCB  model to the nominal term structure and at-the-money caps-floors volatilities, and found the economic expectation volatility functions \(b_{X}(t)\) and  \(b_{I}(t)\): one chooses these functions  to ensure that the model at-the-money cap-floors prices match the ones observed in the market. If one wanted to use the CTCB model to price nominal rate derivatives, the calibration process could be ended here. Alternatively, one  specifies the function  \(b_{X}(t)\) based on historic data and only calibrate \(b_{I}(t).\) 
\item A first consequence of the above result is that, exploiting the standard result
\(\sigma_{P}(t,T)=-\sigma_{n}(t)B(t,T)\) in the Hull-White model, we can write explicitly the bond volatilities: these are needed either if one needs to simulate Libor rates \(F(t,T_{i-1},T_{i})=(P(t,T_{i-1})/P(t,T_{i})-1)/(T_{i}-T_{i-1})\) or  when building the drift adjustment to move to the \(T^{*}\)-forward measure. Making all dependencies explicit we write \begin{equation}
\sigma_{P}(t,T)=-\delta\frac{[-h_{x}b_{X}(t)-h_{p}b_{I}(t)]}{(e^{\delta (t+\Omega)}-e^{\delta t})}(\delta^{-1})(1-e^{-\delta^{(T-t)}})=\frac{[h_{x}b_{X}(t)+h_{p}b_{I}(t)]}{(e^{\delta (t+\Omega)}-e^{\delta t})}(1-e^{-\delta^{(T-t)}}).
\end{equation}

\item   The function \(A(t,T) \) can  be made explicit using a standard Hull-White result:\begin{equation}
\log A(t,T)=\log A(0,T)-\log A(0,t)-B(t,T)\frac{\partial\log A(0,t)}{\partial t}-\frac{1}{2}\left[B(t,T) \frac {\partial  B(0,t)}{\partial t} \right]^{2}\int^{t}_{0}\left[\frac{\sigma_{n}(s)}{\frac {\partial  B(0,s)}{\partial s}}  \right]^{2}ds.
\end{equation}

\item 
We  calibrate to the inflation volatilities implied by the market. We recall that the total variance of the quantity \(\log\left( I(T)/I(0) \right)\) is \(\int^{T}_{0}g_{4}(s)\cdot g_{4}(s)ds\)
where \(g_{4}(t)=(T-t)b_{I}(t)+s_{I}(t)\). Therefore one  finds the function \(s_{I}(t)\), under the constraint that we know already the function \(b_{I}(t)\), using the closed forms for inflation zero-coupon options that we found in the previous section.
  \item 
At this stage one has enough information to calibrate the model to the inflation breakeven strikes from zero-coupon inflation swaps, remembering that the expectation of the quantity \(\log\left( I(T)/I(0) \right)\) is \(\int^{T}_{0}g_{3}(s)ds\)
where we recall the definitions of \(g_{1}(t)=-s_{I}(t)\cdot(\lambda(t)-s_{P}(t,T^{*}))\), \(g_{2}(t)=a_{I}(t)-b_{I}(t)\cdot(\lambda(t)-s_{P}(t,T^{*}))\), and \(g_{3}(t)=m_{I}(0)+(T-t)g_{2}(t)+g_{1}(t)-^{1}_{2}s_{I}(t)\cdot\:s_{I}(t)\).  Therefore we have  found the market prices of risk function \(\lambda(t)\) and the inflation expectation drift function \(a_{I}(t)\).  Alternatively, one  specifies the market price of risk \(\lambda(t)\) based on historic data and calibrates only \(a_{I}(t)\).
The former alternative is well suited for relative value analysis, i.e. the trader, based on a view of the economy and the observed market prices, implies the market prices of risks implied by market prices, and gauges the illiquidity spots or the inconsistencies between the market participants' risk preferences. The latter  is more suited to replicate market prices, i.e. the trader makes an assumption on the market risk aversion and implies the implied paths for the price index and GDP growth expectations.
\item 
We can write the Hull-White equivalent mean reversion level by exploiting the standard  result:\begin{equation}
\theta(t)=\lambda(t)\sigma_{n}(t)-\delta\frac {\partial \log A(0,t)}{\partial t}-\frac {\partial ^{2}\log A(0,t)}{\partial t^{2}}+\left[ \frac {\partial  B(0,t)}{\partial t} \right]^{2}\int^{t}_{0}\left[\frac{\sigma_{n}(s)}{\frac {\partial  B(0,s)}{\partial s}}  \right]^{2}ds.
\end{equation}
This result is used to find the growth expectation drift function \(a_{X}(t) \) remembering that the mean reversion level in the Hull-White equivalent model was given by \(\theta(t)=[-h_{p}a_{I}(t)-h_{x}a_{X}(t)]/\zeta (t)\).
\item The previous two points can be compacted into one, if one assumes to know the expectation drifts \(a_{X}(t)\) and \(a_{I}(t)\) and therefore calibrates only the market price of risk \(\lambda(t)\).
\item 
Finally, one recalls the volatility condition \ref{LiqLocRiskless} to calculate \(s_{M}(t)=h_{p} s_{I}(t)+h_{x} s_{X}(t) \): these volatilities may be needed to run a full simulation of the model but are not needed to price derivatives. 
\end{enumerate}
\subsection{Calibration steps: an alternative  strategy} Here we propose a minor change to the calibration strategy proposed above that can  be introduced in order to ensure full calibration of the model. In fact, we are calibrating the functions \(b_{X}(t)\)  and \(b_{I}(t)\)  first, based on the market prices of nominal caps and floors (step 4): in a second step (step 7) we find the function  \(s_{I}(t)\)  that calibrates the market prices of inflation zero-coupon options. In this step there can be a problem, given that the total variance is \(\int^{T}_{0}g_{4}(s)\cdot g_{4}(s)ds\), where \(g_{4}(t)=(T-t)b_{I}(t)+s_{I}(t)\): the function   \(s_{I}(t)\) in some cases can only increase the total variance given  \(b_{I}(t)\), and the function  \(b_{I}(t)\) is multiplied by \(T-t\), which can lead to excessive implied variance at long maturities. In practice, in some cases  the model may not  calibrate to inflation zero-coupon options, because it can not reduce the implied variance below a certain threshold. If inflation volatilities are too low, full calibration to the inflation option prices may not be achieved.

To
overcome this problem, we suggest to calibrate the functions \(b_{I}(t)\) and \(s_{I}(t)\) to inflation zero-coupon options across all maturities  as a first step, and then to use the function  \(b_{X}(t)\)  to calibrate the nominal caps and floors: the advantage is that the calibration is guaranteed in both inflation and caps and floors volatilities. However the trader can not freely mark the output expectation volatilities  \(b_{X}(t)\), which was possible in the approach proposed originally. We do not have an explicit preference for either approach: the choice  depends on whether one wants to control the output expectation volatilities  \(b_{X}(t)\)   or guarantee a full calibration to option prices. We decided to show both strategies as the first one is perfectly suited, up to step 6, to calibrate only to the nominal term structure, which can be done to price interest rates derivatives.
\subsection{Variance split and calibration to correlations}
In the steps of the previous section we found some model volatilities, namely \(b_{I}(t)\), \(b_{X}(t)\), \(s_{I}(t)\), and \(s_{X}(t)\). Because these processes are multidimensional with dimension \(n\), given some option prices one wants to calibrate to, there are multiple ways to split the total variance  into its components. We regard this fact as an opportunity to calibrate to an instantaneous correlation structure that the trader can choose.

Let us take the model volatility process \(b_{I}(t)\): all we say for it can be exactly extended to the remaining three processes. We introduce some weights, called \(w^{i}_{b_{I}(t)}\) with \(i=1,2, ...,n\) and such that \(\sum^{n}_{i=1}[w^{i}_{b_{I}(t)}]^{2}=1\). One  writes \(v^{}_{b_{I}(t)}=\sum^{n}_{i=1}[b_{I}^{i}(t)]^{2}\), where \(b_{I}^{i}(t)\) is the \(i\)-th component of  \(b_{I}(t)\): in practice the calibration process proposed in the previous section   only yields the total variance \(v^{}_{b_{I}(t)}\) that matches market option prices. One  defines \([b_{I}^{i}(t)]^{2}=v^{}_{b_{I}(t)}[w^{i}_{b_{I}(t)}]^{2}\) and  the total variance is split according to some pre-defined weights: this is done for all four model volatilities, yielding the total variances \(v^{}_{b_{I}(t)}\), \(v^{}_{b_{X}(t)}\), \(v^{}_{s_{I}(t)}\), and
 \(v^{}_{s_{X}(t)}\), assuming that one knows the weights
\(w^{i}_{b_{I}(t)}\), \(w^{i}_{b_{X}(t)}\), \(w^{i}_{s_{I}(t)}\), and \(w^{i}_{s_{X}(t)}\).  

These four sets of weights can be determined in a way to target a given instantaneous correlation level. The variables for which we want to impose a correlation structure are the changes in the short rate \(dn(t)\), and relative changes in  the price index \(dI(t)/I(t)\). Perhaps one may also be interested to impose a correlation structure that includes the relative  changes of the real GDP \(dX(t)/X(t)\). We assume we know the market-implied  3 x 3 correlation matrix.

We want to find the  weights
\(w^{i}_{b_{I}(t)}\), \(w^{i}_{b_{X}(t)}\), \(w^{i}_{s_{I}(t)}\), and \(w^{i}_{s_{X}(t)}\) so that the instantaneous model correlations are as close as possible to the market-implied correlations: clearly there is a trade-off between the accuracy of this fit and the dimensionality \(n\) of the Brownian motion \(\{W(t)\}_{t\geq0}\). A  high enough dimensionality \(n\) can ensure a perfect fit, however this would make the model overparametrised and difficult to manage.
The accuracy is measured as the square difference between the market implied correlation \(\rho^{MKT}_{a(t),b(t)}(t)\) of the generic variables \(a(t)\) and \(b(t)\) and the model correlations  \(\rho^{MOD}_{a(t),b(t)}(t)\): here \(a(t)\in \mathcal{V}=\{dn(t),dI(t)/I(t),dX(t)/X(t)\}\) and  \(b(t)\in\mathcal{V}\).
 
We know the model volatility functions for the 3 variables in closed form from the previous section, for the short rate change, for the price index relative change, and for the output relative change respectively:
\begin{multline*}
\\Model Vol(n(t))=
\frac{-\delta[h_{x}b_{X}(t)+h_{p}b_{I}(t)]}{(e^{\delta (t+\Omega)}-e^{\delta t})}
 \\
 Model Vol(I(t))=(T-t)b_{I}(t)+ s_{I}(t) 
  \\
 Model Vol(X(t))=(T-t)b_{X}(t)+ s_{X}(t).
 \\
\end{multline*}
In general, for two generic driftless scalar real processes \(Y(t)  \) and \(Z(t)\), four real constants \(a,b,c,f\), two \(n\)-dimensional vector volatility deterministic real processes \(\{s_{1}(t)\}_{t\geq0}\) and \(\{s_{2}(t)\}_{t\geq0}\), and for an \(n\)-dimensional Brownian motion \(\{W(t)\}_{t\geq0}\) with independent components, we can assume the following dynamic equations:\begin{multline*}
\\
dY(t)=(as_{1}(t)+bs_{2}(t))\cdot dW(t)
 \\
 dZ(t)=(cs_{1}(t)+fs_{2}(t))\cdot dW(t).
 \\
\end{multline*}We drop the time dependency to make the notation lighter and write the above as a sum of component-by-component products. The Brownian motion differential components are denoted by \(dW_{i}\), while the single volatility components are denoted by \( s^{i}_{1} \) and \( s^{i}_{2} \):
\begin{multline*}
\\
dY=a\sum^{n}_{i=1}s^{i}_{1}dW_{i}+b\sum^{n}_{i=1}s^{i}_{2}dW_{i}
\\
dZ=c\sum^{n}_{i=1}s^{i}_{1}dW_{i}+f\sum^{n}_{i=1}s^{i}_{2}dW_{i}.
\\
\end{multline*}
We substitute the single components using the total variance technique proposed above,   writing \([s^{i}_{1}]^{2}=v_{1}[w^{i}_{1}]^{2}\) and \([s^{i}_{2}]^{2}=v_{2}[w^{i}_{2}]^{2}\):\begin{multline*}
\\
dY=a\sum^{n}_{i=1}v_{1}^{\frac{1}{2}}[w^{i}_{1}]dW_{i}+b\sum^{n}_{i=1}v_{2}^{\frac{1}{2}}[w^{i}_{2}]dW_{i}
\\
dZ=c\sum^{n}_{i=1}v_{1}^{\frac{1}{2}}[w^{i}_{1}]dW_{i}+f\sum^{n}_{i=1}v_{2}^{\frac{1}{2}}[w^{i}_{2}]dW_{i}.
\\
\end{multline*}

We want to write the instantaneous correlation between \(dY(t)\) and \(dZ(t)\), written as 
\begin{displaymath}
\rho_{dY(t),dZ(t)}(t)=\frac{\left\langle dY(t),dZ(t)\right\rangle}{[\left\langle dY(t),dY(t) \right\rangle\left\langle dZ(t),dZ(t) \right\rangle]^{\frac{1}{2}}}.
\end{displaymath}
By doing the calculations and thanks to the independence of the components of the Brownian motion, one gets:\begin{displaymath}
\left\langle dY,dZ\right\rangle=\left[ a\sum^{n}_{i=1}v_{1}^{\frac{1}{2}}[w^{i}_{1}]dW_{i}+b\sum^{n}_{i=1}v_{2}^{\frac{1}{2}}[w^{i}_{2}]dW_{i} \right]\left[ c\sum^{n}_{i=1}v_{1}^{\frac{1}{2}}[w^{i}_{1}]dW_{i}+f\sum^{n}_{i=1}v_{2}^{\frac{1}{2}}[w^{i}_{2}]dW_{i} \right]=
\end{displaymath}

\begin{displaymath}
(af+bc)v_{1}^{\frac{1}{2}}v_{2}^{\frac{1}{2}}\sum^{n}_{i=1}w^{i}_{1}w^{i}_{2}dt+acv_{1}\sum^{n}_{i=1}[w^{i}_{1}]^{2}dt+bfv_{2}\sum^{n}_{i=1}[w^{i}_{2}]^{2}dt=\left\{ (af+bc)v_{1}^{\frac{1}{2}}v_{2}^{\frac{1}{2}}\sum^{n}_{i=1}w^{i}_{1}w^{i}_{2}+acv_{1}+bfv_{2} \right\}dt.
\end{displaymath}
For the denominator terms one writes similarly:\begin{displaymath}
\left\langle dY,dY\right\rangle=\left[ a\sum^{n}_{i=1}v_{1}^{\frac{1}{2}}[w^{i}_{1}]dW_{i}+b\sum^{n}_{i=1}v_{2}^{\frac{1}{2}}[w^{i}_{2}]dW_{i} \right]\left[ a\sum^{n}_{i=1}v_{1}^{\frac{1}{2}}[w^{i}_{1}]dW_{i}+b\sum^{n}_{i=1}v_{2}^{\frac{1}{2}}[w^{i}_{2}]dW_{i} \right]=
\end{displaymath}\begin{displaymath}
(a^{2}v_{1}+b^{2}v_{2})dt+2abv_{1}^{\frac{1}{2}}v_{2}^{\frac{1}{2}}\sum^{n}_{i=1}w^{i}_{1}w^{i}_{2}dt
\end{displaymath}
and
\begin{displaymath}
\left\langle dZ,dZ\right\rangle=\left[ c\sum^{n}_{i=1}v_{1}^{\frac{1}{2}}[w^{i}_{1}]dW_{i}+f\sum^{n}_{i=1}v_{2}^{\frac{1}{2}}[w^{i}_{2}]dW_{i} \right]\left[ c\sum^{n}_{i=1}v_{1}^{\frac{1}{2}}[w^{i}_{1}]dW_{i}+f\sum^{n}_{i=1}v_{2}^{\frac{1}{2}}[w^{i}_{2}]dW_{i} \right]=
\end{displaymath}\begin{displaymath}
(c^{2}v_{1}+f^{2}v_{2})dt+2cfv_{1}^{\frac{1}{2}}v_{2}^{\frac{1}{2}}\sum^{n}_{i=1}w^{i}_{1}w^{i}_{2}dt.
\end{displaymath}One  therefore writes:\begin{displaymath}
\rho_{dY(t),dZ(t)}(t)=\frac{(af+bc)v_{1}^{\frac{1}{2}}v_{2}^{\frac{1}{2}}\sum^{n}_{i=1}w^{i}_{1}w^{i}_{2}+acv_{1}+bfv_{2}}{[(a^{2}v_{1}+b^{2}v_{2})+2abv_{1}^{\frac{1}{2}}v_{2}^{\frac{1}{2}}\sum^{n}_{i=1}w^{i}_{1}w^{i}_{2}]^{\frac{1}{2}}[(c^{2}v_{1}+f^{2}v_{2})+2cfv_{1}^{\frac{1}{2}}v_{2}^{\frac{1}{2}}\sum^{n}_{i=1}w^{i}_{1}w^{i}_{2}]^{\frac{1}{2}}}.
\end{displaymath}
It is clear that the above generic parametrisation is a slight simplification of the format of all SDEs in the CTCB model, and therefore can be used as a general framework (the simplification is that in the above example for clarity we have assumed only two model volatility functions \(s_{1}(t)\) and \(s_{2}(t)\), while the CTCB has four, namely \(b_{I}(t)\), \(b_{X}(t)\), \(s_{I}(t)\), and \(s_{X}(t)\)).
Here we notice that, known some model parameters \(a,b,c,f\)
and the total variances \(v_{1}\) and \(v_{2}\)  from the previous calibration step, one  chooses the weights \(w^{i}_{1}\) and  \(w^{i}_{2}\) to target a specific correlation level.

For example, to get \(\rho^{MKT}_{dn(t),d  I(t)/  I(t)}(t)\), one writes:
  \begin{displaymath}
\\ \rho^{MKT}_{dn(t),d  I(t)/  I(t)}(t)=\frac{\left\langle \sigma_{n}(t)\cdot dW(t),[(T-t)b_{I}(t)+s_{ I}(t,T)]\cdot dW(t)\right\rangle}{[\left\langle \sigma_{n}(t)\cdot dW(t),\langle \sigma_{n}(t)\cdot dW(t)\right\rangle\left\langle [(T-t)b_{I}(t)+s_{ I}(t,T)]\cdot dW(t),[(T-t)b_{I}(t)+s_{ I}(t,T)]\cdot dW(t)\right\rangle]^{\frac{1}{2}}}
\end{displaymath}
By doing the calculations one gets to the final result. 

This example shows that all  model correlations can be computed in closed form as a function of the known model parameters and the unknown model volatilities weights
\(w^{i}_{b_{I}(t)}\), \(w^{i}_{b_{X}(t)}\), \(w^{i}_{s_{I}(t)}\), and \(w^{i}_{s_{X}(t)}\).

The non-linear optimisation problem can be formalised as follows:

\begin{equation}
\min\sum_{a(t)\in \mathcal{V}}\;\sum_{b(t)\in \mathcal{V}, b(t)\neq a(t)}[\rho^{MKT}_{a(t),b(t)}(t,w^{i}_{b_{I}(t)}, w^{i}_{b_{X}(t)}, w^{i}_{s_{I}(t)}, w^{i}_{s_{X}(t)})-\rho^{MOD}_{a(t),b(t)}(t)]^{2}
\end{equation}
  under the constrains:  \(\sum^{n}_{i=1}[w^{i}_{b_{I}(t)}]^{2}=1\), \(\sum^{n}_{i=1}[w^{i}_{b_{X}(t)}]^{2}=1\), \(\sum^{n}_{i=1}[w^{i}_{s_{I}(t)}]^{2}=1\), \(\sum^{n}_{i=1}[w^{i}_{s_{X}(t)}]^{2}=1\).
\section{At-the-money calibration results }
\subsection{Technical assumptions}
We make some operational assumptions to deal with the data, and they are not to be considered part of the core model construction; however we make explicit here. In general, when making choices, we assume we want to maximise the calibration accuracy for pricing purposes, as a market maker would do.\begin{enumerate}
\item We assume that all model functions \(b_{I}(t)\), \(s_{I}(t)\), \(a_{I}(t)\), \(b_{X}(t)\), \(s_{X}(t)\), and  \(a_{X}(t)\)   as step functions, where the discontinuities are located at the quoted maturities.  \item We linearly interpolate the market observables at equally spaced time steps, where the time interval is one year. The market observables are the nominal interest curve, the inflation zero-coupon curve, the prices of at-the-money caplets, and the prices of  at-the-money zero-coupon inflation options. \item  At-the-money caplets are not directly traded in the market, but are  recovered as differences between the PV of the  at-the-money caps of two maturities. \item The prices of zero-coupon inflation options are not quoted for  at-the-money strikes but for fixed strikes, therefore a second linear interpolation across strikes is done for each maturity.\item We assume that the market prices of risk are constant and equal to zero for all components: therefore one implies the risk-neutral paths for the expected inflation and growth rate.\item We calibrate inflation options first and then nominal caplets, by keeping the function \(b_{X}(t) \) constant. Therefore we use the \textquotedblleft alternative strategy" detailed in the previous section 17.2.
\item The dimensionality of the driving
Brownian motion is 3. The choice appears to be a good compromise between model simplicity and calibration flexibility.
\item For the zero-finding routine, we used Newton's method with maximum 5,000 iterations and absolute price difference tolerance of 0.00000001.\item The integrals such as the ones in \ref{ExampleHWInterval}  are calculated using the rectangles method with a time interval of 0.01 years.
\item The weights used in the correlation targeting step to allocate the variance between the different components of the noise source are assumed to be constant over time.
\item The correlations assumed are: -60\% for interest rates/inflation, -60\% for interest rates/growth, and 70\%\ for inflation/growth; they are chosen by following standard economic theory. Higher interest rates reduce growth and inflation. Higher growth normally brings about higher inflation.
\end{enumerate}
\subsection{Economic assumptions}
We made the following assumptions regarding the static model parameters.
\\

\begin{tabular}{|c|c|c|}\hline
Parameter & Level \\\hline
\(\delta\) & 0.05  \\\hline
\(h_{P}\) & 1.75  \\\hline
\(h_{X}\)& 2.5 \\\hline
\(\bar p\) & 2\%\ \\\hline
\(\bar x\) & 2\%\ \\\hline
\(\Omega\)& 5  \\\hline
\end{tabular}
\begin{displaymath}
\end{displaymath}
This model parametrisation is certainly subjective, however it reflects our view that the European Central Bank (ECB) under governor Draghi is attaching more importance to  growth than  inflation (therefore \(h_{X}>h_{P}\)) in  the last years. The ECB's official inflation target is 2\%, and it is consensus between economists that the long term growth rate of a developed economy should be around 2\%: hence we set \(\bar  p=0.02\) and  \(\bar  x=0.02\).
 Finally, up to 2012 the ECB had no tradition of quantitative easing on long maturities (like, for example, the FED): therefore we cap the maturity of the instruments used for monetary policy to 5 years (\(\Omega=5\)). The choice of  \(\delta\) has been made as follows: in the Hull-White model, this parameter is the product between the long-term equilibrium level for the short interest rate and the adjustment speed. Because interest rates are at historic lows, we assume  a much higher equilibrium level at 4\%: further, an acceptable adjustment speed is 1.25, therefore yielding \(\delta=0.05\).
To ensure the stability of the calibration, these parameters have been shocked and the model was recalibrated satisfactorily.\subsection{Market data} We calibrate the model to the European inflation market as of 7th December 2012 (data show below).
\\

\begin{tabular}{|c|c|c|c|c|}\hline Maturity (years)  & Nominal IR  & Inflation ZC B/E  & ATM Caplet PV  & ATM ZC Infl. Option PV \\\hline 1  & 0.0022  & 0.0152  & 0.0007  & 0.0039 \\\hline 2  & 0.0026  & 0.016  & 0.0017  & 0.0086 \\\hline 3  & 0.0045  & 0.0163  & 0.0044  & 0.0147 \\\hline 4  & 0.0063  & 0.0166  & 0.0055  & 0.0234 \\\hline 5  & 0.0081  & 0.017  & 0.0076  & 0.0317 \\\hline 6  & 0.01  & 0.0173  & 0.0094  & 0.0402 \\\hline 7  & 0.0118  & 0.0176  & 0.0108  & 0.0483 \\\hline 8  & 0.0136  & 0.0182  & 0.0119  & 0.0594 \\\hline 9  & 0.0152  & 0.0189  & 0.0127  & 0.0696 \\\hline 10  & 0.0168  & 0.0195  & 0.0134  & 0.079 \\\hline \end{tabular}
\subsection{Correlation targeting}
Correlation targeting has been has been achieved by finding the variance  weights
\(w^{i}_{b_{I}(t)}\), \(w^{i}_{b_{X}(t)}\), \(w^{i}_{s_{I}(t)}\), and \(w^{i}_{s_{X}(t)}\) under  constraints. Given that the dimension of the  Brownian motion is 3, we have to find 8 weights (2 weights for each function, given that the third is calculated from the request that the sum of their squares has to be 1). We assumed that the weights are constant over time.
The result of the numerical optimisation is:
\\

\begin{tabular}{|c|c|c|c|}\hline
 & \textit{i=}1 & \textit{i=}2 & \textit{i=}3 \\\hline
\(w^{i}_{b_{I}(t)}\) & 0.20285 & 0.13219 & 0.97024 \\\hline
\(w^{i}_{b_{X}(t)}\) & -0.95101 & -0.02865 & 0.30781 \\\hline
 \(w^{i}_{s_{I}(t)}\) & 0.14035 & 0.10000 & 0.98503 \\\hline
\(w^{i}_{s_{X}(t)}\)  & 0.85195 & -0.07168 & 0.51868 \\\hline
\end{tabular}
 \begin{displaymath}
\end{displaymath}
Interestingly the weights for \(b_{I}(t)\) and \(s_{I}(t)\) have the same sign across  the 3 components, which is  consistent with the original idea of the DSGE\ macroeconomic model, i.e.  inflation depends heavily on its expectations. Instead, the first component shows different signs for  \(b_{X}(t)\) and \(s_{X}(t)\), which is consistent with the idea of productivity shocks, i.e. that the growth expectations can differ from realised growth rates.   \subsection{Results} The following model parameters have been found for the price index processes:
\\

\begin{tabular}{|c|c|c|c|c|c|c|c|}\hline Maturity (years)  & \(b_{I}^{1}(t)\)  &   \(b_{2}^{2}(t)\) &  \(b_{3}^{3}(t)\)  &  \(s_{I}^{1}(t)\)  & \(s_{I}^{2}(t)\)   & \(s_{I}^{3}(t)\)   & \(a_{I}(t)\) \\\hline 1  & -0.000269  & 0.000314  & 0.000911  & 0.000273  & 0.005481  & 0.007732  & -0.000059 \\\hline 2  & -0.000269  & 0.000314  & 0.000911  & 0.000512  & 0.010292  & 0.014518  & 0.001543 \\\hline 3  & -0.000269  & 0.000314  & 0.000911  & 0.000783  & 0.015726  & 0.022182  & -0.000174 \\\hline 4  & -0.000269  & 0.000314  & 0.000911  & 0.001167  & 0.023437  & 0.03306  & 0.000697 \\\hline 5  & -0.000269  & 0.000314  & 0.000911  & 0.001317  & 0.026467  & 0.037333  & 0.00026 \\\hline 6  & -0.000269  & 0.000314  & 0.000911  & 0.001461  & 0.029347  & 0.041396  & 0.00016 \\\hline 7  & -0.000269  & 0.000314  & 0.000911  & 0.001517  & 0.030486  & 0.043003  & 0.00043 \\\hline 8  & -0.000269  & 0.000314  & 0.000911  & 0.001919  & 0.038553  & 0.054382  & 0.003361 \\\hline 9  & -0.000269  & 0.000314  & 0.000911  & 0.001895  & 0.038063  & 0.053691  & 0.000545 \\\hline 10  & -0.000269  & 0.000314  & 0.000911  & 0.001803  & 0.036218  & 0.051088  & 0.00151 \\\hline \end{tabular}
\begin{displaymath}
\end{displaymath}
The following model parameters have been found for the GDP processes:
\\

\begin{tabular}{|c|c|c|c|c|c|c|c|}\hline Maturity (years)  & \(b_{X}^{1}(t)\)  & \(b_{X}^{2}(t)\)  & \(b_{X}^{3}(t)\)  &  \(s_{X}^{1}(t)\)  &  \(s_{X}^{2}(t)\) & \(s_{X}^{3}(t)\)  &  \(a_{X}(t)\) \\\hline 1  & -0.003463  & -0.000807  & 0.001562  & 0.009875  & 0.000464  & 0.001507  & -0.009337 \\\hline 2  & -0.008657  & -0.002016  & 0.003905  & 0.009875  & 0.000464  & 0.001507  & -0.015211 \\\hline 3  & -0.024641  & -0.005739  & 0.011115  & 0.009875  & 0.000464  & 0.001507  & -0.008325 \\\hline 4  & -0.022408  & -0.005219  & 0.010107  & 0.009875  & 0.000464  & 0.001507  & -0.013803 \\\hline 5  & -0.038067  & -0.008867  & 0.01717  & 0.009875  & 0.000464  & 0.001507  & -0.013651 \\\hline 6  & -0.045838  & -0.010677  & 0.020675  & 0.009875  & 0.000464  & 0.001507  & -0.013985 \\\hline 7  & -0.049767  & -0.011592  & 0.022448  & 0.009875  & 0.000464  & 0.001507  & -0.013004 \\\hline 8  & -0.053907  & -0.012556  & 0.024315  & 0.009875  & 0.000464  & 0.001507  & -0.015206 \\\hline 9  & -0.05733  & -0.013353  & 0.025859  & 0.009875  & 0.000464  & 0.001507  & -0.012411 \\\hline 10  & -0.05733  & -0.013353  & 0.025859  & 0.009875  & 0.000464  & 0.001507  & -0.009437 \\\hline \end{tabular}
\\

In all cases the absolute calibration error has been below the  threshold
of 0.0000001.
\section{Applications } 
\subsection{Derivatives risk as a function of the central bank reaction function }We price a 2\% zero-coupon inflation cap with 10 years maturity and 1 EUR notional in the CTCB model\ and then shock the central bank reaction function parameters. In this way we assess the impact of a sudden (and  not hedgeable) change in the central bank reaction function (or, more practically, of a new president of the central bank who may have different views compared to the current one). 

In particular we find that inflation delta (defined as the change in PV when the inflation curve is shifted up by 1 basis point) is not sensitive  to the central bank reaction function parameters (we shock separately the parameters \(h_{P}\) and \(h_{X}\) by 0.5 and in both cases the inflation delta stays at 0.04447): this is expected as the sensitivity of an inflation claim to inflation should mainly depend on the inflation level and not by the central bank reaction function. 
 \subsection{Inflation book macro-hedging in the CTCB model}Let us assume that an investment bank has sold a low strike inflation floor, which is a popular hedge against deflation: for example a macro hedge fund may want to buy protection against a deflation scenario.  This trade would probably make good margin for the bank, given the relative low liquidity of low strike inflation options, however this would expose the bank to a significant downside risk that is difficult to recycle. An option for the bank would be to buy a nominal interest rates floor, given that this market is much more liquid than the inflation options market: the idea would be that in a low inflation environment interest rates would go down, therefore making money on the long nominal interest rates hedge while losing on the short inflation client trade. Investment banks use different models to price nominal rates  and inflation trades, and  the decision on the amount of nominal hedge to buy to offset the short inflation position is  taken in a very informal  way. This can lead to significant losses due to model risk.
We argue that one of the advantages of the CTCB model is that it offers a global representation of the economy and allows consistent pricing of interest rates and inflation trades with no ambiguity: this is because this hedging problem boils down to how the central bank can affect the nominal yield curve given a deflationary scenario.  

For example, one  uses the calibrated  CTCB model to run a Monte Carlo simulation over the maturity of the inflation client trade. One selects  the paths where inflation has gone down enough for the short client trade to be in the money, and obtains a conditional distribution for the forward Libor rates given the inflation decrease: by pricing nominal floors in these scenarios, the trader can assess what nominal rates strikes are best used to hedge a deflationary scenario, choose the cheapest strikes, and, most importantly, calculate a scenario-driven hedge ratio. We stress that this example is not a pricing application, and therefore there is  some profit and loss volatility during the life of the trade, as we are simply hedging a deflationary scenario that may not materialise in the end: however, we think that this methodology would help the trader to  macro-hedge his inflation book in a way that is consistent with his view of the economy and with no model bias, given that the same model would be used to price the inflation client trade and the interest rates  macro hedge.
\subsection{Stress testing in the CTCB model} In recent years regulators have increasingly requested  financial institutions to run stress tests: for example, the FED has  introduced the CCAR in late 2010
(Comprehensive Capital Analysis and Review). One of the challenges that  financial institutions had to face was how to convert the market moves observed in the market and in the economy into the model parameter. Because the CTCB model takes the economy as an input, the economic shocks can be easily taken as an input and the model   answers  the questions asked by regulators: there is no need to shock model parameters like volatilities, given that the model calibration  delivers the new set of parameters that  fit to the stressed economic conditions.  

\section{Conclusions  }In this article we have proposed a continuous-time model inspired by the DSGE model
to price inflation derivatives: in the recent years the debate around the central bank behaviour towards growth and inflation as been a pivotal topic, and the central bank reaction function is an essential ingredient in our model. 

Our modelling choice  offers new perspectives on inflation securities pricing and at the same time is extremely tractable, providing closed form solutions for the most common inflation and interest rates payoffs. Interestingly, we have showed that the short rate behaves according to the well-established Hull-White model. This makes the model calibration much simpler: we proposed a separable calibration strategy.

The good model tractability does not prevent one to use it for many pricing and risk applications, including stress tests and macro-hedging.

We are working to extend this framework
in the following directions:\begin{enumerate}
\item 
Calibrating the model to inflation and interest rates options smiles;\item Introducing credit risk and a multi-curve setting in this model;\item 
Using this model to calculate counterparty and funding adjustments for OTC derivatives. \end{enumerate}

\end{document}